\documentclass[12pt,a4paper]{article}
\usepackage{amssymb}
\usepackage{amsmath}
\usepackage{latexsym}
\usepackage{cite}
\usepackage{physics}
\begin{document}

\title{\vspace{-1cm}\bf Nonlinear corrections in the quantization\\ of a weakly nonideal Bose gas\\ at zero temperature}

\author{
Mikhail~N.~Smolyakov
\\
{\small{\em Skobeltsyn Institute of Nuclear Physics, Lomonosov Moscow
State University,
}}\\
{\small{\em Moscow 119991, Russia}}}

\date{}
\maketitle

\begin{abstract}
In the present paper, quantization of a weakly nonideal Bose gas at zero temperature along the lines of the well-known Bogolyubov approach is performed. The analysis presented in this paper is based, in addition to the steps of the original Bogolyubov approach, on the use of nonoscillation modes (which are also solutions of the linearized Heisenberg equation) for recovering the canonical commutation relations in the linear approximation, as well as on the calculation of the first nonlinear correction to the solution of the linearized Heisenberg equation which satisfies the canonical commutation relations at the next order. It is shown that, at least in the case of free quasi-particles, consideration of the nonlinear correction automatically solves the problem of nonconserved particle number, which is inherent to the original approach.
\end{abstract}

\section{Introduction}
In the famous paper \cite{Bogolyubov}, N.N.~Bogolyubov proposed an explanation of the superfluidity phenomenon at microscopic level, which was based on the effects produced by collective excitations arising in a quantized weakly nonideal Bose gas at zero temperature. Later, the ideas of Bogolyubov were extensively developed, and the amount of scientific literature on this subject is very large. In particular, in reviews \cite{DGPS,Review,Andersen:2003qj} and books \cite{Pitaevskii3,Pethick,Kvasnikov} one can find a description of many applications of the Bogolyubov approach, including a development of the corresponding perturbation theory, as well as many relevant references.

It is well known that the Bogolyubov approach results in nonconservation of the particle number in the effective theory describing quasi-particles. It is universally accepted that this is a consequence of the approximation used, i.e., of the replacement of the creation and annihilation operators of condensed particles by $c$-numbers. Within the standard approach, there exist at least two main ways to solve this problem, they will be briefly mentioned below (a detailed discussion of these solutions can be found, for example, in \cite{Review}). But in these methods for the particle number conservation, some nondiagonal quadratic parts of the effective Hamiltonian turn out to be, in fact, just skipped (for example, by incorporating them into the background energy of the system). However, there exists another one way --- as will be explicitly demonstrated below, nonconservation of the particle number in the canonical ensemble is due to the fact that, in the case of a stationary time-dependent background solution, the use of only the linear approximation within the second quantization formalism is not sufficient for a correct description of even free quasi-particles. Meanwhile, the use of the first nonlinear correction, together with the use of the additional nonoscillation modes in the linear approximation (which are also solutions of the linearized Heisenberg equation), allows one to modify the Bogolyubov approach and to solve its problems keeping all the key steps and ideas proposed in \cite{Bogolyubov}. The main idea is quite simple --- the terms (zero order)$\times$(first nonlinear correction) provide contribution of the same order as the terms (linear order)$\times$(linear order), thus compensating the unwanted terms which arise when one calculates the operator of the particle number using only the linear approximation. In other words, there exists a {\em different} solution of the standard Heisenberg equation such that the particle number is conserved automatically for this solution, which means that no extra methods are necessary to maintain the particle number conservation in the effective theory describing quasi-particles.\footnote{Another interesting approach to deal with a system with a well-defined number of particles was proposed in \cite{Castin}.} As an additional advantage, with this solution one gets a more accurate fulfillment of the canonical commutation relations. Of course, the solution correctly describes free quasi-particles in the effective theory, but it may lead to a slightly different interaction Hamiltonian in comparison with the standard one.

In the present paper, only the setup, main assumptions and second quantization formalism that were used in the original paper \cite{Bogolyubov} are considered as a starting point. The paper is organized as follows. In Section~2 the basic setup is presented in the notations that will be used throughout the paper. In Section~3 the Bogolyubov approach is briefly discussed. In Section~4 recovering of the canonical commutation relations for the linear part of perturbations is discussed. In Section~5 nonconservation of the particle number in the case of the classical nonlinear Schr\"{o}dinger equation is discussed and it is demonstrated how the use of the first nonlinear correction recovers the particle number conservation. In Section~6 the quantum case of a weakly nonideal Bose gas is considered  along the lines of the Bogolyubov approach, but with the first nonlinear correction taken into account, including the calculation of all necessary commutators. In Section~7 all the results obtained in Section~6 are combined and the operators of the integrals of motion in their final form are presented. The obtained results are discussed in Conclusion. The auxiliary materials are collected in four appendixes.

\section{Setup and equations of motion}
Within the second quantization formalism, let us consider a system with the Hamiltonian \cite{Bogolyubov}
\begin{equation}\label{initHamiltonian}
\hat H=\int d^{3}x\left(\frac{\hbar^{2}}{2m}\sum\limits_{i=1}^{3}\partial_{i}\hat\Psi^{\dagger}(t,\vec x)\partial_{i}\hat\Psi(t,\vec x)+\frac{1}{2}\int d^{3}y\hat\Psi^{\dagger}(t,\vec x)\hat\Psi^{\dagger}(t,\vec y)V(|\vec x-\vec y|)\hat\Psi(t,\vec x)\hat\Psi(t,\vec y)\right),
\end{equation}
where the operators $\hat\Psi(t,\vec x)$ and $\hat\Psi^{\dagger}(t,\vec x)$ are supposed to satisfy the standard canonical commutation relations
\begin{eqnarray}\label{commrel1}
&&[\hat\Psi(t,\vec x),\hat\Psi^{\dagger}(t,\vec y)]=\delta^{(3)}(\vec x-\vec y),\\\label{commrel2}
&&[\hat\Psi(t,\vec x),\hat\Psi(t,\vec y)]=0.
\end{eqnarray}
It is clear that with \eqref{commrel2} the relation $[\hat\Psi^{\dagger}(t,\vec x),\hat\Psi^{\dagger}(t,\vec y)]=0$ is satisfied automatically.

In what follows, the interaction potential of form \cite{Fetter}
\begin{equation}\label{interactpotent}
V(|\vec x-\vec y|)=g\delta^{(3)}(\vec x-\vec y),
\end{equation}
where $g>0$, will be used. This simplest form of the interaction potential was selected just to reduce the amount of calculations that are necessary for obtaining the explicit solution for the first nonlinear correction, which turns out to be rather bulky even in this case, and to present the main steps of the analysis in a simpler manner. Meanwhile, the method that will be discussed below can be applied to the case of a general $V(|\vec x-\vec y|)$ too. Although with potential \eqref{interactpotent} the system suffers from disintegration of quasi-particles \cite{Desintegration1,Desintegration2}, it is not important for the subsequent analysis. Thus, with \eqref{interactpotent} the Hamiltonian takes the form
\begin{equation}\label{initHamiltonian2}
\hat H=\int d^{3}x\left(\frac{\hbar^{2}}{2m}\sum\limits_{i=1}^{3}\partial_{i}\hat\Psi^{\dagger}(t,\vec x)\partial_{i}\hat\Psi(t,\vec x)+\frac{g}{2}\hat\Psi^{\dagger}(t,\vec x)\hat\Psi^{\dagger}(t,\vec x)\hat\Psi(t,\vec x)\hat\Psi(t,\vec x)\right).
\end{equation}
The operator of the particle number is just
\begin{equation}
\hat N=\int d^{3}x\hat\Psi^{\dagger}(t,\vec x)\hat\Psi(t,\vec x).
\end{equation}
For $\hat H$ defined by \eqref{initHamiltonian2} and with \eqref{commrel1}, \eqref{commrel2}, the Heisenberg equation
\begin{equation}\label{Heiseq}
\frac{d\hat\Psi(t,\vec x)}{dt}=\frac{i}{\hbar}[\hat H,\hat\Psi(t,\vec x)]
\end{equation}
leads to the well-known Gross-Pitaevskii equation \cite{Gross,Pitaevskii1} for the operator $\hat\Psi(t,\vec x)$:
\begin{equation}\label{eqmotgeneral}
i\hbar\dot{\hat\Psi}(t,\vec x)=-\frac{\hbar^{2}}{2m}\Delta\hat\Psi(t,\vec x)+g\hat\Psi^{\dagger}(t,\vec x)\hat\Psi(t,\vec x)\hat\Psi(t,\vec x),
\end{equation}
where $\dot{}=\frac{\partial}{\partial t}$ and $\Delta=\sum\limits_{i=1}^{3}\partial_{i}^{2}$. As in the original paper \cite{Bogolyubov}, from hereon let us suppose that the system is placed into a spatial ``box'' of size $L\times L\times L$ with periodic boundary conditions for $\hat\Psi(t,\vec x)$ and $\hat\Psi^{\dagger}(t,\vec x)$.

\section{Bogolyubov approach and its problems}\label{section3}
As suggested in \cite{Bogolyubov}, because of the large occupation number the creation and annihilation operators of condensed particles can be replaced by $c$-numbers, which corresponds to the following classical solution of Eq.~\eqref{eqmotgeneral}
\begin{equation}\label{backgrsol0}
\Psi_{0}(t)=\sqrt{\frac{\omega}{g}}e^{-\frac{i}{\hbar}\omega t}
\end{equation}
with $\omega>0$. In these notations, the energy and the particle number of the background solution have the form
\begin{align}\label{backgrE}
&E_{0}=\frac{\omega^{2}L^{3}}{2g},\\\label{backgrN}
&N_{0}=\frac{\omega L^{3}}{g}.
\end{align}
Now let us represent $\hat\Psi(t,\vec x)$ as
\begin{equation}\label{linearization}
\hat\Psi(t,\vec x)=\Psi_{0}(t)+e^{-\frac{i}{\hbar}\omega t}\hat\varphi(t,\vec x)=e^{-\frac{i}{\hbar}\omega t}\left(\sqrt{\frac{\omega}{g}}+\hat\varphi(t,\vec x)\right).
\end{equation}
Substituting this representation into Eq.~\eqref{eqmotgeneral} and keeping only the linear in $\hat\varphi(t,\vec x)$ terms, we get
\begin{equation}\label{lineq}
i\hbar\dot{\hat\varphi}(t,\vec x)=-\frac{\hbar^{2}}{2m}\Delta\hat\varphi(t,\vec x)+\omega\left(\hat\varphi(t,\vec x)+\hat\varphi^{\dagger}(t,\vec x)\right).
\end{equation}
The solution of this equation is well known and has the form
\begin{eqnarray}\nonumber
\hat\varphi(t,\vec x)=
\frac{1}{\sqrt{L^{3}}}\sum\limits_{j\neq 0}\left(c_{j}e^{-\frac{i}{\hbar}(\gamma_{j}t-\vec k_{j}\vec x)}\hat a_{j}-d_{j}e^{\frac{i}{\hbar}(\gamma_{j}t-\vec k_{j}\vec x)}\hat a^{\dagger}_{j}\right)\\\label{linsol}
=\frac{1}{\sqrt{L^{3}}}\sum\limits_{j\neq 0}e^{\frac{i}{\hbar}\vec k_{j}\vec x}\left(c_{j}e^{-\frac{i}{\hbar}\gamma_{j}t}\hat a_{j}-d_{j}e^{\frac{i}{\hbar}\gamma_{j}t}\hat a^{\dagger}_{-j}\right)
\end{eqnarray}
with
\begin{equation}\label{cddef}
c_{j}=\frac{\omega}{\sqrt{2\gamma_{j}\left(\frac{{\vec k_{j}}^{2}}{2m}+\omega-\gamma_{j}\right)}},\qquad d_{j}=\sqrt{\frac{\frac{{\vec k_{j}}^{2}}{2m}+\omega-\gamma_{j}}{2\gamma_{j}}},
\end{equation}
\begin{equation}\label{kdef}
\vec k_{j}=\frac{2\pi\hbar}{L}\vec j,\qquad \vec j=(j_{1},j_{2},j_{3}),
\end{equation}
where $j_{1}$, $j_{2}$, $j_{3}$ are integers and $j_{1}^2+j_{2}^2+j_{3}^2\neq 0$,
\begin{equation}\label{gammadef}
\gamma_{j}=\sqrt{
\frac{{\vec k_{j}}^{2}}{2m}\left(\frac{{\vec k_{j}}^{2}}{2m}+2\omega\right)},
\end{equation}
and
\begin{eqnarray}
&&[\hat a_{j},\hat a^{\dagger}_{l}]=\delta_{jl},\\
&&[\hat a_{j},\hat a_{l}]=0.
\end{eqnarray}
The Bogolyubov transformations are already taken into account in \eqref{linsol}, so the operators $\hat a_{j}$ and $\hat a^{\dagger}_{j}$ correspond to quasi-particles. Although $\vec j$ is a vector, for simplicity the mark of a vector is omitted for the subscript ``$j$''.

Substituting \eqref{linearization} into \eqref{initHamiltonian2}, skipping the background energy $E_{0}$, keeping only the terms that are linear and quadratic in $\hat\varphi$ and $\hat\varphi^{\dagger}$, and using Eq.~\eqref{lineq}, we arrive at the operator of the energy of perturbations, which takes the form
\begin{equation}\label{Epquadr}
\hat E_{p}=\omega\hat N_{p}+\frac{i\hbar}{2}\int d^{3}x\left(\hat\varphi^{\dagger}\dot{\hat\varphi}-\dot{\hat\varphi}^{\dagger}\hat\varphi\right).
\end{equation}
The operator of the particle number of perturbations is defined as
\begin{equation}
\hat N_{p}=\hat N-N_{0}=\int d^{3}x\hat\Psi^{\dagger}(t,\vec x)\hat\Psi(t,\vec x)-\int d^{3}x\Psi_{0}^{*}(t)\Psi_{0}(t),
\end{equation}
leading to
\begin{equation}\label{Npquadr}
\hat N_{p}=\int d^{3}x\left(\sqrt{\frac{\omega}{g}}\left(\hat\varphi^{\dagger}+\hat\varphi\right)+\hat\varphi^{\dagger}\hat\varphi\right).
\end{equation}

Finally, substituting \eqref{linsol} into \eqref{Epquadr} and skipping $c$-number terms which arise when one passes from $\hat a_{j}^{}\hat a_{j}^{\dagger}$ to $\hat a_{j}^{\dagger}\hat a_{j}^{}$, we get
\begin{align}\label{Epquadr2}
&\hat E_{p}=\omega\hat N_{p}+\sum\limits_{j\neq 0}\gamma_{j}\hat a^{\dagger}_{j}\hat a_{j},\\\label{Npquadr2}
&\hat N_{p}=\sum\limits_{j\neq 0}\left(\left(c_{j}^{2}+d_{j}^{2}\right)\hat a^{\dagger}_{j}\hat a_{j}-c_{j}d_{j}\left(e^{2\frac{i}{\hbar}\gamma_{j}t}\hat a^{\dagger}_{j}\hat a^{\dagger}_{-j}+e^{-2\frac{i}{\hbar}\gamma_{j}t}\hat a_{j}\hat a_{-j}\right)\right).
\end{align}
In \cite{Bogolyubov} the term corresponding to $\omega\hat N_{p}$ was incorporated into the background energy of the system (look at the transition $N_{0}\to N$ on page 27 of \cite{Bogolyubov} and formulas (12), (13) on page 26 of \cite{Bogolyubov}), and for the physically relevant contribution the well-known result
\begin{equation}
\hat E_{p}\to \sum\limits_{j\neq 0}\gamma_{j}\hat a^{\dagger}_{j}\hat a_{j}
\end{equation}
was obtained.

However, the approach presented above has two drawbacks, which are well known. First, as a consequence of the approximation used to describe the Bose-Einstein condensate, for representation \eqref{linearization} with solution \eqref{linsol} the canonical commutation relation \eqref{commrel1} is not satisfied. Indeed, because of the absence of the operators $\hat a_{0}$ and $\hat a^{\dagger}_{0}$ in \eqref{linsol}, we get
\begin{equation}\label{CCRviolat}
[\hat\Psi(t,\vec x),\hat\Psi^{\dagger}(t,\vec y)]=\delta^{(3)}(\vec x-\vec y)-\frac{1}{L^{3}}.
\end{equation}
In such a case, formally Eq.~\eqref{eqmotgeneral} does not follow from Eq.~\eqref{Heiseq}.

Second, the quadratic in perturbations particle number \eqref{Npquadr2}, and consequently the Hamiltonian \eqref{Epquadr2}, explicitly depend on time. From the mathematical point of view, it looks strange for a dynamical system with a correct solution of the corresponding equation of motion --- indeed, the relation $\dot{\hat N}=0$ is fulfilled whenever Eq.~\eqref{eqmotgeneral} holds even if the canonical commutation relations are violated. Of course, as usual, one can suppose that at some moment of time, say, at $t=0$
\begin{equation}\label{linsolfixedT}
\hat\Psi(0,\vec x)=\sqrt{\frac{\omega}{g}}+\hat\varphi(0,\vec x),
\end{equation}
and the subsequent evolution of $\hat\Psi(t,\vec x)$ is governed by Eq.~\eqref{eqmotgeneral}. In this case, instead of \eqref{Npquadr2} we get
\begin{equation}\label{Npquadr3}
\hat N_{p}=\sum\limits_{j\neq 0}\left(\left(c_{j}^{2}+d_{j}^{2}\right)\hat a^{\dagger}_{j}\hat a_{j}-c_{j}d_{j}\left(\hat a^{\dagger}_{j}\hat a^{\dagger}_{-j}+\hat a_{j}\hat a_{-j}\right)\right).
\end{equation}
Although result \eqref{Npquadr3} does not depend on time, it is nondiagonal, indicating nonconservation of the number of particles even in a free theory \cite{Review}, which looks rather unnatural from the physical point of view. Moreover, it is not clear whether the commutator $[\hat\Psi(t,\vec x),\hat\Psi^{\dagger}(t,\vec y)]$ is conserved over time and corresponds to \eqref{CCRviolat}, the same is valid for $[\hat\Psi(t,\vec x),\hat\Psi(t,\vec y)]$ (if we choose Eq.~\eqref{Heiseq} with $\hat H$ obtained by substituting \eqref{linsolfixedT} into \eqref{initHamiltonian2} instead of Eq.~\eqref{eqmotgeneral}, we can be sure that the canonical commutation relations are conserved over time for \eqref{linsolfixedT}, but cannot be sure that $\dot{\hat N}=0$ because it is expected that $[\hat H,\hat N]\neq 0$ if the canonical commutation relations are violated).

As was mentioned in Introduction, within the standard approach there exist at least two main solutions of the problem of nonconserved particle number (a more detailed discussion and the corresponding references can be found in Section~2.2 of \cite{Review}). The first solution was proposed by Bogolyubov himself. It is based on the introduction of the new creation and annihilation operators of a special form instead of the original ones, but a certain part of the Hamiltonian containing the creation and annihilation operators corresponding to nonzero momenta still turns out to be, in fact, neglected by incorporating it into the total particle number which is supposed to be a constant in the canonical ensemble. The second solution is based just on the use of a grand-canonical ensemble from the very beginning, which, in our notations, for the case of a free theory leads to $\hat H-\omega\hat N\to \sum_{j\neq 0}\gamma_{j}\hat a^{\dagger}_{j}\hat a_{j}$ with $\omega$ playing the role of a chemical potential because $\mu=\left(\frac{\partial E_{0}}{\partial N_{0}}\right)\Bigl|_{L}=\omega$. However, there exists a different solution of the problems of the Bogolyubov approach, which retains all propositions of \cite{Bogolyubov} but provides conserved integrals of motion at least for free quasi-particles while keeping all the operator parts of the Hamiltonian corresponding to nonzero momenta. Let us start with recovering the canonical commutation relations in the linear approximation.

\section{Recovering the canonical commutation relations for the linear part of perturbations}\label{section4}
Let us look more precisely at Eq.~\eqref{lineq}. Apart from the oscillation modes presented in solution \eqref{linsol}, this equation provides two more solutions, satisfying the periodic boundary conditions in the ``box'', which have the form
\begin{equation}\label{quantmode0}
\hat\varphi_{\textrm{no\,1}}(t,\vec x)=\hat A,\qquad \hat A^{\dagger}=-\hat A,
\end{equation}
\begin{equation}\label{quantmode}
\hat\varphi_{\textrm{no\,2}}(t,\vec x)=\left(\frac{1}{2}-\frac{i}{\hbar}\omega t\right)\hat B,\qquad \hat B^{\dagger}=\hat B.
\end{equation}
In the classical case, the first mode corresponds to the global $U(1)$ symmetry of the theory ($\hat\Psi(t,\vec x)\to e^{i\alpha}\hat\Psi(t,\vec x)$), whereas the second mode corresponds to a change of the frequency $\omega$ of the background solution \eqref{backgrsol0} and is just $\sim\frac{\partial\Psi_{0}(t)}{\partial\omega}$. Let us choose
\begin{eqnarray}\label{quantmode1}
&&\hat A=\frac{1}{2q\sqrt{L^{3}}}\left(\hat a_{0}-\hat a_{0}^{\dagger}\right),\\\label{quantmode2}
&&\hat B=\frac{q}{\sqrt{L^{3}}}\left(\hat a_{0}+\hat a_{0}^{\dagger}\right),
\end{eqnarray}
where $q\neq 0$ is an arbitrary dimensionless real constant and
\begin{eqnarray}
&&[\hat a_{0},\hat a^{\dagger}_{0}]=1,\\
&&[\hat a_{0},\hat a_{j}]=0,\qquad j\neq 0,\\
&&[\hat a_{0},\hat a^{\dagger}_{j}]=0,\qquad j\neq 0
\end{eqnarray}
(it is clear that one can use $\hat a_{0}\to \hat a_{0}e^{i\alpha}$ and $\hat a_{0}^{\dagger}\to \hat a_{0}^{\dagger}e^{-i\alpha}$ with any real $\alpha$ in \eqref{quantmode1} and \eqref{quantmode2}). Now, instead of \eqref{linsol}, let us take
\begin{eqnarray}\nonumber
\hat\varphi(t,\vec x)=
\frac{1}{\sqrt{L^{3}}}\sum\limits_{j\neq 0}\left(c_{j}e^{-\frac{i}{\hbar}(\gamma_{j}t-\vec k_{j}\vec x)}\hat a_{j}-d_{j}e^{\frac{i}{\hbar}(\gamma_{j}t-\vec k_{j}\vec x)}\hat a^{\dagger}_{j}\right)\\\label{linsol2}+\frac{1}{2q\sqrt{L^{3}}}\left(\hat a_{0}-\hat a_{0}^{\dagger}\right)+\frac{q}{\sqrt{L^{3}}}\left(\frac{1}{2}-\frac{i}{\hbar}\omega t\right)\left(\hat a_{0}+\hat a_{0}^{\dagger}\right),
\end{eqnarray}
which satisfies the linear Eq.~\eqref{lineq}. It is easy to check that for \eqref{linsol2}
\begin{eqnarray}\label{CCR01}
&&[\hat\varphi(t,\vec x),\hat\varphi(t,\vec y)]=0,\\\label{CCR02}
&&[\hat\varphi(t,\vec x),\hat\varphi^{\dagger}(t,\vec y)]=\delta^{(3)}(\vec x-\vec y),
\end{eqnarray}
and consequently the canonical commutation relations \eqref{commrel1} and \eqref{commrel2} are satisfied for $\hat\Psi(0,\vec x)$ defined by \eqref{linsolfixedT} with \eqref{linsol2}. Then, with $\hat\Psi(t,\vec x)$ defined by the standard formula
\begin{equation}\label{Heisevol}
\hat\Psi(t,\vec x)=e^{\frac{i}{\hbar}\hat Ht}\hat\Psi(0,\vec x)e^{-\frac{i}{\hbar}\hat Ht}
\end{equation}
following from Eq.~\eqref{Heiseq}, where $\hat H$ is obtained by substituting $\hat\Psi(0,\vec x)$ into \eqref{initHamiltonian2}, the canonical commutation relations \eqref{commrel1} and \eqref{commrel2} are satisfied for any $t$. It means that now Eqs.~\eqref{Heiseq} and \eqref{eqmotgeneral} are equivalent, $[\hat H,\hat N]=0$ and we get a self-consistent theory.

Let us briefly discuss solution \eqref{linsol2}. First, one should not worry about the linear in $t$ term in \eqref{linsol2} that is coming from \eqref{quantmode}. This term does not indicate any instability, it is just the first term of the decomposition of $\Psi_{0}(t,\omega+{\scriptstyle\triangle}\omega)-\Psi_{0}(t,\omega)$, whereas the latter expression is obviously bounded. Second, it is clear that the spectra of the operators $\hat a_{0}-\hat a_{0}^{\dagger}$ and $\hat a_{0}+\hat a_{0}^{\dagger}$ are continuous. Indeed, one can represent these operators in the standard form as
\begin{eqnarray}\label{aPZ}
&&\hat a_{0}=\frac{1}{\sqrt{2}}\left(\hat Z+i\hat P\right),\\\label{adagPZ}
&&\hat a^{\dagger}_{0}=\frac{1}{\sqrt{2}}\left(\hat Z-i\hat P\right)
\end{eqnarray}
with the operators $\hat Z$ and $\hat P$ defined as $\hat Z=Z$ and $\hat P=-i\frac{\partial}{\partial Z}$. In this case, $\hat a_{0}+\hat a^{\dagger}_{0}=\sqrt{2}\hat Z$ and $\hat a_{0}-\hat a^{\dagger}_{0}=\sqrt{2}i\hat P$. The eigenvalues and eigenfunctions of these operators are well-known and have the form
\begin{eqnarray}
&&\hat Z\delta(Z-Z_{0})=Z_{0}\delta(Z-Z_{0}),\\
&&\hat P\left(\frac{1}{\sqrt{2\pi}}e^{iP_{0}Z}\right)=P_{0}\left(\frac{1}{\sqrt{2\pi}}e^{iP_{0}Z}\right).
\end{eqnarray}
Substituting \eqref{linsol2} into \eqref{Npquadr} and keeping only the linear in $\hat a_{0}-\hat a_{0}^{\dagger}$ and $\hat a_{0}+\hat a_{0}^{\dagger}$ terms, one gets simply
\begin{equation}\label{Npomegaquant}
\hat N_{p}^{(\hat a_{0},\hat a_{0}^{\dagger})}=\sqrt{\frac{\omega L^{3}}{g}}\,q\left(\hat a_{0}+\hat a_{0}^{\dagger}\right).
\end{equation}
At the classical level this contribution corresponds to
\begin{equation}\label{Npomegaclass}
N_{p}^{({\scriptscriptstyle\triangle}\omega)}=\frac{dN_{0}}{d\omega}{\scriptstyle\triangle}\omega,
\end{equation}
i.e., to a change of the frequency $\omega$ of the background solution $\Psi_{0}(t)$ (this topic will be discussed in more detail later). That is why the value of the parameter $q$ is not important --- one can choose the state corresponding to any necessary eigenvalue of the operator $\hat a_{0}+\hat a_{0}^{\dagger}$, providing, correspondingly, any necessary value for the eigenvalue in $\sqrt{\frac{\omega L^{3}}{g}}\,q\left(\hat a_{0}+\hat a_{0}^{\dagger}\right)\ket{Z_{0}}=\sqrt{\frac{2\omega L^{3}}{g}}\,q Z_{0}\ket{Z_{0}}$. A remarkable feature of \eqref{Npomegaquant} is that the quantum mode \eqref{quantmode} with \eqref{quantmode2} allows one to get a classical behavior exactly where it is necessary.

It should be necessarily mentioned that the idea to use nonoscillation modes for providing correct canonical commutation relations for fields is not new. A fully analogous procedure was used in \cite{Creutz} for quantization of extended objects of the soliton type (see also \cite{Rajaraman:1982is} and references therein for discussion of the soliton quantization in relativistic scalar field theories). The only difference with the approach presented above is that in \cite{Creutz} the translational mode and the mode corresponding to the Lorentz symmetry (also containing linear in $t$ term) were used. In the case of Bose-Einstein condensates, such modes were discussed in \cite{ZM1,ZM2}.

Note that $\hat\Psi(0,\vec x)$ defined by \eqref{linsolfixedT} with \eqref{linsol2} may lead to the interaction Hamiltonian that differs from the one based on the use of only the oscillation modes because of the terms describing interaction of quasi-particles with the modes containing the operators $\hat a_{0}^{}$ and $\hat a_{0}^{\dagger}$.

However, even though now the canonical commutation relations are satisfied and the theory is self-consistent, we still get contribution \eqref{Npquadr3} in the operator of the particle number. Using the nonoscillation mode discussed in this section, one can make a naive attempt to remove this contribution from the theory. Indeed, since mode \eqref{quantmode} changes the particle number of the theory, let us add to \eqref{linsol2} the term
\begin{equation}
{\scriptstyle\triangle}\hat\varphi(t)=-\frac{1}{L^{3}}\sqrt{\frac{g}{\omega}}\left(\frac{1}{2}-\frac{i}{\hbar}\omega t\right)\sum\limits_{j\neq 0}\left(\left(c_{j}^{2}+d_{j}^{2}\right)\hat a^{\dagger}_{j}\hat a_{j}-c_{j}d_{j}\left(\hat a^{\dagger}_{j}\hat a^{\dagger}_{-j}+\hat a_{j}\hat a_{-j}\right)\right),
\end{equation}
which is also a solution of the linear Eq.~\eqref{lineq}. Now let us suppose that at time $t=0$
\begin{eqnarray}\nonumber
\hat\Psi(0,\vec x)=\sqrt{\frac{\omega}{g}}+
\frac{1}{\sqrt{L^{3}}}\sum\limits_{j\neq 0}\left(c_{j}e^{\frac{i}{\hbar}\vec k_{j}\vec x}\hat a_{j}-d_{j}e^{-\frac{i}{\hbar}\vec k_{j}\vec x}\hat a^{\dagger}_{j}\right)\\\label{linsolatt}+\frac{1}{2q\sqrt{L^{3}}}\left(\hat a_{0}-\hat a_{0}^{\dagger}\right)+\frac{q}{2\sqrt{L^{3}}}\left(\hat a_{0}+\hat a_{0}^{\dagger}\right)
+{\scriptstyle\triangle}\hat\varphi(0),
\end{eqnarray}
and the subsequent dynamics of the system is governed by Eq.~\eqref{eqmotgeneral} (as was mentioned in the previous section, with Eq.~\eqref{eqmotgeneral} we can be sure that at least $\dot{\hat N}=0$ even if the canonical commutation relations are violated, which is not the case of Eq.~\eqref{Heiseq}). Now let us substitute \eqref{linsolatt} into \eqref{Npquadr} and keep the terms quadratic in $\hat a_{j}$ and $\hat a^{\dagger}_{j}$, where $j\neq 0$. The result is just
\begin{equation}
\hat N_{p}^{(2)}=0,
\end{equation}
which, at first glance, solves the problem of nonconserved particle number, at least in a free theory. However, one should also check that the canonical commutation relations are satisfied. It is easy to check that for $\hat\Psi(0,\vec x)$ defined by \eqref{linsolatt}
\begin{equation}\label{CCR1violat}
[\hat\Psi(0,\vec x),\hat\Psi(0,\vec y)]=\frac{1}{2L^{3}}\sqrt{\frac{g}{\omega}}\left(\hat\varphi_{\textrm{osc}}^{}(0,\vec y)-\hat\varphi_{\textrm{osc}}^{}(0,\vec x)\right)\neq 0
\end{equation}
and
\begin{equation}\label{CCR2violat}
[\hat\Psi(0,\vec x),\hat\Psi^{\dagger}(0,\vec y)]=\delta^{(3)}(\vec x-\vec y)-\frac{1}{2L^{3}}\sqrt{\frac{g}{\omega}}\left(\hat\varphi_{\textrm{osc}}^{\dagger}(0,\vec y)+\hat\varphi_{\textrm{osc}}^{}(0,\vec x)\right)\neq\delta^{(3)}(\vec x-\vec y),
\end{equation}
where $\hat\varphi_{\textrm{osc}}^{}(0,\vec x)$ is defined by \eqref{linsol}. One can see that the canonical commutation relations are not satisfied, which implies that formally Eqs.~\eqref{Heiseq} and \eqref{eqmotgeneral} are not equivalent and the theory again becomes non-self-consistent. Although the violation in \eqref{CCR1violat}, \eqref{CCR2violat} looks better than the one in \eqref{CCRviolat},\footnote{Indeed, since the number of particles in the condensate $N_{0}\gg 1$, from \eqref{backgrN} it follows that $\sqrt{\frac{g}{\omega L^{3}}}\ll 1$, whereas $\hat\varphi_{\textrm{osc}}^{}(0,\vec x)$ can be roughly estimated as $\hat\varphi_{\textrm{osc}}^{}(0,\vec x)\sim\frac{1}{\sqrt{L^{3}}}$ (of course, the operator nature of $\hat\varphi_{\textrm{osc}}^{}(0,\vec x)$ is discarded for such an estimate). Thus, for the degree of violation in \eqref{CCR1violat} and \eqref{CCR2violat} we get $\frac{1}{L^{3}}\sqrt{\frac{g}{\omega}}\frac{1}{\sqrt{L^{3}}}=\frac{1}{\sqrt{N_{0}}}\frac{1}{L^{3}}\ll\frac{1}{L^{3}}$ instead of $\frac{1}{L^{3}}$ in \eqref{CCRviolat}. Note that it is the inequality $N_{0}\gg 1$ that in fact justifies the applicability of the linear approximation in \cite{Bogolyubov}: for the estimate $\hat\varphi(t,\vec x)\sim\frac{1}{\sqrt{L^{3}}}$, the condition $\sqrt{\frac{\omega}{g}}\gg|\hat\varphi(t,\vec x)|$ is just $\sqrt{\frac{g}{\omega L^{3}}}=\frac{1}{\sqrt{N_{0}}}\ll 1$.} even taking into account that the commutators in \eqref{CCR1violat} and \eqref{CCR2violat} are not conserved over time, still it is not the best solution. As was already mentioned, there exists another approach to solve the problem of nonconserved particle number. This approach is much more complicated, but it leads to more accurate and consistent results. To demonstrate its main idea, it is illustrative first to consider the classical theory.

\section{Solving the problem of nonconserved particle number in the classical theory}
Let us take a classical system in a spatial ``box'' of size $L\times L\times L$, obeying the nonlinear Schr\"{o}dinger equation
\begin{equation}\label{NSE}
i\dot\Phi=-\Delta\Phi+F(\Phi^{*}\Phi)\Phi.
\end{equation}
The particle number (norm) of this system is
\begin{equation}
N=\int\Phi^{*}\Phi\,d^{3}x
\end{equation}
and the energy of the system is
\begin{equation}
E=\int\left(\sum\limits_{i=1}^{3}\partial_{i}\Phi^{*}\partial_{i}\Phi+\int\limits_{0}^{\Phi^{*}\Phi}F(s)ds\right)d^{3}x.
\end{equation}
Suppose we have an exact stationary solution of Eq.~\eqref{NSE} of the form
\begin{equation}\label{backgrsol}
\Phi_{0}(t,\vec x)=e^{-i\omega t}f(\vec x),
\end{equation}
where the function $f(\vec x)$ is real. Now we consider a small perturbation against this background solution such that
\begin{equation}\label{psipert}
\Phi(t,\vec x)=e^{-i\omega t}\left(f(\vec x)+\alpha\phi(t,\vec x)\right),
\end{equation}
where $\alpha\ll 1$ is a small dimensionless parameter which is introduced for convenience. Now let us substitute \eqref{psipert} into Eq.~\eqref{NSE} and keep the terms that are linear and {\em quadratic} in $\phi(t,\vec x)$ \cite{{Smolyakov}}:
\begin{equation}\label{eqpert}
i\dot\phi=-\Delta\phi+(U-\omega)\phi+fW(\phi+\phi^{*})+\alpha\left(W(\phi^{2}+2\phi^{*}\phi)+J(\phi+\phi^{*})^{2}\right)+\alpha^{2}\left(...\right)+...\,\,,
\end{equation}
where
\begin{equation}
U(\vec x)=F\left(f^{2}(\vec x)\right),\qquad W(\vec x)=\frac{dF(s)}{ds}\biggl|_{s=f^{2}(\vec x)}f(\vec x),\qquad J(\vec x)=\frac{1}{2}\frac{d^{2}F(s)}{ds^{2}}\biggl|_{s=f^{2}(\vec x)}f^{3}(\vec x).
\end{equation}
Of course, in the general case one expects an infinite series in $\alpha$ in the r.h.s. of Eq.~\eqref{eqpert}. Analogously, for the particle number and the energy of the perturbation one gets
\begin{align}\label{Npert}
&N_{p}=N[\Phi(t,\vec x)]-N[\Phi_{0}(t,\vec x)]=\int d^{3}x\left(\alpha f(\phi+\phi^{*})+\alpha^{2}\phi^{*}\phi\right),\\\label{Epert}
&E_{p}=E[\Phi(t,\vec x)]-E[\Phi_{0}(t,\vec x)]=\omega N_{p}+\frac{i\alpha^{2}}{2}\int d^{3}x\left(\phi^{*}\dot\phi-\dot\phi^{*}\phi\right)+...\,\,.
\end{align}
Note that formula \eqref{Npert} is exact, whereas in formula \eqref{Epert} one also expects an infinite series in $\alpha$ in the general case. In derivation of formula \eqref{Epert} the nonlinear Eq.~\eqref{eqpert} was used to eliminate the terms with spatial derivatives $\partial_{i}$.

A solution of Eq.~\eqref{eqpert} can be represented in the form
\begin{equation}\label{classsolfull}
\phi(t,\vec x)=\phi_{\textrm{lin}}(t,\vec x)+\sum\limits_{j=1}^{\infty}\alpha^{j}\phi_{\textrm{nl}}^{(j)}(t,\vec x).
\end{equation}
Substituting representation \eqref{classsolfull} into Eq.~\eqref{eqpert}, keeping the terms of the order of $\alpha^{0}=1$ and $\alpha$, and defining $\phi_{\textrm{nl}}^{(1)}=\phi_{\textrm{nl}}$, the following set of equations can be obtained:
\begin{align}\label{linclasseq}
&i\dot\phi_{\textrm{lin}}+\Delta\phi_{\textrm{lin}}-(U-\omega)\phi_{\textrm{lin}}-fW(\phi_{\textrm{lin}}+\phi_{\textrm{lin}}^{*})=0,\\
&i\dot\phi_{\textrm{nl}}+\Delta\phi_{\textrm{nl}}-(U-\omega)\phi_{\textrm{nl}}-fW(\phi_{\textrm{nl}}+\phi_{\textrm{nl}}^{*})
=W(\phi_{\textrm{lin}}^{2}+2\phi_{\textrm{lin}}^{*}\phi_{\textrm{lin}})+J(\phi_{\textrm{lin}}+\phi_{\textrm{lin}}^{*})^{2}.
\end{align}
Analogously, substituting representation \eqref{classsolfull} into \eqref{Npert} and keeping the terms of the order of $\alpha$ and $\alpha^{2}$, one gets
\begin{equation}\label{Npert2}
N_{p}=\int d^{3}x\left(\alpha f(\phi_{\textrm{lin}}+\phi_{\textrm{lin}}^{*})+\alpha^{2} f(\phi_{\textrm{nl}}+\phi_{\textrm{nl}}^{*})+\alpha^{2}\phi_{\textrm{lin}}^{*}\phi_{\textrm{lin}}\right).
\end{equation}
The first term $\sim\alpha^{2}$ in \eqref{Npert2} originates from the nonlinear part of the solution for the perturbation $\phi$ (i.e., from $\phi_{\textrm{nl}}$), but comes from the terms of \eqref{Npert} that are linear in $\phi$ and $\phi^{*}$; whereas the second term $\sim\alpha^{2}$ in \eqref{Npert2} originates from the linear part of the solution for the perturbation $\phi$ (i.e., from $\phi_{\textrm{lin}}$), but comes from the term of \eqref{Npert} that is quadratic in $|\phi|$.

It was shown in \cite{Smolyakov} that for a perturbation whose linear part consists of only the oscillation modes
\begin{equation}\label{classpert}
\phi_{\textrm{lin}}(t,\vec x)=\frac{1}{2}\sum\limits_{n}\left((\xi_{n}(\vec x)+\eta_{n}(\vec x))e^{-i\gamma_{n}t}+(\xi^{*}_{n}(\vec x)-\eta^{*}_{n}(\vec x))e^{i\gamma_{n}t}\right),
\end{equation}
the particle number and the energy of the whole perturbation (i.e., with the first nonlinear correction) take the form
\begin{align}\label{PNfinal}
&N_{p}=\alpha^{2}\sum\limits_{n}\int\left(\frac{1}{2}\left(\xi_{n}^{*}\xi_{n}+\eta_{n}^{*}\eta_{n}\right)-\frac{df}{d\omega}W\left(3\xi_{n}^{*}\xi_{n}+\eta_{n}^{*}\eta_{n}\right)
-4\frac{df}{d\omega}J\xi_{n}^{*}\xi_{n}\right)d^{3}x,\\\label{Efinal}
&E_{p}=\omega N_{p}+\frac{\alpha^{2}}{2}\sum\limits_{n}\left(\gamma_{n}\int\left(\xi_{n}\eta_{n}^{*}+\xi_{n}^{*}\eta_{n}\right)d^{3}x\right).
\end{align}
Note that \eqref{PNfinal} and \eqref{Efinal} are expressed only through the parameters of the linear part of the perturbation (see \eqref{classpert}) and the additivity property is valid for \eqref{PNfinal} and \eqref{Efinal} even though the first {\em nonlinear} correction is taken into account in \eqref{PNfinal} and \eqref{Efinal}.

Let us apply these results to the case
\begin{equation}
F(\Phi^{*}\Phi)=g\Phi^{*}\Phi,
\end{equation}
where $g>0$, which corresponds to \eqref{eqmotgeneral}. The stationary background solution takes the standard form
\begin{equation}\label{GPsol}
\Phi_{0}(t,\vec x)=e^{-i\omega t}\sqrt{\frac{\omega}{g}}
\end{equation}
with $\omega>0$. Let us take the linear part of the perturbation, satisfying Eq.~\eqref{linclasseq}, in the form of a superposition of the plane waves (exactly as in \eqref{linsol}, but at the classical level)
\begin{equation}\label{pertGP}
\phi_{\textrm{lin}}(t,\vec x)=\frac{1}{\sqrt{L^{3}}}\sum\limits_{j\neq 0}\left(c_{j}e^{-i(\gamma_{j}t-\vec k_{j}\vec x)}a_{j}-d_{j}e^{i(\gamma_{j}t-\vec k_{j}\vec x)}a^{*}_{j}\right),
\end{equation}
where $c_{j}$, $d_{j}$, $\vec k_{j}$ and $\gamma_{j}$ are defined by \eqref{cddef}, \eqref{kdef} and \eqref{gammadef} with $\hbar=1$ and $m=\frac{1}{2}$. If only the linear part of the perturbation is taken into account, the particle number of such a perturbation takes the form
\begin{eqnarray}\nonumber
N_{p}^{\textrm{lin}}=\int d^{3}x\left(\alpha f(\phi_{\textrm{lin}}+\phi_{\textrm{lin}}^{*})+\alpha^{2}\phi_{\textrm{lin}}^{*}\phi_{\textrm{lin}}\right)\\\label{Nplin}=
\alpha^{2}\sum\limits_{j\neq 0}\left(\left(c_{j}^{2}+d_{j}^{2}\right)a_{j}^{*}a_{j}-c_{j}d_{j}\left(a_{j}^{*}a_{-j}^{*}e^{2i\gamma_{j}t}+a_{j}a_{-j}e^{-2i\gamma_{j}t}\right)\right),
\end{eqnarray}
compare with \eqref{Npquadr2}. However, for \eqref{Npert2} formula \eqref{PNfinal} provides
\begin{equation}\label{Np}
N_{p}=-\alpha^{2}\sum\limits_{j\neq 0}(c_{j}-d_{j})^{2}a_{j}^{*}a_{j}.
\end{equation}
Now $N_{p}$ does not depend on time, as it should be, and is diagonal. Moreover, since $N_{p}$ in \eqref{Np} is time-independent, we can ``tune'' it using one of the nonoscillation modes. Indeed, let us add
\begin{equation}\label{classextrasol}
\alpha C\left(\frac{df}{d\omega}-itf\right)=\alpha C\left(\frac{1}{2\sqrt{\omega g}}-it\sqrt{\frac{\omega}{g}}\right)
\end{equation}
with
\begin{equation}
C=\frac{g}{L^{3}}\sum\limits_{j\neq 0}(c_{j}-d_{j})^{2}a_{j}^{*}a_{j},
\end{equation}
which is a solution of the linear Eq.~\eqref{linclasseq}, to the solution $\phi(t,\vec x)=\phi_{\textrm{lin}}(t,\vec x)+\alpha\phi_{\textrm{nl}}(t,\vec x)$. Substituting \eqref{classextrasol} into \eqref{Npert} and keeping the term $\sim\alpha^{2}$, with \eqref{Np} one can get
\begin{equation}
N_{p}=\alpha^{2}\int d^{3}x\sqrt{\frac{\omega}{g}}\frac{C}{\sqrt{\omega g}}-\alpha^{2}\sum\limits_{j\neq 0}(c_{j}-d_{j})^{2}a_{j}^{*}a_{j}=0,
\end{equation}
and consequently
\begin{equation}\label{Ep}
E_{p}=\alpha^{2}\sum\limits_{j\neq 0}\gamma_{j}a_{j}^{*}a_{j}.
\end{equation}
Note that the results presented in this section are valid for any $t$ (i.e., we do not fix $t=0$), including ``tuning'' of the particle number, contrary to the simple attempt to solve the problem of nonconserved particle number discussed in Section~4. The same approach will be used in the quantum case, but in a more explicit manner.

It should be mentioned that the necessity for the first nonlinear correction for obtaining correct values of the integrals of motion of perturbations is not a distinctive property of the systems described by the nonlinear Schr\"{o}dinger equation. A fully analogous situation occurs for perturbations against stationary time-dependent background solutions (namely, nontopological solitons) in relativistic scalar field theories \cite{Smolyakov:2017axd,Smolyakov:2019cld}. Meanwhile, in the case of static background solutions in relativistic scalar field theories it is not necessary to consider nonlinear correction for obtaining correct values of the energy of perturbations against such a solution in the lowest nonzero order (i.e., $\sim\alpha^{2}$ in our notations) --- nonlinear corrections contribute only to the higher order terms of the energy of perturbations (i.e., to the terms $\sim\alpha^{3}$ and higher). For example, for the kink soliton solution the analysis of perturbations can be found in \cite{Rajaraman:1982is}.

\section{Solving the problem of nonconserved particle number in the quantum theory}
\subsection{Nonlinear perturbations and commutation relations}
Now let us turn to the quantum case and represent the operator $\hat\Psi(t,\vec x)$ as (see \eqref{backgrsol0} and \eqref{linearization})
\begin{equation}\label{representInit}
\hat\Psi(t,\vec x)=e^{-\frac{i}{\hbar}\omega t}\left(\sqrt{\frac{\omega}{g}}+\hat\psi(t,\vec x)\right),
\end{equation}
where $\hat\psi(t,\vec x)$ is supposed to satisfy the full nonlinear equation of motion
\begin{equation}\label{eqnlcquant}
i\hbar\dot{\hat\psi}(t,\vec x)=-\frac{\hbar^{2}}{2m}\Delta\hat\psi(t,\vec x)+\omega\left(\hat\psi(t,\vec x)+\hat\psi^{\dagger}(t,\vec x)\right)
+\sqrt{\omega g}\left(2\hat\psi^{\dagger}(t,\vec x)\hat\psi(t,\vec x)+\hat\psi^{2}(t,\vec x)\right)+...\,\,,
\end{equation}
where ``$...$'' denotes the terms of the higher orders in $\hat\psi^{\dagger}(t,\vec x)$ and $\hat\psi(t,\vec x)$. The operator $\hat\psi(t,\vec x)$ can be represented as the infinite series
\begin{equation}\label{seriesPhiinf}
\hat\psi(t,\vec x)=\hat\varphi(t,\vec x)+\sum\limits_{n=1}^{\infty}\beta^{n}\hat\phi_{n}(t,\vec x)
\end{equation}
with $\beta=\frac{1}{\sqrt{N_{0}}}=\sqrt{\frac{g}{\omega L^{3}}}\ll 1$, and $\hat\varphi(t,\vec x)$ can be represented either as
\begin{equation}\label{linsolquant}
\hat\varphi(t,\vec x)=\hat\varphi_{\textrm{no}}(t)+\sum\limits_{j\neq 0}\hat\varphi_{j}(t,\vec x)
\end{equation}
or as
\begin{equation}\label{linsolquant1}
\hat\varphi(t,\vec x)=\hat\varphi_{\textrm{no}}(t)+\sum\limits_{j\neq 0}\hat{\tilde\varphi}_{j}(t,\vec x),
\end{equation}
where
\begin{align}\label{philinnodef}
&\hat\varphi_{\textrm{no}}(t)=\frac{1}{2q\sqrt{L^{3}}}\left(\hat a_{0}-\hat a_{0}^{\dagger}\right)+\frac{q}{\sqrt{L^{3}}}\left(\frac{1}{2}-\frac{i}{\hbar}\omega t\right)\left(\hat a_{0}+\hat a_{0}^{\dagger}\right),\\
&\hat\varphi_{j}(t,\vec x)=\frac{1}{\sqrt{L^{3}}}\left(c_{j}e^{-\frac{i}{\hbar}(\gamma_{j}t-\vec k_{j}\vec x)}\hat a_{j}-d_{j}e^{\frac{i}{\hbar}(\gamma_{j}t-\vec k_{j}\vec x)}\hat a^{\dagger}_{j}\right),\\
&\hat{\tilde\varphi}_{j}(t,\vec x)=\frac{1}{\sqrt{L^{3}}}e^{\frac{i}{\hbar}\vec k_{j}\vec x}\left(c_{j}e^{-\frac{i}{\hbar}\gamma_{j}t}\hat a_{j}-d_{j}e^{\frac{i}{\hbar}\gamma_{j}t}\hat a^{\dagger}_{-j}\right),
\end{align}
see \eqref{linsol}. Here $\hat\varphi_{\textrm{no}}(t)$ denotes the two nonoscillation modes, $\hat\varphi_{j}(t,\vec x)$ denotes the $j$-th oscillation mode, and $\hat{\tilde\varphi}_{j}(t,\vec x)$ is a mixture of some parts of the $j$-th and $-j$-th oscillation modes. Of course, representations \eqref{linsolquant} and \eqref{linsolquant1} are equivalent and lead to the same results, but sometimes it is more convenient to use for calculations representation \eqref{linsolquant}, whereas sometimes representation \eqref{linsolquant1} slightly simplifies the calculations (because of the extraction of the overall factor $e^{\frac{i}{\hbar}\vec k_{j}\vec x}$).

Now we should substitute \eqref{seriesPhiinf} into Eq.~\eqref{eqnlcquant}, combine the terms of the same order in $\beta^{n}$ and get an infinite number of equations of motion for $\hat\phi_{n}(t,\vec x)$. Of course, it is impossible to solve the whole task analytically, so in what follows only the first nonlinear correction $\hat\phi_{1}(t,\vec x)$ will be considered (as will be shown below, even this simplified task demands very bulky, although straightforward, calculations). In such an approximation the problem is similar to the case of the classical nonlinear Schr\"{o}dinger equation discussed in the previous section, except one important point --- in the quantum case it is not sufficient just to calculate $\hat N_{p}$ and $\hat E_{p}$, it is also necessary to calculate $\hat\phi_{1}(t,\vec x)$ explicitly in order to check that the canonical commutation relations are satisfied at least up and including the terms $\sim\beta$.

Thus, let us define $\hat\phi_{1}(t,\vec x)$ as $\hat\phi(t,\vec x)$ (just to eliminate the unnecessary in this approximation subscript ``1''), represent the operator $\hat\psi(t,\vec x)$ as
\begin{equation}\label{representNL}
\hat\psi(t,\vec x)=\hat\varphi(t,\vec x)+\beta\hat\phi(t,\vec x)
\end{equation}
and substitute this representation into Eq.~\eqref{eqnlcquant}. Using $\sqrt{\omega g}=\omega\sqrt{L^{3}}\beta$ and combining all terms of the order of $\beta$, we arrive at the equation
\begin{equation}\label{eqnlcquant2}
i\hbar\dot{\hat\phi}+\frac{\hbar^{2}}{2m}\Delta\hat\phi-\omega\left(\hat\phi+\hat\phi^{\dagger}\right)
=\omega\sqrt{L^{3}}\left(2\hat\varphi^{\dagger}\hat\varphi+\hat\varphi^{2}\right).
\end{equation}
Now, using representation \eqref{representInit}, for the particle number we get the exact result
\begin{equation}\label{Npquadrpsi}
\hat N_{p}=\int d^{3}x\left(\sqrt{\frac{\omega}{g}}\left(\hat\psi^{\dagger}+\hat\psi\right)+\hat\psi^{\dagger}\hat\psi\right)
\end{equation}
(compare with \eqref{Npquadr}). Substituting representation \eqref{representInit} into \eqref{initHamiltonian2}, using the {\em nonlinear} Eq.~\eqref{eqnlcquant} to eliminate the terms with $\partial_{i}$ and keeping only the terms that are linear and quadratic in $\hat\psi$ and $\hat\psi^{\dagger}$, we get
\begin{equation}\label{Epquadrpsi}
\hat E_{p}=\omega\hat N_{p}+\frac{i\hbar}{2}\int d^{3}x\left(\hat\psi^{\dagger}\dot{\hat\psi}-\dot{\hat\psi}^{\dagger}\hat\psi\right)
\end{equation}
(compare with \eqref{Epquadr}). Finally, substituting representation \eqref{representNL} into \eqref{Npquadrpsi} and \eqref{Epquadrpsi}, using $\sqrt{\frac{\omega}{g}}=\frac{1}{\beta\sqrt{L^{3}}}$, and keeping the terms of the order of $\beta^{-1}$ and $\beta^{0}=1$, we obtain
\begin{eqnarray}\label{NpquadrQNL}
&&\hat N_{p}=\int d^{3}x\left(\sqrt{\frac{\omega}{g}}\left(\hat\varphi^{\dagger}+\hat\varphi\right)+\hat\varphi^{\dagger}\hat\varphi
+\frac{1}{\sqrt{L^{3}}}\left(\hat\phi^{\dagger}+\hat\phi\right)\right),
\\\label{EpquadrQNL}
&&\hat E_{p}=\omega\hat N_{p}+\frac{i\hbar}{2}\int d^{3}x\left(\hat\varphi^{\dagger}\dot{\hat\varphi}-\dot{\hat\varphi}^{\dagger}\hat\varphi\right).
\end{eqnarray}

Now let us turn to the canonical commutation relations. Since here only the first nonlinear correction is taken into account, it is impossible to check that the canonical commutation relations are satisfied exactly. For the approximation used and with \eqref{CCR01}, \eqref{CCR02} for $\hat\varphi$ and $\hat\varphi^{\dagger}$, they turn into
\begin{align}
&[\hat\psi(t,\vec x),\hat\psi(t,\vec y)]\to \beta[\hat\phi(t,\vec x),\hat\varphi(t,\vec y)]+\beta[\hat\varphi(t,\vec x),\hat\phi(t,\vec y)]=0,\\
&[\hat\psi(t,\vec x),\hat\psi^{\dagger}(t,\vec y)]\to\delta^{(3)}(\vec x-\vec y)+\beta[\hat\phi(t,\vec x),\hat\varphi^{\dagger}(t,\vec y)]+\beta[\hat\varphi(t,\vec x),\hat\phi^{\dagger}(t,\vec y)]=\delta^{(3)}(\vec x-\vec y),
\end{align}
leading to
\begin{align}\label{CCR1}
&[\hat\phi(t,\vec x),\hat\varphi(t,\vec y)]+[\hat\varphi(t,\vec x),\hat\phi(t,\vec y)]=0,\\\label{CCR2}
&[\hat\phi(t,\vec x),\hat\varphi^{\dagger}(t,\vec y)]+[\hat\varphi(t,\vec x),\hat\phi^{\dagger}(t,\vec y)]=0.
\end{align}
It is clear that it is senseless to consider the terms $\beta^{2}[\hat\phi(t,\vec x),\hat\phi(t,\vec y)]$ and $\beta^{2}[\hat\phi(t,\vec x),\hat\phi^{\dagger}(t,\vec y)]$ --- in order to take into account all contributions $\sim \beta^{2}$ to the canonical commutation relations, it is necessary to calculate the second nonlinear correction $\hat\phi_{2}(t,\vec x)$, which exceeds the scope of the present analysis. Thus, if commutation relations \eqref{CCR1} and \eqref{CCR2} are satisfied, it indicates that the canonical commutation relations are satisfied at least up to the terms $\sim\beta$.

Now let us turn to Eq.~\eqref{eqnlcquant2}. The r.h.s. of this equation can be represented as
\begin{eqnarray}\nonumber
2\hat\varphi^{\dagger}\hat\varphi+\hat\varphi^{2}
=2\hat\varphi_{\textrm{no}}^{\dagger}\hat\varphi_{\textrm{no}}^{}+\hat\varphi_{\textrm{no}}^{2}
+\sum\limits_{j\neq 0}\left(2\hat{\tilde\varphi}_{j}^{\dagger}\hat\varphi_{\textrm{no}}^{}
+2\hat\varphi_{\textrm{no}}^{\dagger}\hat{\tilde\varphi}_{j}^{}
+\hat{\tilde\varphi}_{j}^{}\hat\varphi_{\textrm{no}}^{}+\hat\varphi_{\textrm{no}}^{}\hat{\tilde\varphi}_{j}^{}
\right)\\
+\frac{1}{2}\sum\limits_{j\neq 0}\sum\limits_{l\neq 0}\left(
2\hat\varphi_{j}^{\dagger}\hat\varphi_{l}^{}+2\hat\varphi_{l}^{\dagger}\hat\varphi_{j}^{}
+\hat\varphi_{j}^{}\hat\varphi_{l}^{}+\hat\varphi_{l}^{}\hat\varphi_{j}^{}\right),
\end{eqnarray}
where the factor $\frac{1}{2}$ in the third group of terms is due to the double counting of the modes. In the first sum in the r.h.s. of this formula representation \eqref{linsolquant1} is used, whereas in the second sum representation \eqref{linsolquant} is used. It is also convenient to represent the operator $\hat\phi(t,\vec x)$ as
\begin{equation}
\hat\phi(t,\vec x)=\hat\phi_{\textrm{no}}(t)+\hat\phi_{\times}(t,\vec x)+\hat\phi_{\textrm{o}}(t,\vec x),
\end{equation}
where the operators $\hat\phi_{\textrm{no}}(t)$, $\hat\phi_{\times}(t,\vec x)$ and $\hat\phi_{\textrm{o}}(t,\vec x)$ satisfy the equations
\begin{align}\label{eqquant1}
&i\hbar\dot{\hat\phi}_{\textrm{no}}^{}-\omega\left(\hat\phi_{\textrm{no}}^{}+\hat\phi_{\textrm{no}}^{\dagger}\right)
=\omega\sqrt{L^{3}}\left(2\hat\varphi_{\textrm{no}}^{\dagger}\hat\varphi_{\textrm{no}}^{}+\hat\varphi_{\textrm{no}}^{2}\right),\\\label{eqquant2}
&i\hbar\dot{\hat\phi}_{\times}^{}+\frac{\hbar^{2}}{2m}\Delta\hat\phi_{\times}^{}-\omega\left(\hat\phi_{\times}^{}+\hat\phi_{\times}^{\dagger}\right)
=\omega\sqrt{L^{3}}\sum\limits_{j\neq 0}\left(2\hat{\tilde\varphi}_{j}^{\dagger}\hat\varphi_{\textrm{no}}^{}
+2\hat\varphi_{\textrm{no}}^{\dagger}\hat{\tilde\varphi}_{j}^{}
+\hat{\tilde\varphi}_{j}^{}\hat\varphi_{\textrm{no}}^{}+\hat\varphi_{\textrm{no}}^{}\hat{\tilde\varphi}_{j}^{}
\right),\\\label{eqquant3}
&i\hbar\dot{\hat\phi}_{\textrm{o}}^{}+\frac{\hbar^{2}}{2m}\Delta\hat\phi_{\textrm{o}}^{}-\omega\left(\hat\phi_{\textrm{o}}^{}+\hat\phi_{\textrm{o}}^{\dagger}\right)
=\omega\sqrt{L^{3}}\frac{1}{2}\sum\limits_{j\neq 0}\sum\limits_{l\neq 0}\left(
2\hat\varphi_{j}^{\dagger}\hat\varphi_{l}^{}+2\hat\varphi_{l}^{\dagger}\hat\varphi_{j}^{}
+\hat\varphi_{j}^{}\hat\varphi_{l}^{}+\hat\varphi_{l}^{}\hat\varphi_{j}^{}\right).
\end{align}

Now we are ready to solve Eqs.~\eqref{eqquant1}--\eqref{eqquant3}. Below I will not present all the calculations --- some of them are rather bulky, although entirely straightforward. However, all the key steps of the analysis will be presented.

\subsection{Nonoscillation modes}
Let us start with the simplest case of the nonoscillation modes. In the explicit form, Eq.~\eqref{eqquant1} looks like
\begin{align}\nonumber
i\hbar\dot{\hat\phi}_{\textrm{no}}^{}-\omega\left(\hat\phi_{\textrm{no}}^{}+\hat\phi_{\textrm{no}}^{\dagger}\right)
&=\frac{\omega}{\sqrt{L^{3}}}\left(
-\frac{1}{4q^{2}}\left(\hat a_{0}-\hat a_{0}^{\dagger}\right)^{2}+\left(\frac{1}{2}+\frac{i\omega t}{\hbar}\right)\left(\hat a_{0}^{2}-\hat a_{0}^{\dagger\,2}\right)\right.\\\label{eqquant1expl}
&+\left. q^{2}\left(\frac{3}{4}-\frac{i\omega t}{\hbar}+\frac{\omega^{2}t^{2}}{\hbar^{2}}\right)\left(\hat a_{0}+\hat a_{0}^{\dagger}\right)^{2}-1\right),
\end{align}
where the constant $1$ (the last term in the overall brackets) arises after permutation of the operators $\hat a_{0}$ and $\hat a_{0}^{\dagger}$ in some terms. The solution of this inhomogeneous equation can be easily found, one can take it in the form
\begin{equation}\label{phinosol}
\hat\phi_{\textrm{no}}(t)=\frac{\omega}{2\sqrt{L^{3}}}\left(
\frac{1}{4\omega q^{2}}\left(\hat a_{0}-\hat a_{0}^{\dagger}\right)^{2}-\frac{it}{\hbar}\left(\hat a_{0}^{2}-\hat a_{0}^{\dagger\,2}\right)- q^{2}\left(\frac{1}{4\omega}+\frac{it}{\hbar}+\frac{\omega t^{2}}{\hbar^{2}}\right)\left(\hat a_{0}+\hat a_{0}^{\dagger}\right)^{2}+\frac{1}{\omega}\right).
\end{equation}
Since solution \eqref{phinosol} depends only on $t$, whereas the terms with the operators $\hat a_{0}$ and $\hat a_{0}^{\dagger}$ in $\hat\varphi(t,\vec x)$ (i.e., $\hat\varphi_{\textrm{no}}(t)$) also depend only on $t$, commutation relation \eqref{CCR1} is satisfied automatically. As for commutation relation \eqref{CCR2}, straightforward calculations lead to
\begin{equation}
[\hat\phi_{\textrm{no}}(t),\hat\varphi^{\dagger}(t,\vec y)]+[\hat\varphi(t,\vec x),\hat\phi_{\textrm{no}}^{\dagger}(t)]=-\frac{q}{2L^{3}}\left(\hat a_{0}+\hat a_{0}^{\dagger}\right),
\end{equation}
which is not equal to zero. However, this problem can be easily solved by adding the term
\begin{equation}\label{deltaphinosol}
{\scriptstyle\triangle}\hat\phi_{\textrm{no}}=\frac{1}{4\sqrt{L^{3}}}\left(\hat a_{0}^{2}-\hat a_{0}^{\dagger\,2}\right)
\end{equation}
to solution \eqref{phinosol}. In this case, commutation relation \eqref{CCR1} is still satisfied automatically, whereas for commutation relation \eqref{CCR2} we get
\begin{equation}
[\hat\phi_{\textrm{no}}(t)+{\scriptstyle\triangle}\hat\phi_{\textrm{no}},\hat\varphi^{\dagger}(t,\vec y)]+[\hat\varphi(t,\vec x),\hat\phi_{\textrm{no}}^{\dagger}(t)+{\scriptstyle\triangle}\hat\phi_{\textrm{no}}^{\dagger}]=0.
\end{equation}
The extra term \eqref{deltaphinosol} is a solution of the homogeneous part of Eq.~\eqref{eqquant1expl}, it is just mode \eqref{quantmode0}, but with the operator $\hat A$ quadratic in $\hat a_{0}$ and $\hat a_{0}^{\dagger}$. Although this mode is a solution of the linear equation of motion, it provides contribution $\sim\beta$ to $\hat\psi(t,\vec x)$ and thus it is natural to incorporate this extra term into $\hat\phi_{\textrm{no}}(t)$. As will be shown below, the use of solutions of the homogeneous parts of the corresponding equations will allow one to ensure the fulfillment of commutation relations \eqref{CCR1}, \eqref{CCR2}.

Finally, redefining $\hat\phi_{\textrm{no}}(t)+{\scriptstyle\triangle}\hat\phi_{\textrm{no}}\to\hat\phi_{\textrm{no}}(t)$, we get
\begin{align}\nonumber
\hat\phi_{\textrm{no}}(t)&=\frac{\omega}{2\sqrt{L^{3}}}\left(
\frac{1}{4\omega q^{2}}\left(\hat a_{0}-\hat a_{0}^{\dagger}\right)^{2}+\left(\frac{1}{2\omega}-\frac{it}{\hbar}\right)\left(\hat a_{0}^{2}-\hat a_{0}^{\dagger\,2}\right)\right.\\
\label{phinosolfull}&\left.-q^{2}\left(\frac{1}{4\omega}+\frac{it}{\hbar}+\frac{\omega t^{2}}{\hbar^{2}}\right)\left(\hat a_{0}+\hat a_{0}^{\dagger}\right)^{2}+\frac{1}{\omega}\right).
\end{align}
Now let us substitute $\hat\varphi_{\textrm{no}}(t)$ defined by \eqref{philinnodef} and $\hat\phi_{\textrm{no}}(t)$ defined by \eqref{phinosolfull} into \eqref{NpquadrQNL} and \eqref{EpquadrQNL}. The result looks as follows:
\begin{align}\label{NpNO}
&\hat N_{p}^{(\textrm{no})}=\frac{1}{2}+\sqrt{\frac{L^{3}\omega}{g}}\,q\left(\hat a_{0}+\hat a_{0}^{\dagger}\right),
\\\label{EpNO}
&\hat E_{p}^{(\textrm{no})}=\omega\hat N_{p}-\frac{\omega}{2}+\frac{\omega q^{2}}{2}\left(\hat a_{0}+\hat a_{0}^{\dagger}\right)^{2}=\omega\sqrt{\frac{L^{3}\omega}{g}}\,q\left(\hat a_{0}+\hat a_{0}^{\dagger}\right)+\frac{\omega q^{2}}{2}\left(\hat a_{0}+\hat a_{0}^{\dagger}\right)^{2}.
\end{align}
To better understand the latter formulas, let us redefine the dimensionless parameter $q$ as
\begin{equation}
q=\sqrt{\frac{L^{3}}{\omega g}}\,\tilde q,
\end{equation}
where $\tilde q$ has dimensionality of $\omega$. In this case, formulas \eqref{NpNO} and \eqref{EpNO} take the form
\begin{align}\label{NpNO1}
&\hat N_{p}^{(\textrm{no})}=\frac{1}{2}+\frac{L^{3}}{g}\,\tilde q\left(\hat a_{0}+\hat a_{0}^{\dagger}\right),
\\\label{EpNO1}
&\hat E_{p}^{(\textrm{no})}=\omega\frac{L^{3}}{g}\,\tilde q\left(\hat a_{0}+\hat a_{0}^{\dagger}\right)+\frac{L^{3}}{2g}\,{\tilde q}^{2}\left(\hat a_{0}+\hat a_{0}^{\dagger}\right)^{2}.
\end{align}
From \eqref{backgrE} and \eqref{backgrN} it follows that
\begin{equation}
\frac{L^{3}}{g}=\frac{dN_{0}}{d\omega}=\frac{1}{\omega}\frac{dE_{0}}{d\omega}=\frac{d^{2}E_{0}}{d\omega^{2}}.
\end{equation}
Thus, formulas \eqref{NpNO1}, \eqref{EpNO1} can be rewritten as
\begin{align}\label{NpNO2}
&\hat N_{p}^{(\textrm{no})}=\frac{1}{2}+\frac{dN_{0}}{d\omega}\,\tilde q\left(\hat a_{0}+\hat a_{0}^{\dagger}\right),
\\\label{EpNO2}
&\hat E_{p}^{(\textrm{no})}=\frac{dE_{0}}{d\omega}\,\tilde q\left(\hat a_{0}+\hat a_{0}^{\dagger}\right)+\frac{1}{2}\frac{d^{2}E_{0}}{d\omega^{2}}\,{\tilde q}^{2}\left(\hat a_{0}+\hat a_{0}^{\dagger}\right)^{2}.
\end{align}
This is exactly what is expected for the nonoscillation mode, which in the classical theory corresponds to a change of the frequency $\omega$ of the background solution \eqref{backgrsol0}, and consequently to a change of the particle number and the energy of the system. In the quantum theory, this mode also changes the particle number and the energy of the system by adding (or removing) particles to the condensate.\footnote{One should not worry about the nonstandard operator form of \eqref{NpNO2} and \eqref{EpNO2}. For example, according to \eqref{aPZ} and \eqref{adagPZ}, the Hamiltonian of a free particle in the one-dimensional quantum mechanics $\frac{\hat P^{2}}{2}$ can be represented in a somewhat similar form $-\frac{1}{4}\left(\hat a^{\dagger}-\hat a\right)^{2}$ (here mass of the particle and $\hbar$ are omitted), which obviously does not cause any problems.} The absence of the term $\sim\left(\hat a_{0}+\hat a_{0}^{\dagger}\right)^{2}$ in \eqref{NpNO2} simply reflects the fact that $\frac{d^{2}N_{0}}{d\omega^{2}}=0$ for the background solution \eqref{backgrsol0}.

\subsection{Overlap terms between nonoscillation and oscillation modes}
Let us represent $\hat\phi_{\times}(t,\vec x)$ as
\begin{equation}
\hat\phi_{\times}(t,\vec x)=\sum\limits_{j\neq 0}\hat\phi^{\times}_{j}(t,\vec x)
\end{equation}
such that each $\hat\phi^{\times}_{j}(t,\vec x)$ satisfies the equation
\begin{align}\nonumber
&i\hbar\dot{\hat\phi}^{\times}_{j}+\frac{\hbar^{2}}{2m}\Delta\hat\phi^{\times}_{j}-\omega\left(\hat\phi^{\times}_{j}+\hat\phi_{-j}^{\times\dagger}\right)=\frac{\omega}{\sqrt{L^{3}}}e^{\frac{i}{\hbar}\vec k_{j}\vec x}\\\nonumber&\times\left[e^{-\frac{i}{\hbar}\gamma_{j}t}\left(q(2c_{j}-d_{j})\left(\hat a_{0}+\hat a_{0}^{\dagger}\right)\hat a_{j}-\frac{1}{q}d_{j}\left(\hat a_{0}-\hat a_{0}^{\dagger}\right)\hat a_{j}\right)
+t\,e^{-\frac{i}{\hbar}\gamma_{j}t}\frac{2i\omega}{\hbar}qd_{j}\left(\hat a_{0}+\hat a_{0}^{\dagger}\right)\hat a_{j}\right.\\\label{eqquant2expl}&\left.
+e^{\frac{i}{\hbar}\gamma_{j}t}\left(q(c_{j}-2d_{j})\left(\hat a_{0}+\hat a_{0}^{\dagger}\right)\hat a_{-j}^{\dagger}+\frac{1}{q}c_{j}\left(\hat a_{0}-\hat a_{0}^{\dagger}\right)\hat a_{-j}^{\dagger}\right)
-t\,e^{\frac{i}{\hbar}\gamma_{j}t}\frac{2i\omega}{\hbar}qc_{j}\left(\hat a_{0}+\hat a_{0}^{\dagger}\right)\hat a_{-j}^{\dagger}\right],
\end{align}
see  Eq.~\eqref{eqquant2}. The existence of the terms $\sim t\,e^{\mp\frac{i}{\hbar}\gamma_{j}t}$ suggests the following form of the solution of the inhomogeneous Eq.~\eqref{eqquant2expl}:
\begin{equation}\label{substX}
\hat\phi^{\times}_{j}(t,\vec x)=\frac{\omega}{\sqrt{L^{3}}}e^{\frac{i}{\hbar}\vec k_{j}\vec x}\left(e^{-\frac{i}{\hbar}\gamma_{j}t}\left(\hat A_{j}+t\hat B_{j}\right)+e^{\frac{i}{\hbar}\gamma_{j}t}\left(\hat C_{j}+t\hat D_{j}\right)\right).
\end{equation}
Substituting \eqref{substX} into \eqref{eqquant2expl} and combining the terms with the same dependence on time, we get the following system of equations:
\begin{align}
&\left(\gamma_{j}-\frac{{\vec k_{j}}^{2}}{2m}-\omega\right)\hat A_{j}+i\hbar\hat B_{j}-\omega \hat C_{-j}^{\dagger}=q(2c_{j}-d_{j})\left(\hat a_{0}+\hat a_{0}^{\dagger}\right)\hat a_{j}-\frac{1}{q}d_{j}\left(\hat a_{0}-\hat a_{0}^{\dagger}\right)\hat a_{j},\\
&\left(-\gamma_{j}-\frac{{\vec k_{j}}^{2}}{2m}-\omega\right)\hat C_{j}+i\hbar\hat D_{j}-\omega \hat A_{-j}^{\dagger}=q(c_{j}-2d_{j})\left(\hat a_{0}+\hat a_{0}^{\dagger}\right)\hat a_{-j}^{\dagger}+\frac{1}{q}c_{j}\left(\hat a_{0}-\hat a_{0}^{\dagger}\right)\hat a_{-j}^{\dagger},\\
&\left(\gamma_{j}-\frac{{\vec k_{j}}^{2}}{2m}-\omega\right)\hat B_{j}-\omega\hat D_{-j}^{\dagger}=\frac{2i\omega}{\hbar}qd_{j}\left(\hat a_{0}+\hat a_{0}^{\dagger}\right)\hat a_{j},\\
&\left(-\gamma_{j}-\frac{{\vec k_{j}}^{2}}{2m}-\omega\right)\hat D_{j}-\omega\hat B_{-j}^{\dagger}=-\frac{2i\omega}{\hbar}qc_{j}\left(\hat a_{0}+\hat a_{0}^{\dagger}\right)\hat a_{-j}^{\dagger}.
\end{align}
It is convenient to take the solution of this system of equation in the form
\begin{align}\label{Xcoeff1}
&\hat A_{j}=\frac{1}{q\omega}c_{j}\left(\hat a_{0}-\hat a_{0}^{\dagger}\right)\hat a_{j},\\\label{Xcoeff2}
&\hat B_{j}=-\frac{iq}{\hbar}\left(1+\frac{{\vec k_{j}}^{2}}{2m\gamma_{j}}\right)c_{j}\left(\hat a_{0}+\hat a_{0}^{\dagger}\right)\hat a_{j},\\\label{Xcoeff3}
&\hat C_{j}=\frac{q}{\omega^{2}}\left(\gamma_{j}-\frac{{\vec k_{j}}^{2}}{2m}-\omega\right)\frac{{\vec k_{j}}^{2}}{2m\gamma_{j}}c_{j}\left(\hat a_{0}+\hat a_{0}^{\dagger}\right)\hat a_{-j}^{\dagger},\\\label{Xcoeff4}
&\hat D_{j}=\frac{iq}{\hbar}\left(1-2\frac{\gamma_{j}}{\omega}+\frac{{\vec k_{j}}^{2}}{2m\gamma_{j}}\left(1+2\frac{\gamma_{j}}{\omega}\right)\right)c_{j}\left(\hat a_{0}+\hat a_{0}^{\dagger}\right)\hat a_{-j}^{\dagger}.
\end{align}
It is not difficult to check that for solution \eqref{substX} with \eqref{Xcoeff1}--\eqref{Xcoeff4} commutation relations \eqref{CCR1} and \eqref{CCR2} are not satisfied. In particular, the sum of the commutators $[\hat\phi^{\times}_{j}(t,\vec x),\hat\varphi^{\dagger}(t,\vec y)]+[\hat\varphi(t,\vec x),\hat\phi^{\times\dagger}_{j}(t,\vec y)]$ contains the time-independent term
\begin{equation}
\sim\left(\hat a_{0}+\hat a_{0}^{\dagger}\right)e^{\frac{i}{\hbar}\vec k_{j}(\vec x-\vec y)}.
\end{equation}
In order to eliminate this term, let us add to $\hat\phi^{\times}_{j}(t,\vec x)$ the combination
\begin{equation}\label{deltaphiX}
{\scriptstyle\triangle}\hat\phi^{\times}_{j}(t,\vec x)=\frac{\omega}{\sqrt{L^{3}}}e^{\frac{i}{\hbar}\vec k_{j}\vec x}\left(e^{-\frac{i}{\hbar}\gamma_{j}t}c_{j}\hat Q_{j}-e^{\frac{i}{\hbar}\gamma_{j}t}d_{j}\hat Q_{-j}^{\dagger}\right)
\end{equation}
with
\begin{equation}\label{deltaphiXQ}
\hat Q_{j}=\frac{q d_{j}^{2}}{\omega}\frac{{\vec k_{j}}^{2}}{2m\gamma_{j}}\left(\hat a_{0}+\hat a_{0}^{\dagger}\right)\hat a_{j}+\frac{1}{2q\omega}\left(\frac{{\vec k_{j}}^{2}}{2m\gamma_{j}}-1\right)\left(\hat a_{0}-\hat a_{0}^{\dagger}\right)\hat a_{j},
\end{equation}
which is a solution of the homogeneous part of Eq.~\eqref{eqquant2expl}. Now, redefining $\hat\phi_{\times}(t,\vec x)$ as
\begin{equation}\label{phiX2def}
\hat\phi_{\times}(t,\vec x)=\sum\limits_{j\neq 0}\left(\hat\phi^{\times}_{j}(t,\vec x)+{\scriptstyle\triangle}\hat\phi^{\times}_{j}(t,\vec x)\right),
\end{equation}
one can get
\begin{align}\nonumber
&[\hat\phi_{\times}(t,\vec x),\hat\varphi(t,\vec y)]+[\hat\varphi(t,\vec x),\hat\phi_{\times}(t,\vec y)]\\\label{CCR1X}
=\frac{\omega}{L^{3}}\sum\limits_{j\neq 0}&\left(e^{\frac{i}{\hbar}\vec k_{j}\vec x}-e^{\frac{i}{\hbar}\vec k_{j}\vec y}\right)\left(
e^{-\frac{i}{\hbar}\gamma_{j}t}\frac{c_{j}{\vec k_{j}}^{2}}{4m\gamma_{j}^{2}}\left(2\frac{\gamma_{j}}{\omega}+1\right)\hat a_{j}
+e^{\frac{i}{\hbar}\gamma_{j}t}\frac{d_{j}{\vec k_{j}}^{2}}{4m\gamma_{j}^{2}}\left(2\frac{\gamma_{j}}{\omega}-1\right)\hat a_{-j}^{\dagger}
\right)
\end{align}
and
\begin{align}\nonumber
&[\hat\phi_{\times}(t,\vec x),\hat\varphi^{\dagger}(t,\vec y)]+[\hat\varphi(t,\vec x),\hat\phi_{\times}^{\dagger}(t,\vec y)]=\frac{\omega}{L^{3}}\sum\limits_{j\neq 0}\frac{\left(\left(\frac{{\vec k_{j}}^{2}}{2m}\right)^{2}+\omega\frac{{\vec k_{j}}^{2}}{2m}+\gamma_{j}^{2}\right)}{2\omega\gamma_{j}^{2}}\\\label{CCR2X} \times&\left[e^{\frac{i}{\hbar}\vec k_{j}\vec x}\left(
e^{-\frac{i}{\hbar}\gamma_{j}t}c_{j}\hat a_{j}-e^{\frac{i}{\hbar}\gamma_{j}t}d_{j}\hat a_{-j}^{\dagger}\right)+e^{-\frac{i}{\hbar}\vec k_{j}\vec y}\left(
e^{\frac{i}{\hbar}\gamma_{j}t}c_{j}\hat a_{j}^{\dagger}-e^{-\frac{i}{\hbar}\gamma_{j}t}d_{j}\hat a_{-j}\right)\right],
\end{align}
which are still not equal to zero (although now \eqref{CCR2X} does not contain time-independent terms). However, as will be shown below, these nonzero contributions are fully compensated by the other nonzero contributions coming from the oscillation modes.

\subsection{Oscillation modes}
\subsubsection{Equations of motion}
The case of the oscillation modes is the most complicated one, so the analysis will be divided into several parts. To begin with, let us consider the r.h.s. of Eq.~\eqref{eqquant3} and represent it in the explicit form as
\begin{align}\nonumber
&\frac{\omega\sqrt{L^{3}}}{2}\sum\limits_{j\neq 0}\sum\limits_{l\neq 0}\left(
2\hat\varphi_{j}^{\dagger}\hat\varphi_{l}^{}+2\hat\varphi_{l}^{\dagger}\hat\varphi_{j}^{}
+\hat\varphi_{j}^{}\hat\varphi_{l}^{}+\hat\varphi_{l}^{}\hat\varphi_{j}^{}\right)=\frac{\omega}{\sqrt{L^{3}}}\sum\limits_{j\neq 0}\left(\left(2d_{j}^{2}-c_{j}d_{j}\right)\hat a_{j}^{}\hat a_{j}^{\dagger}\right.\\\nonumber&\left.+\left(2c_{j}^{2}-c_{j}d_{j}\right)\hat a_{j}^{\dagger}\hat a_{j}^{}
+e^{-\frac{2i}{\hbar}\gamma_{j}t}\left(c_{j}^{2}-2c_{j}d_{j}\right)\hat a_{j}^{}\hat a_{-j}^{}
+e^{\frac{2i}{\hbar}\gamma_{j}t}\left(d_{j}^{2}-2c_{j}d_{j}\right)\hat a_{j}^{\dagger}\hat a_{-j}^{\dagger}\right)
\\\nonumber
&+\frac{\omega}{2\sqrt{L^{3}}}\underset{j\neq -l}{\sum\limits_{j\neq 0}\sum\limits_{l\neq 0}}\left(q_{j,l}^{(1)}e^{\frac{i}{\hbar}(\vec k_{j}+\vec k_{l})\vec x}e^{-\frac{i}{\hbar}(\gamma_{j}+\gamma_{l})t}\hat a_{j}^{}\hat a_{l}^{}+q_{j,l}^{(2)}e^{-\frac{i}{\hbar}(\vec k_{j}+\vec k_{l})\vec x}e^{\frac{i}{\hbar}(\gamma_{j}+\gamma_{l})t}\hat a_{j}^{\dagger}\hat a_{l}^{\dagger}\right)\\\label{oscrhs}
&+\frac{\omega}{2\sqrt{L^{3}}}\underset{j\neq l}{\sum\limits_{j\neq 0}\sum\limits_{l\neq 0}}\left(
q_{j,l}^{(3)}e^{\frac{i}{\hbar}(\vec k_{j}-\vec k_{l})\vec x}e^{-\frac{i}{\hbar}(\gamma_{j}-\gamma_{l})t}\hat a_{j}^{}\hat a_{l}^{\dagger}+q_{j,l}^{(4)}e^{-\frac{i}{\hbar}(\vec k_{j}-\vec k_{l})\vec x}e^{\frac{i}{\hbar}(\gamma_{j}-\gamma_{l})t}\hat a_{j}^{\dagger}\hat a_{l}^{}\right),
\end{align}
where
\begin{align}\label{qdef1}
&q_{j,l}^{(1)}=2\left(c_{j}c_{l}-c_{j}d_{l}-d_{j}c_{l}\right),\\\label{qdef2}
&q_{j,l}^{(2)}=2\left(d_{j}d_{l}-c_{j}d_{l}-d_{j}c_{l}\right),\\\label{qdef3}
&q_{j,l}^{(3)}=2\left(d_{j}d_{l}+c_{j}c_{l}-c_{j}d_{l}\right),\\\label{qdef4}
&q_{j,l}^{(4)}=2\left(d_{j}d_{l}+c_{j}c_{l}-d_{j}c_{l}\right).
\end{align}
In representation \eqref{oscrhs}, the terms that do not depend on the coordinate $\vec x$ (the first sum), the terms that depend on $e^{\pm\frac{i}{\hbar}(\vec k_{j}+\vec k_{l})\vec x}$ (the second sum), and the terms that depend on $e^{\pm\frac{i}{\hbar}(\vec k_{j}-\vec k_{l})\vec x}$ (the third sum) are separated.

Let us represent the operator $\hat\phi_{\textrm{o}}(t,\vec x)$ as
\begin{equation}\label{oscrepresent1}
\hat\phi_{\textrm{o}}(t,\vec x)=\hat\phi_{\textrm{t}}(t)+\frac{1}{2}\underset{\substack{j\neq 0, l\neq 0\\j\neq-l}}{\sum\sum}\hat\phi_{j,l}^{\scriptscriptstyle+}(t,\vec x)+\frac{1}{2}\underset{\substack{j\neq 0, l\neq 0\\j\neq l}}{\sum\sum}\hat\phi_{j,l}^{\scriptscriptstyle-}(t,\vec x)
\end{equation}
such that the operators $\hat\phi_{\textrm{t}}(t)$, $\hat\phi_{j,l}^{\scriptscriptstyle+}(t,\vec x)$ and $\hat\phi_{j,l}^{\scriptscriptstyle-}(t,\vec x)$ satisfy the equations
\begin{align}\nonumber
&i\hbar\dot{\hat\phi}_{\textrm{t}}^{}-\omega\left(\hat\phi_{\textrm{t}}^{}+\hat\phi_{\textrm{t}}^{\dagger}\right)=\frac{\omega}{\sqrt{L^{3}}}\sum\limits_{j\neq 0}\left(\left(2d_{j}^{2}-c_{j}d_{j}\right)\hat a_{j}^{}\hat a_{j}^{\dagger}\right.\\\label{eqonlyt}&\left.+\left(2c_{j}^{2}-c_{j}d_{j}\right)\hat a_{j}^{\dagger}\hat a_{j}^{}
+e^{-\frac{2i}{\hbar}\gamma_{j}t}\left(c_{j}^{2}-2c_{j}d_{j}\right)\hat a_{j}^{}\hat a_{-j}^{}
+e^{\frac{2i}{\hbar}\gamma_{j}t}\left(d_{j}^{2}-2c_{j}d_{j}\right)\hat a_{j}^{\dagger}\hat a_{-j}^{\dagger}\right),\\\nonumber
&i\hbar\dot{\hat\phi}_{j,l}^{\scriptscriptstyle+}+\frac{\hbar^{2}}{2m}\Delta\hat\phi_{j,l}^{\scriptscriptstyle+}-\omega\left(\hat\phi_{j,l}^{\scriptscriptstyle+}
+\hat\phi_{j,l}^{\scriptscriptstyle+\dagger}\right)\\\label{eqosc+}
&=\frac{\omega}{\sqrt{L^{3}}}\left(q_{j,l}^{(1)}e^{\frac{i}{\hbar}(\vec k_{j}+\vec k_{l})\vec x}e^{-\frac{i}{\hbar}(\gamma_{j}+\gamma_{l})t}\hat a_{j}^{}\hat a_{l}^{}+q_{j,l}^{(2)}e^{-\frac{i}{\hbar}(\vec k_{j}+\vec k_{l})\vec x}e^{\frac{i}{\hbar}(\gamma_{j}+\gamma_{l})t}\hat a_{j}^{\dagger}\hat a_{l}^{\dagger}\right),\\\nonumber
&i\hbar\dot{\hat\phi}_{j,l}^{\scriptscriptstyle-}+\frac{\hbar^{2}}{2m}\Delta\hat\phi_{j,l}^{\scriptscriptstyle-}-\omega\left(\hat\phi_{j,l}^{\scriptscriptstyle-}
+\hat\phi_{j,l}^{\scriptscriptstyle-\dagger}\right)\\\label{eqosc-}
&=\frac{\omega}{\sqrt{L^{3}}}\left(q_{j,l}^{(3)}e^{\frac{i}{\hbar}(\vec k_{j}-\vec k_{l})\vec x}e^{-\frac{i}{\hbar}(\gamma_{j}-\gamma_{l})t}\hat a_{j}^{}\hat a_{l}^{\dagger}+q_{j,l}^{(4)}e^{-\frac{i}{\hbar}(\vec k_{j}-\vec k_{l})\vec x}e^{\frac{i}{\hbar}(\gamma_{j}-\gamma_{l})t}\hat a_{j}^{\dagger}\hat a_{l}^{}\right).
\end{align}
Now let us turn to solving Eqs.~\eqref{eqonlyt}--\eqref{eqosc-}.

\subsubsection{Solution and commutation relations for $\hat\phi_{\textrm{t}}(t)$}\label{sectCompensation}
The case of $\hat\phi_{\textrm{t}}(t)$ is the simplest one among the oscillation modes. The solution of Eq.~\eqref{eqonlyt} can be easily found and has the form
\begin{align}\nonumber
\hat\phi_{\textrm{t}}(t)=\frac{\omega}{\sqrt{L^{3}}}\sum\limits_{l\neq 0}&\left(-\frac{\frac{{\vec k_{j}}^{2}}{2m}+\frac{\omega}{2}+\gamma_{j}}{2\omega\gamma_{j}}
\hat a_{j}^{\dagger}\hat a_{j}^{}-\frac{\frac{{\vec k_{j}}^{2}}{2m}+\frac{\omega}{2}-\gamma_{j}}{2\omega\gamma_{j}}\hat a_{j}^{}\hat a_{j}^{\dagger}\right.\\\label{phitosc}&\left.
+\frac{\gamma_{j}+\frac{{\vec k_{j}}^{2}}{2m}}{4\gamma_{j}^{2}}e^{-\frac{2i}{\hbar}\gamma_{j}t}\hat a_{j}^{}\hat a_{-j}^{}
+\frac{\gamma_{j}-\frac{{\vec k_{j}}^{2}}{2m}}{4\gamma_{j}^{2}}e^{\frac{2i}{\hbar}\gamma_{j}t}\hat a_{j}^{\dagger}\hat a_{-j}^{\dagger}\right).
\end{align}
After straightforward calculations, for the commutation relations one can obtain
\begin{align}\nonumber
&[\hat\phi_{\textrm{t}}(t),\hat\varphi(t,\vec y)]+[\hat\varphi(t,\vec x),\hat\phi_{\textrm{t}}(t)]\\\label{CCR1ot}
=\frac{\omega}{L^{3}}\sum\limits_{j\neq 0}&\left(e^{\frac{i}{\hbar}\vec k_{j}\vec y}-e^{\frac{i}{\hbar}\vec k_{j}\vec x}\right)\left(
e^{-\frac{i}{\hbar}\gamma_{j}t}\frac{c_{j}{\vec k_{j}}^{2}}{4m\gamma_{j}^{2}}\left(2\frac{\gamma_{j}}{\omega}+1\right)\hat a_{j}
+e^{\frac{i}{\hbar}\gamma_{j}t}\frac{d_{j}{\vec k_{j}}^{2}}{4m\gamma_{j}^{2}}\left(2\frac{\gamma_{j}}{\omega}-1\right)\hat a_{-j}^{\dagger}
\right)
\end{align}
and
\begin{align}\nonumber
&[\hat\phi_{\textrm{t}}(t),\hat\varphi^{\dagger}(t,\vec y)]+[\hat\varphi(t,\vec x),\hat\phi_{\textrm{t}}^{\dagger}(t)]=-\frac{\omega}{L^{3}}\sum\limits_{j\neq 0}\frac{\left(\left(\frac{{\vec k_{j}}^{2}}{2m}\right)^{2}+\omega\frac{{\vec k_{j}}^{2}}{2m}+\gamma_{j}^{2}\right)}{2\omega\gamma_{j}^{2}}\\\label{CCR2ot} \times&\left[e^{-\frac{i}{\hbar}\vec k_{j}\vec y}\left(
e^{\frac{i}{\hbar}\gamma_{j}t}c_{j}\hat a_{j}^{\dagger}-e^{-\frac{i}{\hbar}\gamma_{j}t}d_{j}\hat a_{-j}\right)+e^{\frac{i}{\hbar}\vec k_{j}\vec x}\left(
e^{-\frac{i}{\hbar}\gamma_{j}t}c_{j}\hat a_{j}-e^{\frac{i}{\hbar}\gamma_{j}t}d_{j}\hat a_{-j}^{\dagger}\right)\right].
\end{align}
It turns out that \eqref{CCR1ot} and \eqref{CCR2ot} are just \eqref{CCR1X} and \eqref{CCR2X} with the opposite sign. Thus, \eqref{CCR1ot} fully compensates \eqref{CCR1X}, whereas \eqref{CCR2ot} fully compensates \eqref{CCR2X}. A remarkable feature of this compensation is that \eqref{CCR1ot} and \eqref{CCR2ot} contain contribution of only the oscillation modes, whereas \eqref{CCR1X} and \eqref{CCR2X} contain contributions of the oscillation and {\em nonoscillation} modes. In other words, the nonoscillation modes not only recover the canonical commutation relations in the linear approximation, but also provide their fulfillment for the terms of the order of $\beta$. If there were no nonoscillation modes in the theory, such a compensation would be impossible.

\subsubsection{Solutions for $\hat\phi_{j,l}^{\scriptscriptstyle\pm}(t,\vec x)$}
For the most $j$ and $l$ the relations
\begin{align}\label{notransform1}
(\gamma_{j}+\gamma_{l})^{2}-\frac{(\vec k_{j}+\vec k_{l})^{2}}{2m}\left(\frac{(\vec k_{j}+\vec k_{l})^{2}}{2m}+2\omega\right)\neq 0,\\\label{notransform2}
(\gamma_{j}-\gamma_{l})^{2}-\frac{(\vec k_{j}-\vec k_{l})^{2}}{2m}\left(\frac{(\vec k_{j}-\vec k_{l})^{2}}{2m}+2\omega\right)\neq 0
\end{align}
hold. In such a case, solutions of Eqs.~\eqref{eqosc+},~\eqref{eqosc-} can be chosen in the form
\begin{align}\label{phijldef+}
&\hat\phi_{j,l}^{\scriptscriptstyle+}(t,\vec x)=\frac{\omega}{\sqrt{L^{3}}}\left(M_{j,l}^{\star\scriptscriptstyle+}e^{\frac{i}{\hbar}(\vec k_{j}+\vec k_{l})\vec x}e^{-\frac{i}{\hbar}(\gamma_{j}+\gamma_{l})t}\hat a_{j}^{}\hat a_{l}^{}+N_{j,l}^{\star\scriptscriptstyle+}e^{-\frac{i}{\hbar}(\vec k_{j}+\vec k_{l})\vec x}e^{\frac{i}{\hbar}(\gamma_{j}+\gamma_{l})t}\hat a_{j}^{\dagger}\hat a_{l}^{\dagger}\right),\\\label{phijldef-}
&\hat\phi_{j,l}^{\scriptscriptstyle-}(t,\vec x)=\frac{\omega}{\sqrt{L^{3}}}\left(M_{j,l}^{\star\scriptscriptstyle-}e^{\frac{i}{\hbar}(\vec k_{j}-\vec k_{l})\vec x}e^{-\frac{i}{\hbar}(\gamma_{j}-\gamma_{l})t}\hat a_{j}^{}\hat a_{l}^{\dagger}+N_{j,l}^{\star\scriptscriptstyle-}e^{-\frac{i}{\hbar}(\vec k_{j}-\vec k_{l})\vec x}e^{\frac{i}{\hbar}(\gamma_{j}-\gamma_{l})t}\hat a_{j}^{\dagger}\hat a_{l}^{}\right).
\end{align}
Substituting the latter representations into Eqs.~\eqref{eqosc+},~\eqref{eqosc-}, for the coefficients in $\hat\phi_{j,l}^{\scriptscriptstyle+}(t,\vec x)$ we get the system of equations
\begin{align}\label{tildeNeq+}
&M_{j,l}^{\star\scriptscriptstyle+}\left(\frac{(\vec k_{j}+\vec k_{l})^{2}}{2m}+\omega-(\gamma_{j}+\gamma_{l})\right)+\omega \left(N_{j,l}^{\star\scriptscriptstyle+}\right)^{*}=-q_{j,l}^{(1)},\\
&M_{j,l}^{\star\scriptscriptstyle+}\omega+\left(\frac{(\vec k_{j}+\vec k_{l})^{2}}{2m}+\omega+(\gamma_{j}+\gamma_{l})\right) \left(N_{j,l}^{\star\scriptscriptstyle+}\right)^{*}=-q_{j,l}^{(2)},
\end{align}
leading to
\begin{align}\label{coefftildeM+}
&M_{j,l}^{\star\scriptscriptstyle+}=\frac{\left(\frac{(\vec k_{j}+\vec k_{l})^{2}}{2m}+(\gamma_{j}+\gamma_{l})\right)q_{j,l}^{(1)}+\omega\left( q_{j,l}^{(1)}-q_{j,l}^{(2)}\right)}
{(\gamma_{j}+\gamma_{l})^{2}+\omega^{2}-\left(\frac{(\vec k_{j}+\vec k_{l})^{2}}{2m}+\omega\right)^{2}},\\\label{coefftildeN+}
&N_{j,l}^{\star\scriptscriptstyle+}=\frac{\left(\frac{(\vec k_{j}+\vec k_{l})^{2}}{2m}-(\gamma_{j}+\gamma_{l})\right)q_{j,l}^{(2)}-\omega \left( q_{j,l}^{(1)}-q_{j,l}^{(2)}\right)}
{(\gamma_{j}+\gamma_{l})^{2}+\omega^{2}-\left(\frac{(\vec k_{j}+\vec k_{l})^{2}}{2m}+\omega\right)^{2}};
\end{align}
whereas for the coefficients in $\hat\phi_{j,l}^{\scriptscriptstyle-}(t,\vec x)$ we get the system of equations
\begin{align}\label{tildeNeq-}
&M_{j,l}^{\star\scriptscriptstyle-}\left(\frac{(\vec k_{j}-\vec k_{l})^{2}}{2m}+\omega-(\gamma_{j}-\gamma_{l})\right)+\omega \left(N_{j,l}^{\star\scriptscriptstyle-}\right)^{*}=-q_{j,l}^{(3)},\\
&M_{j,l}^{\star\scriptscriptstyle-}\omega+\left(\frac{(\vec k_{j}-\vec k_{l})^{2}}{2m}+\omega+(\gamma_{j}-\gamma_{l})\right) \left(N_{j,l}^{\star\scriptscriptstyle-}\right)^{*}=-q_{j,l}^{(4)},
\end{align}
leading to
\begin{align}\label{coefftildeM-}
&M_{j,l}^{\star\scriptscriptstyle-}=\frac{\left(\frac{(\vec k_{j}-\vec k_{l})^{2}}{2m}+(\gamma_{j}-\gamma_{l})\right)q_{j,l}^{(3)}+\omega\left( q_{j,l}^{(3)}-q_{j,l}^{(4)}\right)}
{(\gamma_{j}-\gamma_{l})^{2}+\omega^{2}-\left(\frac{(\vec k_{j}-\vec k_{l})^{2}}{2m}+\omega\right)^{2}},\\\label{coefftildeN-}
&N_{j,l}^{\star\scriptscriptstyle-}=\frac{\left(\frac{(\vec k_{j}-\vec k_{l})^{2}}{2m}-(\gamma_{j}-\gamma_{l})\right)q_{j,l}^{(4)}-\omega\left( q_{j,l}^{(3)}-q_{j,l}^{(4)}\right)}
{(\gamma_{j}-\gamma_{l})^{2}+\omega^{2}-\left(\frac{(\vec k_{j}-\vec k_{l})^{2}}{2m}+\omega\right)^{2}}.
\end{align}
Coefficients \eqref{coefftildeM+}, \eqref{coefftildeN+}, \eqref{coefftildeM-} and \eqref{coefftildeN-} have the following properties:
\begin{equation}\label{starMNsymmetry}
M_{l,j}^{\star\scriptscriptstyle+}=M_{j,l}^{\star\scriptscriptstyle+},\quad N_{l,j}^{\star\scriptscriptstyle+}=N_{j,l}^{\star\scriptscriptstyle+},\quad  N_{l,j}^{\star\scriptscriptstyle-}=M_{j,l}^{\star\scriptscriptstyle-}.
\end{equation}

However, there may exist some $j$ and $l$ such that for $k=j+l$ (which automatically implies $\vec k_{k}=\vec k_{j}+\vec k_{l}$) the frequency $\gamma_{k}$ is such that $\gamma_{k}=\gamma_{j+l}=\gamma_{j}+\gamma_{l}$ (analogously, for $\gamma_{j-l}=|\gamma_{j}-\gamma_{l}|$), i.e., it is possible that for some $j$ and $l$ the equality
\begin{equation}\label{transform1}
(\gamma_{j}+\gamma_{l})^{2}-\frac{(\vec k_{j}+\vec k_{l})^{2}}{2m}\left(\frac{(\vec k_{j}+\vec k_{l})^{2}}{2m}+2\omega\right)=0
\end{equation}
is fulfilled or the equality
\begin{equation}\label{transform2}
(\gamma_{j}-\gamma_{l})^{2}-\frac{(\vec k_{j}-\vec k_{l})^{2}}{2m}\left(\frac{(\vec k_{j}-\vec k_{l})^{2}}{2m}+2\omega\right)=0
\end{equation}
is fulfilled. A simple example of such a situation in the limit $L\to\infty$ is presented in Appendix~A. In such a case, solutions of Eqs.~\eqref{eqosc+},~\eqref{eqosc-} should be chosen in a more complicated form
\begin{align}\label{phijldef+1}
\hat\phi_{j,l}^{\scriptscriptstyle+}(t,\vec x)=\frac{\omega}{L^{\frac{3}{2}}}\left(\left(M_{j,l}^{\scriptscriptstyle+}+tL_{j,l}^{\scriptscriptstyle+}\right)e^{\frac{i}{\hbar}(\vec k_{j}+\vec k_{l})\vec x}e^{-\frac{i}{\hbar}(\gamma_{j}+\gamma_{l})t}\hat a_{j}^{}\hat a_{l}^{}+\left(N_{j,l}^{\scriptscriptstyle+}+tJ_{j,l}^{\scriptscriptstyle+}\right)e^{-\frac{i}{\hbar}(\vec k_{j}+\vec k_{l})\vec x}e^{\frac{i}{\hbar}(\gamma_{j}+\gamma_{l})t}\hat a_{j}^{\dagger}\hat a_{l}^{\dagger}\right),\\\label{phijldef-1}
\hat\phi_{j,l}^{\scriptscriptstyle-}(t,\vec x)=\frac{\omega}{L^{\frac{3}{2}}}\left(\left(M_{j,l}^{\scriptscriptstyle-}+tL_{j,l}^{\scriptscriptstyle-}\right)e^{\frac{i}{\hbar}(\vec k_{j}-\vec k_{l})\vec x}e^{-\frac{i}{\hbar}(\gamma_{j}-\gamma_{l})t}\hat a_{j}^{}\hat a_{l}^{\dagger}+\left(N_{j,l}^{\scriptscriptstyle-}+tJ_{j,l}^{\scriptscriptstyle-}\right)e^{-\frac{i}{\hbar}(\vec k_{j}-\vec k_{l})\vec x}e^{\frac{i}{\hbar}(\gamma_{j}-\gamma_{l})t}\hat a_{j}^{\dagger}\hat a_{l}^{}\right).
\end{align}
Substituting these representations into Eqs.~\eqref{eqosc+},~\eqref{eqosc-}, for the coefficients in $\hat\phi_{j,l}^{\scriptscriptstyle+}(t,\vec x)$ defined by \eqref{phijldef+1} we get the system of equations
\begin{align}\label{LJeq+}
&L_{j,l}^{\scriptscriptstyle+}\left(\frac{(\vec k_{j}+\vec k_{l})^{2}}{2m}+\omega-(\gamma_{j}+\gamma_{l})\right)+\omega \left(J_{j,l}^{\scriptscriptstyle+}\right)^{*}=0,\\
&L_{j,l}^{\scriptscriptstyle+}\omega+\left(\frac{(\vec k_{j}+\vec k_{l})^{2}}{2m}+\omega+(\gamma_{j}+\gamma_{l})\right)\left(J_{j,l}^{\scriptscriptstyle+}\right)^{*}=0,\\
&M_{j,l}^{\scriptscriptstyle+}\left(\frac{(\vec k_{j}+\vec k_{l})^{2}}{2m}+\omega-(\gamma_{j}+\gamma_{l})\right)+\omega\left(N_{j,l}^{\scriptscriptstyle+}\right)^{*}-i\hbar L_{j,l}^{\scriptscriptstyle+}=-q_{j,l}^{(1)},\\
&M_{j,l}^{\scriptscriptstyle+}\omega+\left(\frac{(\vec k_{j}+\vec k_{l})^{2}}{2m}+\omega+(\gamma_{j}+\gamma_{l})\right)\left(N_{j,l}^{\scriptscriptstyle+}\right)^{*}+i\hbar\left(J_{j,l}^{\scriptscriptstyle+}\right)^{*}=-q_{j,l}^{(2)},
\end{align}
leading to
\begin{align}\label{L+def}
&L_{j,l}^{\scriptscriptstyle+}=\frac{-i}{2\hbar(\gamma_{j}+\gamma_{l})}\left(\left(\frac{(\vec k_{j}+\vec k_{l})^{2}}{2m}+(\gamma_{j}+\gamma_{l})\right)q_{j,l}^{(1)}+\omega\left(q_{j,l}^{(1)}-q_{j,l}^{(2)}\right)\right),\\\label{J+def}
&J_{j,l}^{\scriptscriptstyle+}=\frac{i}{2\hbar(\gamma_{j}+\gamma_{l})}\left(\left(\frac{(\vec k_{j}+\vec k_{l})^{2}}{2m}-(\gamma_{j}+\gamma_{l})\right)q_{j,l}^{(2)}-\omega\left(q_{j,l}^{(1)}-q_{j,l}^{(2)}\right)\right),\\\label{M+def}
&M_{j,l}^{\scriptscriptstyle+}=K_{j,l}^{\scriptscriptstyle+}+c_{j+l}Q_{j,l}^{\scriptscriptstyle+},\\
\label{N+def}
&N_{j,l}^{\scriptscriptstyle+}=-d_{j+l}Q_{j,l}^{\scriptscriptstyle+},
\end{align}
where
\begin{equation}\label{Kab+}
K_{j,l}^{\scriptscriptstyle+}=\frac{-1}{2\omega(\gamma_{j}+\gamma_{l})}\left(\left(\frac{(\vec k_{j}+\vec k_{l})^{2}}{2m}+(\gamma_{j}+\gamma_{l})\right)q_{j,l}^{(2)}-\omega\left(q_{j,l}^{(1)}-q_{j,l}^{(2)}\right)\right)
\end{equation}
and the coefficients $Q_{j,l}^{\scriptscriptstyle+}$ are not specified and can be arbitrary. Note that
\begin{equation}\label{K+symmetry}
K_{j,l}^{\scriptscriptstyle+}=K_{l,j}^{\scriptscriptstyle+}.
\end{equation}
For the coefficients in $\hat\phi_{j,l}^{\scriptscriptstyle-}(t,\vec x)$ defined by \eqref{phijldef-1} we get the system of equations
\begin{align}\label{LJeq-}
&L_{j,l}^{\scriptscriptstyle-}\left(\frac{(\vec k_{j}-\vec k_{l})^{2}}{2m}+\omega-(\gamma_{j}-\gamma_{l})\right)+\omega\left(J_{j,l}^{\scriptscriptstyle-}\right)^{*}=0,\\
&L_{j,l}^{\scriptscriptstyle-}\omega+\left(\frac{(\vec k_{j}-\vec k_{l})^{2}}{2m}+\omega+(\gamma_{j}-\gamma_{l})\right)\left(J_{j,l}^{\scriptscriptstyle-}\right)^{*}=0,\\
&M_{j,l}^{\scriptscriptstyle-}\left(\frac{(\vec k_{j}-\vec k_{l})^{2}}{2m}+\omega-(\gamma_{j}-\gamma_{l})\right)+\omega\left(N_{j,l}^{\scriptscriptstyle-}\right)^{*}-i\hbar L_{j,l}^{\scriptscriptstyle-}=-q_{j,l}^{(3)},\\
&M_{j,l}^{\scriptscriptstyle-}\omega+\left(\frac{(\vec k_{j}-\vec k_{l})^{2}}{2m}+\omega+(\gamma_{j}-\gamma_{l})\right)\left(N_{j,l}^{\scriptscriptstyle-}\right)^{*}+i\hbar\left(J_{j,l}^{\scriptscriptstyle-}\right)^{*}=-q_{j,l}^{(4)},
\end{align}
leading to
\begin{align}\label{L-def}
&L_{j,l}^{\scriptscriptstyle-}=\frac{-i}{2\hbar(\gamma_{j}-\gamma_{l})}\left(\left(\frac{(\vec k_{j}-\vec k_{l})^{2}}{2m}+(\gamma_{j}-\gamma_{l})\right)q_{j,l}^{(3)}+\omega\left(q_{j,l}^{(3)}-q_{j,l}^{(4)}\right)\right),\\\label{J-def}
&J_{j,l}^{\scriptscriptstyle-}=\frac{i}{2\hbar(\gamma_{j}-\gamma_{l})}\left(\left(\frac{(\vec k_{j}-\vec k_{l})^{2}}{2m}-(\gamma_{j}-\gamma_{l})\right)q_{j,l}^{(4)}-\omega\left(q_{j,l}^{(3)}-q_{j,l}^{(4)}\right)\right),\\\label{M-def}
&M_{j,l}^{\scriptscriptstyle-}=K_{j,l}^{\scriptscriptstyle-}+c_{j-l}Q_{j,l}^{\scriptscriptstyle-},\\
\label{N-def}
&N_{j,l}^{\scriptscriptstyle-}=-d_{j-l}Q_{j,l}^{\scriptscriptstyle-},
\end{align}
where
\begin{equation}\label{Kab-}
K_{j,l}^{\scriptscriptstyle-}=\frac{-1}{2\omega(\gamma_{j}-\gamma_{l})}\left(\left(\frac{(\vec k_{j}-\vec k_{l})^{2}}{2m}+(\gamma_{j}-\gamma_{l})\right)q_{j,l}^{(4)}-\omega\left(q_{j,l}^{(3)}-q_{j,l}^{(4)}\right)\right)
\end{equation}
and the coefficients $Q_{j,l}^{\scriptscriptstyle-}$ are not specified and can also be arbitrary.

At this step it is convenient to represent the sums in \eqref{oscrepresent1} as
\begin{align}\nonumber
&\frac{1}{2}\underset{\substack{j\neq 0, l\neq 0\\j\neq-l}}{\sum\sum}\hat\phi_{j,l}^{\scriptscriptstyle+}(t,\vec x)+\frac{1}{2}\underset{\substack{j\neq 0, l\neq 0\\j\neq l}}{\sum\sum}\hat\phi_{j,l}^{\scriptscriptstyle-}(t,\vec x)\\\label{oscrepresent2}&=\frac{1}{2}\underset{\substack{j\neq 0, l\neq 0\\j\neq-l}}{\sum\sum}\hat{\tilde\phi}_{j,l}^{\scriptscriptstyle+}(t,\vec x)+
\frac{1}{2}\underset{\substack{j\neq 0, l\neq 0\\j\neq l}}{\sum\sum}\hat{\tilde\phi}_{j,l}^{\scriptscriptstyle-}(t,\vec x)+
\frac{1}{2}\underset{\substack{[j,l]}}{\sum\sum}\hat\rho_{j,l}^{\scriptscriptstyle+}(t,\vec x)+\frac{1}{2}\underset{\substack{(j,l)}}{\sum\sum}\hat\rho_{j,l}^{\scriptscriptstyle-}(t,\vec x).
\end{align}
Here
\begin{align}
&\hat{\tilde\phi}_{j,l}^{\scriptscriptstyle+}(t,\vec x)=\frac{\omega}{\sqrt{L^{3}}}\left(\tilde M_{j,l}^{\scriptscriptstyle+}e^{\frac{i}{\hbar}(\vec k_{j}+\vec k_{l})\vec x}e^{-\frac{i}{\hbar}(\gamma_{j}+\gamma_{l})t}\hat a_{j}^{}\hat a_{l}^{}+\tilde N_{j,l}^{\scriptscriptstyle+}e^{-\frac{i}{\hbar}(\vec k_{j}+\vec k_{l})\vec x}e^{\frac{i}{\hbar}(\gamma_{j}+\gamma_{l})t}\hat a_{j}^{\dagger}\hat a_{l}^{\dagger}\right),\\
&\hat{\tilde\phi}_{j,l}^{\scriptscriptstyle-}(t,\vec x)=\frac{\omega}{\sqrt{L^{3}}}\left(\tilde M_{j,l}^{\scriptscriptstyle-}e^{\frac{i}{\hbar}(\vec k_{j}-\vec k_{l})\vec x}e^{-\frac{i}{\hbar}(\gamma_{j}-\gamma_{l})t}\hat a_{j}^{}\hat a_{l}^{\dagger}+\tilde N_{j,l}^{\scriptscriptstyle-}e^{-\frac{i}{\hbar}(\vec k_{j}-\vec k_{l})\vec x}e^{\frac{i}{\hbar}(\gamma_{j}-\gamma_{l})t}\hat a_{j}^{\dagger}\hat a_{l}^{}\right),
\end{align}
where
\begin{equation}\label{tildeMN1def}
\tilde M_{j,l}^{\scriptscriptstyle+}=M_{j,l}^{\star\scriptscriptstyle+},\qquad \tilde N_{j,l}^{\scriptscriptstyle+}=N_{j,l}^{\star\scriptscriptstyle+}
\end{equation}
if $j$ and $l$ are such that relation \eqref{transform1} {\em is not} fulfilled;
\begin{equation}\label{tildeMN2def}
\tilde M_{j,l}^{\scriptscriptstyle-}=M_{j,l}^{\star\scriptscriptstyle-},\qquad \tilde N_{j,l}^{\scriptscriptstyle-}=N_{j,l}^{\star\scriptscriptstyle-}
\end{equation}
if $j$ and $l$ are such that relation \eqref{transform2} {\em is not} fulfilled;
\begin{equation}\label{tildeMN3def}
\tilde M_{j,l}^{\scriptscriptstyle+}=M_{j,l}^{\scriptscriptstyle+},\qquad \tilde N_{j,l}^{\scriptscriptstyle+}=N_{j,l}^{\scriptscriptstyle+}
\end{equation}
if $j$ and $l$ are such that relation \eqref{transform1} {\em is} fulfilled; and
\begin{equation}\label{tildeMN4def}
\tilde M_{j,l}^{\scriptscriptstyle-}=M_{j,l}^{\scriptscriptstyle-},\qquad \tilde N_{j,l}^{\scriptscriptstyle-}=N_{j,l}^{\scriptscriptstyle-}
\end{equation}
if $j$ and $l$ are such that relation \eqref{transform2} {\em is} fulfilled. The operators $\hat\rho_{j,l}^{\scriptscriptstyle\pm}(t,\vec x)$ are defined by
\begin{align}
\hat\rho_{j,l}^{\scriptscriptstyle+}(t,\vec x)=\frac{\omega t}{\sqrt{L^{3}}}\left(L_{j,l}^{\scriptscriptstyle+}e^{\frac{i}{\hbar}(\vec k_{j}+\vec k_{l})\vec x}e^{-\frac{i}{\hbar}(\gamma_{j}+\gamma_{l})t}\hat a_{j}^{}\hat a_{l}^{}+J_{j,l}^{\scriptscriptstyle+}e^{-\frac{i}{\hbar}(\vec k_{j}+\vec k_{l})\vec x}e^{\frac{i}{\hbar}(\gamma_{j}+\gamma_{l})t}\hat a_{j}^{\dagger}\hat a_{l}^{\dagger}\right),\\
\hat\rho_{j,l}^{\scriptscriptstyle-}(t,\vec x)=\frac{\omega t}{\sqrt{L^{3}}}\left(L_{j,l}^{\scriptscriptstyle-}e^{\frac{i}{\hbar}(\vec k_{j}-\vec k_{l})\vec x}e^{-\frac{i}{\hbar}(\gamma_{j}-\gamma_{l})t}\hat a_{j}^{}\hat a_{l}^{\dagger}+J_{j,l}^{\scriptscriptstyle-}e^{-\frac{i}{\hbar}(\vec k_{j}-\vec k_{l})\vec x}e^{\frac{i}{\hbar}(\gamma_{j}-\gamma_{l})t}\hat a_{j}^{\dagger}\hat a_{l}^{}\right)
\end{align}
and both contain the overall factor $t$. The notation $[j,l]$ in \eqref{oscrepresent2} denotes the set of $j$ and $l$ for which relation \eqref{transform1} is fulfilled, whereas the notation $(j,l)$ in \eqref{oscrepresent2} denotes the set of $j$ and $l$ for which relation \eqref{transform2} is fulfilled.

Now we are ready to calculate the commutation relations.

\subsubsection{Commutation relations for $\hat{\tilde\phi}_{j,l}^{\scriptscriptstyle\pm}(t,\vec x)$}
Let us start with the first commutation relation and define
\begin{equation}\label{tildephidef}
\hat{\tilde\phi}^{\scriptscriptstyle}(t,\vec x)=\frac{1}{2}\underset{\substack{j\neq 0, l\neq 0\\j\neq-l}}{\sum\sum}\hat{\tilde\phi}_{j,l}^{\scriptscriptstyle+}(t,\vec x)+\frac{1}{2}\underset{\substack{j\neq 0, l\neq 0\\j\neq l}}{\sum\sum}\hat{\tilde\phi}_{j,l}^{\scriptscriptstyle-}(t,\vec x),
\end{equation}
see \eqref{oscrepresent2}. For the operator $\hat{\tilde\phi}^{\scriptscriptstyle}(t,\vec x)$ one can get (see Appendix~B for detailed calculations)
\begin{align}\nonumber
&[\hat{\tilde\phi}(t,\vec x),\hat\varphi(t,\vec y)]+[\hat\varphi(t,\vec x),\hat{\tilde\phi}(t,\vec y)]\\\nonumber
=&\frac{\omega}{L^{3}}\underset{\substack{j\neq 0, l\neq 0\\j\neq-l}}{\sum\sum}\Biggl(\hat a_{j}^{}e^{-\frac{i}{\hbar}\gamma_{j}t}e^{\frac{i}{\hbar}(\vec k_{j}+\vec k_{l})\vec y}e^{-\frac{i}{\hbar}\vec k_{l}\vec x}\\\nonumber\times&\left(\frac{\tilde M_{j,l}^{\scriptscriptstyle+}+\tilde M_{l,j}^{\scriptscriptstyle+}}{2}d_{l}-\frac{\tilde M_{j,-j-l}^{\scriptscriptstyle+}+\tilde M_{-j-l,j}^{\scriptscriptstyle+}}{2}d_{j+l}+\frac{\tilde M_{j,-l}^{\scriptscriptstyle-}+\tilde N_{-l,j}^{\scriptscriptstyle-}}{2}c_{l}-\frac{\tilde M_{j,j+l}^{\scriptscriptstyle-}+\tilde N_{j+l,j}^{\scriptscriptstyle-}}{2}c_{j+l}\right)\\\nonumber
+&\hat a_{j}^{\dagger}e^{\frac{i}{\hbar}\gamma_{j}t}e^{-\frac{i}{\hbar}(\vec k_{j}+\vec k_{l})\vec y}e^{\frac{i}{\hbar}\vec k_{l}\vec x}
\\\label{commtildefinal}
\times&\left(\frac{\tilde N_{j,l}^{\scriptscriptstyle+}+\tilde N_{l,j}^{\scriptscriptstyle+}}{2}c_{l}-\frac{\tilde N_{j,-j-l}^{\scriptscriptstyle+}+\tilde N_{-j-l,j}^{\scriptscriptstyle+}}{2}c_{j+l}+\frac{\tilde N_{j,-l}^{\scriptscriptstyle-}+\tilde M_{-l,j}^{\scriptscriptstyle-}}{2}d_{l}-\frac{\tilde N_{j,j+l}^{\scriptscriptstyle-}+\tilde M_{j+l,j}^{\scriptscriptstyle-}}{2}d_{j+l}\right)\Biggr).
\end{align}
For the second commutation relation one can get (see Appendix~C for detailed calculations)
\begin{align}\nonumber
&[\hat{\tilde\phi}(t,\vec x),\hat\varphi^{\dagger}(t,\vec y)]+[\hat\varphi(t,\vec x),\hat{\tilde\phi}^{\dagger}(t,\vec y)]\\\nonumber
=&\frac{\omega}{L^{3}}\underset{\substack{j\neq 0, l\neq 0\\j\neq-l}}{\sum\sum}\Biggl(\hat a_{j}^{\dagger}e^{\frac{i}{\hbar}\gamma_{j}t}e^{-\frac{i}{\hbar}(\vec k_{j}+\vec k_{l})\vec y}e^{\frac{i}{\hbar}\vec k_{l}\vec x}\\\nonumber\times&\left(\frac{\tilde M_{j,l}^{\scriptscriptstyle+}+\tilde M_{l,j}^{\scriptscriptstyle+}}{2}c_{l}+\frac{\tilde N_{j,-j-l}^{\scriptscriptstyle+}+\tilde N_{-j-l,j}^{\scriptscriptstyle+}}{2}d_{j+l}+\frac{\tilde M_{j,-l}^{\scriptscriptstyle-}+\tilde N_{-l,j}^{\scriptscriptstyle-}}{2}d_{l}+
\frac{\tilde N_{j,j+l}^{\scriptscriptstyle-}+\tilde M_{j+l,j}^{\scriptscriptstyle-}}{2}c_{j+l}\right)
\\\nonumber
+&\hat a_{j}^{}e^{-\frac{i}{\hbar}\gamma_{j}t}e^{\frac{i}{\hbar}(\vec k_{j}+\vec k_{l})\vec x}e^{-\frac{i}{\hbar}\vec k_{l}\vec y}\\\label{commdagger}\times&\left(\frac{\tilde M_{j,l}^{\scriptscriptstyle+}+\tilde M_{l,j}^{\scriptscriptstyle+}}{2}c_{l}+\frac{\tilde N_{j,-j-l}^{\scriptscriptstyle+}+\tilde N_{-j-l,j}^{\scriptscriptstyle+}}{2}d_{j+l}+\frac{\tilde M_{j,-l}^{\scriptscriptstyle-}+\tilde N_{-l,j}^{\scriptscriptstyle-}}{2}d_{l}+
\frac{\tilde N_{j,j+l}^{\scriptscriptstyle-}+\tilde M_{j+l,j}^{\scriptscriptstyle-}}{2}c_{j+l}\right)\Biggr).
\end{align}
Thus, in the combinations of commutators in \eqref{commtildefinal} and \eqref{commdagger} we have the three types of coefficients, which are
\begin{align}\label{commcoeff1}
&\frac{\tilde M_{j,l}^{\scriptscriptstyle+}+\tilde M_{l,j}^{\scriptscriptstyle+}}{2}d_{l}-\frac{\tilde M_{j,-j-l}^{\scriptscriptstyle+}+\tilde M_{-j-l,j}^{\scriptscriptstyle+}}{2}d_{j+l}+\frac{\tilde M_{j,-l}^{\scriptscriptstyle-}+\tilde N_{-l,j}^{\scriptscriptstyle-}}{2}c_{l}-\frac{\tilde M_{j,j+l}^{\scriptscriptstyle-}+\tilde N_{j+l,j}^{\scriptscriptstyle-}}{2}c_{j+l},\\\label{commcoeff2}
&\frac{\tilde N_{j,l}^{\scriptscriptstyle+}+\tilde N_{l,j}^{\scriptscriptstyle+}}{2}c_{l}-\frac{\tilde N_{j,-j-l}^{\scriptscriptstyle+}+\tilde N_{-j-l,j}^{\scriptscriptstyle+}}{2}c_{j+l}+\frac{\tilde N_{j,-l}^{\scriptscriptstyle-}+\tilde M_{-l,j}^{\scriptscriptstyle-}}{2}d_{l}-\frac{\tilde N_{j,j+l}^{\scriptscriptstyle-}+\tilde M_{j+l,j}^{\scriptscriptstyle-}}{2}d_{j+l},\\\label{commcoeff3}
&\frac{\tilde M_{j,l}^{\scriptscriptstyle+}+\tilde M_{l,j}^{\scriptscriptstyle+}}{2}c_{l}+\frac{\tilde N_{j,-j-l}^{\scriptscriptstyle+}+\tilde N_{-j-l,j}^{\scriptscriptstyle+}}{2}d_{j+l}+\frac{\tilde M_{j,-l}^{\scriptscriptstyle-}+\tilde N_{-l,j}^{\scriptscriptstyle-}}{2}d_{l}+
\frac{\tilde N_{j,j+l}^{\scriptscriptstyle-}+\tilde M_{j+l,j}^{\scriptscriptstyle-}}{2}c_{j+l}.
\end{align}
These coefficients should be calculated in order to check whether commutation relations \eqref{CCR1} and \eqref{CCR2} are satisfied for $\hat{\tilde\phi}(t,\vec x)$.

First, let us consider the case in which $j$ and $l$ in \eqref{commcoeff1}--\eqref{commcoeff3} are such that for all the terms in \eqref{commcoeff1}--\eqref{commcoeff3} relations \eqref{transform1}, \eqref{transform2} {\em are not} fulfilled, i.e., the case of \eqref{tildeMN1def}, \eqref{tildeMN2def}. Using relations \eqref{starMNsymmetry}, these coefficients can be rewritten as
\begin{align}\label{commcoeff1star}
&M_{j,l}^{\star\scriptscriptstyle+}d_{l}-M_{j,-j-l}^{\star\scriptscriptstyle+}d_{j+l}+M_{j,-l}^{\star\scriptscriptstyle-}c_{l}- M_{j,j+l}^{\star\scriptscriptstyle-}c_{j+l},\\\label{commcoeff2star}
&N_{j,l}^{\star\scriptscriptstyle+}c_{l}-N_{j,-j-l}^{\star\scriptscriptstyle+}c_{j+l}+N_{j,-l}^{\star\scriptscriptstyle-}d_{l}- N_{j,j+l}^{\star\scriptscriptstyle-}d_{j+l},\\\label{commcoeff3star}
&M_{j,l}^{\star\scriptscriptstyle+}c_{l}+N_{j,-j-l}^{\star\scriptscriptstyle+}d_{j+l}+M_{j,-l}^{\star\scriptscriptstyle-}d_{l}+
N_{j,j+l}^{\star\scriptscriptstyle-}c_{j+l}.
\end{align}
The main idea is to express all the terms in \eqref{commcoeff1star}--\eqref{commcoeff3star} trough $\gamma_{\scriptscriptstyle\#}$ with the subscript ``$\#$'' denoting the corresponding index. Using \eqref{coefftildeM+}, \eqref{coefftildeM-}, definitions \eqref{cddef}, \eqref{qdef1}--\eqref{qdef4}, and the relations
\begin{equation}\label{relationsKtoGamma}
\frac{\vec k_{j}^{2}}{2m}=\sqrt{\gamma_{j}^{2}+\omega^{2}}-\omega,\qquad \frac{\vec k_{l}^{2}}{2m}=\sqrt{\gamma_{l}^{2}+\omega^{2}}-\omega,\qquad
\frac{(\vec k_{j}+\vec k_{l})^{2}}{2m}=\sqrt{\gamma_{j+l}^{2}+\omega^{2}}-\omega
\end{equation}
following from \eqref{gammadef}, after straightforward calculations the terms in \eqref{commcoeff1star} can be represented in the explicit form as
{\small
\begin{align}\nonumber
M_{j,l}^{\star\scriptscriptstyle+}d_{l}=&\frac{c_{j}}{\gamma_{l}\left(\left(\gamma_{j}+\gamma_{l}\right)^{2}-\gamma_{j+l}^{2}\right)}\\\nonumber\times
&\left[\left(\gamma_{j}+\gamma_{l}+\sqrt{\gamma_{j+l}^{2}+\omega^{2}}-\omega\right)\left(\gamma_{j}+\gamma_{l}
-\sqrt{\gamma_{j}^{2}+\omega^{2}}-\sqrt{\gamma_{l}^{2}+\omega^{2}}+\omega\right)\right.\\\label{termcomm1}&\left.+\omega^{2}-\left(\sqrt{\gamma_{j}^{2}+\omega^{2}}-\gamma_{j}\right)
\left(\sqrt{\gamma_{l}^{2}+\omega^{2}}-\gamma_{l}\right)\right],\\\nonumber
M_{j,-j-l}^{\star\scriptscriptstyle+}d_{j+l}=&\frac{c_{j}}{\gamma_{j+l}\left(\left(\gamma_{j}+\gamma_{j+l}\right)^{2}-\gamma_{l}^{2}\right)}\\\nonumber\times
&\left[\left(\gamma_{j}+\gamma_{j+l}+\sqrt{\gamma_{l}^{2}+\omega^{2}}-\omega\right)\left(\gamma_{j}+\gamma_{j+l}
-\sqrt{\gamma_{j}^{2}+\omega^{2}}-\sqrt{\gamma_{j+l}^{2}+\omega^{2}}+\omega\right)\right.\\\label{termcomm2}&\left.+\omega^{2}-\left(\sqrt{\gamma_{j}^{2}+\omega^{2}}-\gamma_{j}\right)
\left(\sqrt{\gamma_{j+l}^{2}+\omega^{2}}-\gamma_{j+l}\right)\right],\\\nonumber
M_{j,-l}^{\star\scriptscriptstyle-}c_{l}=&\frac{c_{j}}{\gamma_{l}\left(\left(\gamma_{j}-\gamma_{l}\right)^{2}-\gamma_{j+l}^{2}\right)\left(\sqrt{\gamma_{l}^{2}+\omega^{2}}-\gamma_{l}\right)}
\\\nonumber\times
&\left[\left(\sqrt{\gamma_{j+l}^{2}+\omega^{2}}+\gamma_{j}-\gamma_{l}-\omega\right)\left(\left(\sqrt{\gamma_{j}^{2}+\omega^{2}}-\gamma_{j}-\omega\right)
\left(\sqrt{\gamma_{l}^{2}+\omega^{2}}-\gamma_{l}\right)+\omega^{2}\right)\right.\\\label{termcomm3}&\left.
+\omega^{2}\left(\sqrt{\gamma_{j}^{2}+\omega^{2}}-\gamma_{j}-\sqrt{\gamma_{l}^{2}+\omega^{2}}+\gamma_{l}\right)\right],\\\nonumber
M_{j,j+l}^{\star\scriptscriptstyle-}c_{j+l}=&\frac{c_{j}}{\gamma_{j+l}\left(\left(\gamma_{j}-\gamma_{j+l}\right)^{2}-\gamma_{l}^{2}\right)\left(\sqrt{\gamma_{j+l}^{2}+\omega^{2}}-\gamma_{j+l}\right)}
\\\nonumber\times
&\left[\left(\sqrt{\gamma_{l}^{2}+\omega^{2}}+\gamma_{j}-\gamma_{j+l}-\omega\right)\left(\left(\sqrt{\gamma_{j}^{2}+\omega^{2}}-\gamma_{j}-\omega\right)
\left(\sqrt{\gamma_{j+l}^{2}+\omega^{2}}-\gamma_{j+l}\right)+\omega^{2}\right)\right.\\\label{termcomm4}&\left.
+\omega^{2}\left(\sqrt{\gamma_{j}^{2}+\omega^{2}}-\gamma_{j}-\sqrt{\gamma_{j+l}^{2}+\omega^{2}}+\gamma_{j+l}\right)\right].
\end{align}}
Although \eqref{termcomm1}--\eqref{termcomm4} look rather bulky, it turns out that
\begin{equation}\label{equality1zero}
M_{j,l}^{\star\scriptscriptstyle+}d_{l}-M_{j,-j-l}^{\star\scriptscriptstyle+}d_{j+l}+M_{j,-l}^{\star\scriptscriptstyle-}c_{l}- M_{j,j+l}^{\star\scriptscriptstyle-}c_{j+l}=0
\end{equation}
for arbitrary nonzero values of $\gamma_{j}$, $\gamma_{l}$ and $\gamma_{j+l}$ (i.e., the explicit values of $\gamma_{j}$, $\gamma_{l}$ and $\gamma_{j+l}$, as well as the form of their dispersion law, are not important --- it is a purely algebraic cancellation).\footnote{The simplest way to check equality \eqref{equality1zero} (and the subsequent equalities of this type) is to use any program package that is capable for performing symbolic computations, namely, for reduction and simplification of analytical expressions. In this case, it is convenient to work with the dimensionless quantities $\frac{\gamma_{\scriptscriptstyle\#}}{\omega}$. In particular, the validity of equality \eqref{equality1zero} and equalities \eqref{equality2zero}, \eqref{equality3zero}, \eqref{equality4zero}, \eqref{equality5zero} was checked using the computer algebra system Maxima (https://maxima.sourceforge.io/), see Supplementary material for the files.

However, just as a demonstration, examples of analytical calculations will be presented later in Appendix~D for the much more simpler expressions that will be obtained in Section~\ref{subsubsectrho}.}

Analogously, for \eqref{commcoeff2star} one can get
\begin{align}\nonumber
&N_{j,l}^{\star\scriptscriptstyle+}c_{l}-N_{j,-j-l}^{\star\scriptscriptstyle+}c_{j+l}+N_{j,-l}^{\star\scriptscriptstyle-}d_{l}- N_{j,j+l}^{\star\scriptscriptstyle-}d_{j+l}\\\nonumber
&=\frac{\gamma_{j}+\gamma_{l}-\sqrt{\gamma_{j+l}^{2}+\omega^{2}}}{\sqrt{\gamma_{l}^{2}+\omega^{2}}-\gamma_{l}}\,M_{j,l}^{\star\scriptscriptstyle+}d_{l}
-\frac{\gamma_{j}+\gamma_{j+l}-\sqrt{\gamma_{l}^{2}+\omega^{2}}}{\sqrt{\gamma_{j+l}^{2}+\omega^{2}}-\gamma_{j+l}}\, M_{j,-j-l}^{\star\scriptscriptstyle+}d_{j+l}\\\nonumber
&+\frac{\gamma_{j}-\gamma_{l}-\sqrt{\gamma_{j+l}^{2}+\omega^{2}}}{\sqrt{\gamma_{l}^{2}+\omega^{2}}+\gamma_{l}}\, M_{j,-l}^{\star\scriptscriptstyle-}c_{l}-\frac{\gamma_{j}-\gamma_{j+l}-\sqrt{\gamma_{l}^{2}+\omega^{2}}}{\sqrt{\gamma_{j+l}^{2}+\omega^{2}}+\gamma_{j+l}}\, M_{j,j+l}^{\star\scriptscriptstyle-}c_{j+l}\\\label{secondstarredcond}
&+\frac{1}{\omega}\left(-q_{j,l}^{(1)}c_{l}+q_{j,j+l}^{(1)}c_{j+l}-q_{j,l}^{(3)}d_{l}+q_{j,j+l}^{(3)}d_{j+l}\right),
\end{align}
where the terms $M_{j,l}^{\star\scriptscriptstyle+}d_{l}$, $M_{j,-j-l}^{\star\scriptscriptstyle+}d_{j+l}$, $M_{j,-l}^{\star\scriptscriptstyle-}c_{l}$ and $M_{j,j+l}^{\star\scriptscriptstyle-}c_{j+l}$ are defined by \eqref{termcomm1}--\eqref{termcomm4}, Eqs. \eqref{tildeNeq+}, \eqref{tildeNeq-} were used to pass from $N_{{\scriptscriptstyle\#},{\scriptscriptstyle\#}}^{\star\scriptscriptstyle\pm}$ to $M_{{\scriptscriptstyle\#},{\scriptscriptstyle\#}}^{\star\scriptscriptstyle\pm}$, as well as definitions \eqref{cddef}, \eqref{qdef1}--\eqref{qdef4} and relations \eqref{relationsKtoGamma}. With $c_{j}^{2}-d_{j}^{2}=1$ for any $j\neq 0$, for the last line of \eqref{secondstarredcond} one simply gets
\begin{equation}
-q_{j,l}^{(1)}c_{l}+q_{j,j+l}^{(1)}c_{j+l}-q_{j,l}^{(3)}d_{l}+q_{j,j+l}^{(3)}d_{j+l}
=2\left(\left(d_{j}-c_{j}\right)\left(c_{l}^{2}-d_{l}^{2}\right)+\left(c_{j}-d_{j}\right)\left(c_{j+l}^{2}-d_{j+l}^{2}\right)\right)=0.
\end{equation}
As for the remaining terms, it also turns out that
\begin{equation}\label{equality2zero}
N_{j,l}^{\star\scriptscriptstyle+}c_{l}-N_{j,-j-l}^{\star\scriptscriptstyle+}c_{j+l}+N_{j,-l}^{\star\scriptscriptstyle-}d_{l}- N_{j,j+l}^{\star\scriptscriptstyle-}d_{j+l}=0
\end{equation}
for arbitrary nonzero values of $\gamma_{j}$, $\gamma_{l}$ and $\gamma_{j+l}$.

And finally, again with the help of Eqs.~\eqref{tildeNeq+}, \eqref{tildeNeq-}, coefficient \eqref{commcoeff3star} can be represented as
\begin{align}\nonumber
&M_{j,l}^{\star\scriptscriptstyle+}c_{l}+N_{j,-j-l}^{\star\scriptscriptstyle+}d_{j+l}+M_{j,-l}^{\star\scriptscriptstyle-}d_{l}+
N_{j,j+l}^{\star\scriptscriptstyle-}c_{j+l}\\\nonumber
&=\frac{\sqrt{\gamma_{l}^{2}+\omega^{2}}+\gamma_{l}}{\omega}\,M_{j,l}^{\star\scriptscriptstyle+}d_{l}
+\frac{\gamma_{j}+\gamma_{j+l}-\sqrt{\gamma_{l}^{2}+\omega^{2}}}{\omega}\,M_{j,-j-l}^{\star\scriptscriptstyle+}d_{j+l}\\\nonumber
&+\frac{\sqrt{\gamma_{l}^{2}+\omega^{2}}-\gamma_{l}}{\omega}\,M_{j,-l}^{\star\scriptscriptstyle-}c_{l}
+\frac{\gamma_{j}-\gamma_{j+l}-\sqrt{\gamma_{l}^{2}+\omega^{2}}}{\omega}\,M_{j,j+l}^{\star\scriptscriptstyle-}c_{j+l}
-\frac{1}{\omega}\left(q_{j,j+l}^{(1)}d_{j+l}+q_{j,j+l}^{(3)}c_{j+l}\right)\\
&=\frac{1}{\omega}\left(\gamma_{l}M_{j,l}^{\star\scriptscriptstyle+}d_{l}
+(\gamma_{j}+\gamma_{j+l})M_{j,-j-l}^{\star\scriptscriptstyle+}d_{j+l}-\gamma_{l}M_{j,-l}^{\star\scriptscriptstyle-}c_{l}
+(\gamma_{j}-\gamma_{j+l})M_{j,j+l}^{\star\scriptscriptstyle-}c_{j+l}-2c_{j}\right),
\end{align}
where the terms $M_{j,l}^{\star\scriptscriptstyle+}d_{l}$, $M_{j,-j-l}^{\star\scriptscriptstyle+}d_{j+l}$, $M_{j,-l}^{\star\scriptscriptstyle-}c_{l}$ and $M_{j,j+l}^{\star\scriptscriptstyle-}c_{j+l}$ are defined by \eqref{termcomm1}--\eqref{termcomm4}, and definitions \eqref{cddef}, \eqref{qdef1}--\eqref{qdef4} and Eq.~\eqref{equality1zero} were used. One can check that
\begin{equation}\label{equality3zero}
M_{j,l}^{\star\scriptscriptstyle+}c_{l}+N_{j,-j-l}^{\star\scriptscriptstyle+}d_{j+l}+M_{j,-l}^{\star\scriptscriptstyle-}d_{l}+
N_{j,j+l}^{\star\scriptscriptstyle-}c_{j+l}=0
\end{equation}
for arbitrary nonzero values of $\gamma_{j}$, $\gamma_{l}$ and $\gamma_{j+l}$ too. Thus, all three expressions \eqref{commcoeff1star}--\eqref{commcoeff3star} are equal to zero.

Now we turn to the case in which some of the terms in coefficients \eqref{commcoeff1}--\eqref{commcoeff3} may correspond to relations \eqref{transform1}, \eqref{transform2}. Suppose that there exist $a$ and $b$ such that $\gamma_{a+b}=\gamma_{a}+\gamma_{b}$, i.e., equality \eqref{transform1} holds for such $j=a$ and $l=b$ (since all the solutions of the equations of motion for the oscillation modes were obtained in the form symmetric in $j$ and $l$, the cases with $j\leftrightarrow l$ are taken into account automatically). Then, equality \eqref{transform2} is satisfied for $j=a$, $l=a+b$ and $j=b$, $l=a+b$. This simple reasoning suggests the combinations of terms corresponding to relations \eqref{transform1}, \eqref{transform2} that may appear in \eqref{commcoeff1}--\eqref{commcoeff3}.

Let us take $j=a$ and $l=b$ in coefficients \eqref{commcoeff1}--\eqref{commcoeff3}. They take the form
\begin{align}\label{commcoeff1sec}
&\frac{M_{a,b}^{\scriptscriptstyle+}+M_{b,a}^{\scriptscriptstyle+}}{2}d_{b}-M_{a,-a-b}^{\star\scriptscriptstyle+}d_{a+b}+M_{a,-b}^{\star\scriptscriptstyle-}c_{b}-
\frac{M_{a,a+b}^{\scriptscriptstyle-}+N_{a+b,a}^{\scriptscriptstyle-}}{2}c_{a+b},\\\label{commcoeff2sec}
&\frac{N_{a,b}^{\scriptscriptstyle+}+N_{b,a}^{\scriptscriptstyle+}}{2}c_{b}-N_{a,-a-b}^{\star\scriptscriptstyle+}c_{a+b}+N_{a,-b}^{\star\scriptscriptstyle-}d_{b}- \frac{N_{a,a+b}^{\scriptscriptstyle-}+M_{a+b,a}^{\scriptscriptstyle-}}{2}d_{a+b},\\\label{commcoeff3sec}
&\frac{M_{a,b}^{\scriptscriptstyle+}+M_{b,a}^{\scriptscriptstyle+}}{2}c_{b}+N_{a,-a-b}^{\star\scriptscriptstyle+}d_{a+b}+M_{a,-b}^{\star\scriptscriptstyle-}d_{b}+
\frac{N_{a,a+b}^{\scriptscriptstyle-}+M_{a+b,a}^{\scriptscriptstyle-}}{2}c_{a+b}.
\end{align}
We see that in each coefficient there are two ``starred'' terms, which correspond to \eqref{coefftildeM+}, \eqref{coefftildeN+} and \eqref{coefftildeM-}, \eqref{coefftildeN-} (namely, the indices $j=a$, $l=-a-b$ and $j=a$, $l=-b$ in these terms are such that relations \eqref{transform1}, \eqref{transform2} are not fulfilled for the parts of the solution corresponding to these indices), and two terms corresponding to \eqref{M+def}, \eqref{N+def} and \eqref{M-def}, \eqref{N-def} (namely, the indices $j=a$, $l=b$ and vice versa, and the indices $j=a$, $l=a+b$ and vice versa in these terms are such that relations \eqref{transform1}, \eqref{transform2} are fulfilled for the parts of the solution corresponding to these indices).

There are also two other cases, in which the terms corresponding to relations \eqref{transform1} and \eqref{transform2} may appear in \eqref{commcoeff1}--\eqref{commcoeff3}. Again, for $a$ and $b$ such that $\gamma_{a+b}=\gamma_{a}+\gamma_{b}$, the substitution $j=a$ and $l=-a-b$ results in
\begin{align}\label{commcoeff1third}
&M_{a,-a-b}^{\star\scriptscriptstyle+}d_{a+b}-\frac{M_{a,b}^{\scriptscriptstyle+}+M_{b,a}^{\scriptscriptstyle+}}{2}d_{b}+
\frac{M_{a,a+b}^{\scriptscriptstyle-}+N_{a+b,a}^{\scriptscriptstyle-}}{2}c_{a+b}-
M_{a,-b}^{\star\scriptscriptstyle-}c_{b},\\\label{commcoeff2third}
&N_{a,-a-b}^{\star\scriptscriptstyle+}c_{a+b}-\frac{N_{a,b}^{\scriptscriptstyle+}+N_{b,a}^{\scriptscriptstyle+}}{2}c_{b}+
\frac{N_{a,a+b}^{\scriptscriptstyle-}+M_{a+b,a}^{\scriptscriptstyle-}}{2}d_{a+b}- N_{a,-b}^{\star\scriptscriptstyle-}d_{b},\\\label{commcoeff3third}
&\frac{N_{a,b}^{\scriptscriptstyle+}+N_{b,a}^{\scriptscriptstyle+}}{2}d_{b}+M_{a,-a-b}^{\star\scriptscriptstyle+}c_{a+b}+
N_{a,-b}^{\star\scriptscriptstyle-}c_{b}+\frac{M_{a,a+b}^{\scriptscriptstyle-}+N_{a+b,a}^{\scriptscriptstyle-}}{2}d_{a+b}.
\end{align}
And finally, the substitution $j=b$ and $l=-a-b$ results in formulas \eqref{commcoeff1third}--\eqref{commcoeff3third} with $a\leftrightarrow b$. All other cases lead to \eqref{commcoeff1sec}--\eqref{commcoeff3sec} or \eqref{commcoeff1third}--\eqref{commcoeff3third}.

Expressions \eqref{commcoeff1third} and \eqref{commcoeff2third} are just expressions \eqref{commcoeff1sec} and \eqref{commcoeff2sec} taken with the opposite sign, but expression \eqref{commcoeff3third} differs from expression \eqref{commcoeff3sec}. Thus, we should calculate expressions \eqref{commcoeff1sec}--\eqref{commcoeff3sec}, \eqref{commcoeff3third}. First, with \eqref{M+def}, \eqref{N+def}, \eqref{M-def}, \eqref{N-def}, and \eqref{K+symmetry}, expressions \eqref{commcoeff1sec}--\eqref{commcoeff3sec}, \eqref{commcoeff3third} can be rewritten as
\begin{align}\label{commcoeff1notstar}
&K_{a,b}^{\scriptscriptstyle+}d_{b}-M_{a,-a-b}^{\star\scriptscriptstyle+}d_{a+b}+M_{a,-b}^{\star\scriptscriptstyle-}c_{b}+
\frac{c_{a+b}}{2}\left(\left(Q_{a,b}^{\scriptscriptstyle+}+Q_{b,a}^{\scriptscriptstyle+}\right)d_{b}-K_{a,a+b}^{\scriptscriptstyle-}-
Q_{a,a+b}^{\scriptscriptstyle-}c_{b}+Q_{a+b,a}^{\scriptscriptstyle-}d_{b}\right),\\\label{commcoeff2notstar}
&-N_{a,-a-b}^{\star\scriptscriptstyle+}c_{a+b}+N_{a,-b}^{\star\scriptscriptstyle-}d_{b}-
\frac{d_{a+b}}{2}\left(\left(Q_{a,b}^{\scriptscriptstyle+}+Q_{b,a}^{\scriptscriptstyle+}\right)c_{b}+K_{a+b,a}^{\scriptscriptstyle-}+
Q_{a+b,a}^{\scriptscriptstyle-}c_{b}-Q_{a,a+b}^{\scriptscriptstyle-}d_{b}\right),\\\label{commcoeff3notstar}
&K_{a,b}^{\scriptscriptstyle+}c_{b}+N_{a,-a-b}^{\star\scriptscriptstyle+}d_{a+b}+M_{a,-b}^{\star\scriptscriptstyle-}d_{b}+
\frac{c_{a+b}}{2}\left(\left(Q_{a,b}^{\scriptscriptstyle+}+Q_{b,a}^{\scriptscriptstyle+}\right)c_{b}+K_{a+b,a}^{\scriptscriptstyle-}+
Q_{a+b,a}^{\scriptscriptstyle-}c_{b}-Q_{a,a+b}^{\scriptscriptstyle-}d_{b}\right),\\\label{commcoeff4notstar}
&M_{a,-a-b}^{\star\scriptscriptstyle+}c_{a+b}+N_{a,-b}^{\star\scriptscriptstyle-}c_{b}-\frac{d_{a+b}}{2}\left(\left(Q_{a,b}^{\scriptscriptstyle+}+
Q_{b,a}^{\scriptscriptstyle+}\right)d_{b}-K_{a,a+b}^{\scriptscriptstyle-}-
Q_{a,a+b}^{\scriptscriptstyle-}c_{b}+Q_{a+b,a}^{\scriptscriptstyle-}d_{b}\right).
\end{align}
The latter expressions contain the three coefficients $Q_{a,b}^{\scriptscriptstyle+}+Q_{b,a}^{\scriptscriptstyle+}$, $Q_{a,a+b}^{\scriptscriptstyle-}$ and $Q_{a+b,a}^{\scriptscriptstyle-}$, which are not defined yet. These three coefficients compose only two different groups of terms in \eqref{commcoeff1notstar}--\eqref{commcoeff4notstar}, which suggests that one of the coefficients is superfluous. It is convenient to discard the combination $Q_{a,b}^{\scriptscriptstyle+}+Q_{b,a}^{\scriptscriptstyle+}$ by setting
\begin{equation}\label{Qabzero}
Q_{a,b}^{\scriptscriptstyle+}=Q_{b,a}^{\scriptscriptstyle+}=0.
\end{equation}

Now let us first take coefficients \eqref{commcoeff1notstar}, \eqref{commcoeff2notstar} and just set
\begin{align}\label{commcoeff1zero}
&K_{a,b}^{\scriptscriptstyle+}d_{b}-M_{a,-a-b}^{\star\scriptscriptstyle+}d_{a+b}+M_{a,-b}^{\star\scriptscriptstyle-}c_{b}-
\frac{c_{a+b}}{2}\left(K_{a,a+b}^{\scriptscriptstyle-}+
Q_{a,a+b}^{\scriptscriptstyle-}c_{b}-Q_{a+b,a}^{\scriptscriptstyle-}d_{b}\right)=0,\\\label{commcoeff2zero}
&-N_{a,-a-b}^{\star\scriptscriptstyle+}c_{a+b}+N_{a,-b}^{\star\scriptscriptstyle-}d_{b}-
\frac{d_{a+b}}{2}\left(K_{a+b,a}^{\scriptscriptstyle-}+
Q_{a+b,a}^{\scriptscriptstyle-}c_{b}-Q_{a,a+b}^{\scriptscriptstyle-}d_{b}\right)=0.
\end{align}
These equations uniquely define $Q_{a+b,a}^{\scriptscriptstyle-}$ and $Q_{a,a+b}^{\scriptscriptstyle-}$, which take the form
\begin{align}\nonumber
Q_{a+b,a}^{\scriptscriptstyle-}=&
d_{b}\left(\frac{2}{c_{a+b}}\left(K_{a,b}^{\scriptscriptstyle+}d_{b}+M_{a,-b}^{\star\scriptscriptstyle-}c_{b}-M_{a,-a-b}^{\star\scriptscriptstyle+}d_{a+b}\right)
-K_{a,a+b}^{\scriptscriptstyle-}\right)\\\label{Qab+def}+&c_{b}
\left(\frac{2}{d_{a+b}}\left(N_{a,-b}^{\star\scriptscriptstyle-}d_{b}-N_{a,-a-b}^{\star\scriptscriptstyle+}c_{a+b}\right)
-K_{a+b,a}^{\scriptscriptstyle-}\right),\\\nonumber
Q_{a,a+b}^{\scriptscriptstyle-}=&
c_{b}\left(\frac{2}{c_{a+b}}\left(K_{a,b}^{\scriptscriptstyle+}d_{b}+M_{a,-b}^{\star\scriptscriptstyle-}c_{b}-M_{a,-a-b}^{\star\scriptscriptstyle+}d_{a+b}\right)
-K_{a,a+b}^{\scriptscriptstyle-}\right)\\\label{Qab-def}+&d_{b}
\left(\frac{2}{d_{a+b}}\left(N_{a,-b}^{\star\scriptscriptstyle-}d_{b}-N_{a,-a-b}^{\star\scriptscriptstyle+}c_{a+b}\right)
-K_{a+b,a}^{\scriptscriptstyle-}\right).
\end{align}
Now let us take expression \eqref{commcoeff3notstar} and substitute \eqref{Qab+def} and \eqref{Qab-def} into it. We get
\begin{align}\nonumber
&K_{a,b}^{\scriptscriptstyle+}c_{b}+N_{a,-a-b}^{\star\scriptscriptstyle+}d_{a+b}+M_{a,-b}^{\star\scriptscriptstyle-}d_{b}+
\frac{c_{a+b}}{2}\left(K_{a+b,a}^{\scriptscriptstyle-}+
Q_{a+b,a}^{\scriptscriptstyle-}c_{b}-Q_{a,a+b}^{\scriptscriptstyle-}d_{b}\right)\\\label{commcoeff3secmod}
&=K_{a,b}^{\scriptscriptstyle+}c_{b}-N_{a,-a-b}^{\star\scriptscriptstyle+}\frac{1}{d_{a+b}}
+N_{a,-b}^{\star\scriptscriptstyle-}\frac{d_{b}c_{a+b}}{d_{a+b}}+M_{a,-b}^{\star\scriptscriptstyle-}d_{b}.
\end{align}
At this stage it is necessary to perform the following steps: first, using Eqs.~\eqref{tildeNeq+} and \eqref{tildeNeq-}, it is convenient to pass from $N_{a,-a-b}^{\star\scriptscriptstyle+}$ and $N_{a,-b}^{\star\scriptscriptstyle-}$ to $M_{a,-a-b}^{\star\scriptscriptstyle+}$ and $M_{a,-b}^{\star\scriptscriptstyle-}$ respectively; second, using definitions \eqref{cddef}, \eqref{qdef1}--\eqref{qdef4} and relations \eqref{relationsKtoGamma}, one should replace all the momenta by the frequencies $\gamma_{\scriptscriptstyle\#}$ in any term of \eqref{commcoeff3secmod}; and third, which is {\em very important}, one should represent $\gamma_{a+b}$ as $\gamma_{a+b}=\gamma_{a}+\gamma_{b}$ in any place where $\gamma_{a+b}$ appears. The calculations corresponding to these steps are completely straightforward, although somewhat bulky. The result looks like
\begin{align}\nonumber
&K_{a,b}^{\scriptscriptstyle+}c_{b}-N_{a,-a-b}^{\star\scriptscriptstyle+}\frac{1}{d_{a+b}}
+N_{a,-b}^{\star\scriptscriptstyle-}\frac{d_{b}c_{a+b}}{d_{a+b}}+M_{a,-b}^{\star\scriptscriptstyle-}d_{b}\\\nonumber
&=K_{a,b}^{\scriptscriptstyle+}c_{b}-\frac{2}
{\omega\left(\sqrt{\left(\gamma_{a}+\gamma_{b}\right)^{2}+\omega^{2}}-\gamma_{a}-\gamma_{b}\right)}\\\nonumber&\times\left[
\left(\gamma_{a}+\gamma_{b}\right)\left(2\gamma_{a}+\gamma_{b}-\sqrt{\gamma_{b}^{2}+\omega^{2}}\right)M_{a,-a-b}^{\star\scriptscriptstyle+}d_{a+b}
+\gamma_{b}\left(\sqrt{\gamma_{b}^{2}+\omega^{2}}-\gamma_{b}\right)M_{a,-b}^{\star\scriptscriptstyle-}c_{b}\right]
\\&+\frac{2c_{a}}{\omega}\left(\frac{\omega^{2}\left(\gamma_{a}+\gamma_{b}+\omega-\sqrt{\gamma_{a}^{2}+\omega^{2}}-\sqrt{\gamma_{b}^{2}+\omega^{2}}\right)}
{2\gamma_{b}\left(\sqrt{\left(\gamma_{a}+\gamma_{b}\right)^{2}+\omega^{2}}-\gamma_{a}-\gamma_{b}\right)\left(\sqrt{\gamma_{b}^{2}+\omega^{2}}-\gamma_{b}\right)}-1\right),
\end{align}
where $M_{a,-a-b}^{\star\scriptscriptstyle+}d_{a+b}$ and $M_{a,-b}^{\star\scriptscriptstyle-}c_{b}$ are defined by \eqref{termcomm2} and \eqref{termcomm3} respectively, but with $\gamma_{a+b}$ replaced by $\gamma_{a}+\gamma_{b}$, and
\begin{align}\nonumber
&K_{a,b}^{\scriptscriptstyle+}c_{b}
=\frac{c_{a}}{2\gamma_{b}\left(\gamma_{a}+\gamma_{b}\right)}\left[\omega\left(\frac{\omega}{\sqrt{\gamma_{b}^{2}+\omega^{2}}-\gamma_{b}}
\left(1-\frac{\sqrt{\gamma_{a}^{2}+\omega^{2}}-\gamma_{a}}{\omega}\right)-1\right)\right.\\\label{K+cbdef}
&\left.-\left(\sqrt{\left(\gamma_{a}+\gamma_{b}\right)^{2}+\omega^{2}}+\gamma_{a}+\gamma_{b}\right)
\left(\frac{\sqrt{\gamma_{a}^{2}+\omega^{2}}-\gamma_{a}}{\omega}\left(1-\frac{\omega}{\sqrt{\gamma_{b}^{2}+\omega^{2}}-\gamma_{b}}\right)-1\right)\right].
\end{align}
It turns out that
\begin{equation}\label{equality4zero}
K_{a,b}^{\scriptscriptstyle+}c_{b}-N_{a,-a-b}^{\star\scriptscriptstyle+}\frac{1}{d_{a+b}}
+N_{a,-b}^{\star\scriptscriptstyle-}\frac{d_{b}c_{a+b}}{d_{a+b}}+M_{a,-b}^{\star\scriptscriptstyle-}d_{b}=0
\end{equation}
for arbitrary nonzero values of $\gamma_{a}$ and $\gamma_{b}$.

Fully analogous steps can be performed for \eqref{commcoeff4notstar}, resulting in
\begin{align}\nonumber
&M_{a,-a-b}^{\star\scriptscriptstyle+}c_{a+b}+N_{a,-b}^{\star\scriptscriptstyle-}c_{b}+\frac{d_{a+b}}{2}\left(K_{a,a+b}^{\scriptscriptstyle-}+
Q_{a,a+b}^{\scriptscriptstyle-}c_{b}-Q_{a+b,a}^{\scriptscriptstyle-}d_{b}\right)\\\nonumber
=&K_{a,b}^{\scriptscriptstyle+}\frac{d_{a+b}d_{b}}{c_{a+b}}+M_{a,-a-b}^{\star\scriptscriptstyle+}\frac{1}{c_{a+b}}
+N_{a,-b}^{\star\scriptscriptstyle-}c_{b}+M_{a,-b}^{\star\scriptscriptstyle-}\frac{d_{a+b}c_{b}}{c_{a+b}}\\\nonumber
=&\frac{\sqrt{\left(\gamma_{a}+\gamma_{b}\right)^{2}+\omega^{2}}-\gamma_{a}-\gamma_{b}}
{\sqrt{\gamma_{b}^{2}+\omega^{2}}+\gamma_{b}}K_{a,b}^{\scriptscriptstyle+}c_{b}+\frac{2\left(\gamma_{a}+\gamma_{b}\right)}{\omega}M_{a,-a-b}^{\star\scriptscriptstyle+}d_{a+b}\\&
-\frac{2\gamma_{b}}{\omega}M_{a,-b}^{\star\scriptscriptstyle-}c_{b}
-\frac{c_{a}\left(\sqrt{\gamma_{a}^{2}+\omega^{2}}-\gamma_{a}+\sqrt{\gamma_{b}^{2}+\omega^{2}}+\gamma_{b}-\omega\right)}{\omega\gamma_{b}}.
\end{align}
Analogously, one can check that for \eqref{termcomm2}, \eqref{termcomm3} (with $\gamma_{a+b}$ replaced by $\gamma_{a}+\gamma_{b}$) and \eqref{K+cbdef}
\begin{equation}\label{equality5zero}
K_{a,b}^{\scriptscriptstyle+}\frac{d_{a+b}d_{b}}{c_{a+b}}+M_{a,-a-b}^{\star\scriptscriptstyle+}\frac{1}{c_{a+b}}
+N_{a,-b}^{\star\scriptscriptstyle-}c_{b}+M_{a,-b}^{\star\scriptscriptstyle-}\frac{d_{a+b}c_{b}}{c_{a+b}}=0
\end{equation}
for arbitrary nonzero values of $\gamma_{a}$ and $\gamma_{b}$.

One can see that all four expressions \eqref{commcoeff1sec}, \eqref{commcoeff2sec}, \eqref{commcoeff3sec} and \eqref{commcoeff3third} are equal to zero. It is clear that in such a case the corresponding expressions with $a\leftrightarrow b$ (i.e., for the substitution $j=b$ and $l=-a-b$) are equal to zero too. Thus, coefficients \eqref{commcoeff1}--\eqref{commcoeff3} are equal to zero for all $j$ and $l$ such that $j\neq -l$, which means that the operator $\hat{\tilde\phi}(t,\vec x)$ satisfies both commutation relations \eqref{CCR1} and \eqref{CCR2}.

\subsubsection{Commutation relations for $\hat\rho_{j,l}^{\scriptscriptstyle\pm}(t,\vec x)$}\label{subsubsectrho}
Suppose we have two modes, one with $j=a$ and another with $l=b$, such that $\gamma_{a+b}=\gamma_{a}+\gamma_{b}$, i.e., for these modes equality \eqref{transform1} holds. Then, as was mentioned above, for the modes with $j=a+b$, $l=b$ and $j=a+b$, $l=a$ equality \eqref{transform2} holds automatically. Thus, these three modes compose a set providing the terms $\sim t$ (as was noted above, since our solution for the operators $\hat\rho_{j,l}^{\scriptscriptstyle\pm}(t,\vec x)$ is symmetric with respect to $j$ and $l$, the cases with $j\leftrightarrow l$ are taken into account automatically). Of course, the modes such that $\gamma_{-a-b}=\gamma_{-a}+\gamma_{-b}$ also compose an analogous set. These reasoning suggest that one can consider only the set of modes defined by the condition $\gamma_{a+b}=\gamma_{a}+\gamma_{b}$, for all other sets one would get fully analogous results.

Thus, the choice $j=a$, $l=b$ provides the term $\hat\rho_{a,b}^{\scriptscriptstyle+}$, the choice $j=a+b$, $l=b$ provides the term $\hat\rho_{a+b,b}^{\scriptscriptstyle-}$, and the choice $j=a+b$, $l=a$ provides the term $\hat\rho_{a+b,a}^{\scriptscriptstyle-}$. Let us take the combination
\begin{equation}\label{rhocombin}
\hat\rho_{a,b}^{\scriptscriptstyle+}(t,\vec x)+\hat\rho_{a+b,b}^{\scriptscriptstyle-}(t,\vec x)+\hat\rho_{a+b,a}^{\scriptscriptstyle-}(t,\vec x).
\end{equation}
The first commutation relation for this combination can be easily obtained by using formula \eqref{commjl+} with $\tilde M_{j,l}^{\scriptscriptstyle+}\to t L_{j,l}^{\scriptscriptstyle+}$, $\tilde N_{j,l}^{\scriptscriptstyle+}\to t J_{j,l}^{\scriptscriptstyle+}$ and formula \eqref{commjl-} with $\tilde M_{j,l}^{\scriptscriptstyle-}\to t L_{j,l}^{\scriptscriptstyle-}$, $\tilde N_{j,l}^{\scriptscriptstyle-}\to t J_{j,l}^{\scriptscriptstyle-}$. One gets
\begin{align}\nonumber
&[\hat\rho_{a,b}^{\scriptscriptstyle+}(t,\vec x)+\hat\rho_{a+b,b}^{\scriptscriptstyle-}(t,\vec x)+\hat\rho_{a+b,a}^{\scriptscriptstyle-}(t,\vec x),\hat\varphi(t,\vec y)]\\\nonumber&+[\hat\varphi(t,\vec x),\hat\rho_{a,b}^{\scriptscriptstyle+}(t,\vec y)+\hat\rho_{a+b,b}^{\scriptscriptstyle-}(t,\vec y)+\hat\rho_{a+b,a}^{\scriptscriptstyle-}(t,\vec y)]\\\nonumber
=-\frac{\omega t}{L^{3}}\Biggl(&\left(e^{\frac{i}{\hbar}(\vec k_{a}+\vec k_{b})\vec x}e^{-\frac{i}{\hbar}\vec k_{b}\vec y}e^{-\frac{i}{\hbar}\gamma_{a}t}\hat a_{a}^{}\left(L_{a,b}^{\scriptscriptstyle+}d_{b}-J_{a+b,a}^{\scriptscriptstyle-}c_{a+b}\right)\right.\\\nonumber&\left.+
e^{-\frac{i}{\hbar}(\vec k_{a}+\vec k_{b})\vec x}e^{\frac{i}{\hbar}\vec k_{b}\vec y}e^{\frac{i}{\hbar}\gamma_{a}t}\hat a_{a}^{\dagger}\left(J_{a,b}^{\scriptscriptstyle+}c_{b}-L_{a+b,a}^{\scriptscriptstyle-}d_{a+b}\right)+\left[a\leftrightarrow b\right]\right)\\\nonumber
&+e^{\frac{i}{\hbar}\vec k_{a}\vec x}e^{\frac{i}{\hbar}\vec k_{b}\vec y}e^{-\frac{i}{\hbar}(\gamma_{a}+\gamma_{b})t}\hat a_{a+b}^{}\left(L_{a+b,b}^{\scriptscriptstyle-}c_{b}-L_{a+b,a}^{\scriptscriptstyle-}c_{a}\right)\\\label{commTterms1}
&+e^{-\frac{i}{\hbar}\vec k_{a}\vec x}e^{-\frac{i}{\hbar}\vec k_{b}\vec y}e^{\frac{i}{\hbar}(\gamma_{a}+\gamma_{b})t}\hat a_{a+b}^{\dagger}\left(J_{a+b,b}^{\scriptscriptstyle-}d_{b}-J_{a+b,a}^{\scriptscriptstyle-}d_{a}\right)
-\left[\vec x\leftrightarrow\vec y\right]\Biggr).
\end{align}
Note that for those $j$ and $l$ for which relation \eqref{transform1} is valid, the relation
\begin{equation}\label{JLtrs1}
J_{j,l}^{\scriptscriptstyle+}=\frac{\sqrt{\left(\gamma_{j}+\gamma_{l}\right)^{2}+\omega^{2}}-(\gamma_{j}+\gamma_{l})}{\omega}L_{j,l}^{\scriptscriptstyle+}
\end{equation}
holds, which follows directly from Eq.~\eqref{LJeq+} and solutions \eqref{L+def}, \eqref{J+def}. Analogously, for those $j$ and $l$ for which relation \eqref{transform2} is valid, the relation
\begin{equation}\label{JLtrs2}
J_{j,l}^{\scriptscriptstyle-}=\frac{\sqrt{\left(\gamma_{j}-\gamma_{l}\right)^{2}+\omega^{2}}-(\gamma_{j}-\gamma_{l})}{\omega}L_{j,l}^{\scriptscriptstyle-}
\end{equation}
holds, which follows directly from Eq.~\eqref{LJeq-} and solutions \eqref{L-def}, \eqref{J-def}. Using these formulas, the sum of the commutators in \eqref{commTterms1} can be rewritten as
\begin{align}\nonumber
&[\hat\rho_{a,b}^{\scriptscriptstyle+}(t,\vec x)+\hat\rho_{a+b,b}^{\scriptscriptstyle-}(t,\vec x)+\hat\rho_{a+b,a}^{\scriptscriptstyle-}(t,\vec x),\hat\varphi(t,\vec y)]\\\nonumber&+[\hat\varphi(t,\vec x),\hat\rho_{a,b}^{\scriptscriptstyle+}(t,\vec y)+\hat\rho_{a+b,b}^{\scriptscriptstyle-}(t,\vec y)+\hat\rho_{a+b,a}^{\scriptscriptstyle-}(t,\vec y)]\\\nonumber
=&-\frac{\omega t}{L^{3}}\Biggl(\Biggl(e^{\frac{i}{\hbar}(\vec k_{a}+\vec k_{b})\vec x}e^{-\frac{i}{\hbar}\vec k_{b}\vec y}e^{-\frac{i}{\hbar}\gamma_{a}t}\hat a_{a}^{}\frac{\sqrt{\gamma_{b}^{2}+\omega^{2}}-\gamma_{b}}{\omega}
\left(L_{a,b}^{\scriptscriptstyle+}c_{b}-L_{a+b,a}^{\scriptscriptstyle-}c_{a+b}\right)\\\nonumber&+
e^{-\frac{i}{\hbar}(\vec k_{a}+\vec k_{b})\vec x}e^{\frac{i}{\hbar}\vec k_{b}\vec y}e^{\frac{i}{\hbar}\gamma_{a}t}\hat a_{a}^{\dagger}
\frac{\sqrt{\left(\gamma_{a}+\gamma_{b}\right)^{2}+\omega^{2}}-(\gamma_{a}+\gamma_{b})}{\omega}
\left(L_{a,b}^{\scriptscriptstyle+}c_{b}-L_{a+b,a}^{\scriptscriptstyle-}c_{a+b}\right)+\left[a\leftrightarrow b\right]\Biggr)\\\nonumber
&+e^{\frac{i}{\hbar}\vec k_{a}\vec x}e^{\frac{i}{\hbar}\vec k_{b}\vec y}e^{-\frac{i}{\hbar}(\gamma_{a}+\gamma_{b})t}\hat a_{a+b}^{}\left(L_{a+b,b}^{\scriptscriptstyle-}c_{b}-L_{a+b,a}^{\scriptscriptstyle-}c_{a}\right)
\\\label{commTterms2}
&+e^{-\frac{i}{\hbar}\vec k_{a}\vec x}e^{-\frac{i}{\hbar}\vec k_{b}\vec y}e^{\frac{i}{\hbar}(\gamma_{a}+\gamma_{b})t}\hat a_{a+b}^{\dagger}
\frac{\sqrt{\gamma_{a}^{2}+\omega^{2}}-\gamma_{a}}
{\sqrt{\gamma_{b}^{2}+\omega^{2}}+\gamma_{b}}\left(L_{a+b,b}^{\scriptscriptstyle-}c_{b}-L_{a+b,a}^{\scriptscriptstyle-}c_{a}\right)
-\left[\vec x\leftrightarrow\vec y\right]\Biggr).
\end{align}
The necessary independent coefficients in \eqref{commTterms2} are
\begin{align}\label{Lcoeff1}
&L_{a,b}^{\scriptscriptstyle+}c_{b}-L_{a+b,a}^{\scriptscriptstyle-}c_{a+b},\\\label{Lcoeff2}
&L_{a+b,b}^{\scriptscriptstyle-}c_{b}-L_{a+b,a}^{\scriptscriptstyle-}c_{a}.
\end{align}

The second commutation relation for combination \eqref{rhocombin} can also be obtained from the formulas used above. However, since $\left(L_{j,l}^{\scriptscriptstyle\pm}\right)^{*}=-L_{j,l}^{\scriptscriptstyle\pm}$ and $\left(J_{j,l}^{\scriptscriptstyle\pm}\right)^{*}=-J_{j,l}^{\scriptscriptstyle\pm}$, the replacements should be made more carefully: in the first and third lines of the r.h.s. of \eqref{commjldagg+} and \eqref{commjldagg-} (i.e., for the terms $\sim e^{\pm\frac{i}{\hbar}(\vec k_{j}\pm\vec k_{l})\vec y}$) one should use $\tilde M_{j,l}^{\scriptscriptstyle\pm}\to -t L_{j,l}^{\scriptscriptstyle\pm}$ and $\tilde N_{j,l}^{\scriptscriptstyle\pm}\to -t J_{j,l}^{\scriptscriptstyle\pm}$, whereas in the second and fourth lines of the r.h.s. of \eqref{commjldagg+} and \eqref{commjldagg-} (i.e., for the terms $\sim e^{\pm\frac{i}{\hbar}(\vec k_{j}\pm\vec k_{l})\vec x}$) one should use $\tilde M_{j,l}^{\scriptscriptstyle\pm}\to t L_{j,l}^{\scriptscriptstyle\pm}$ and $\tilde N_{j,l}^{\scriptscriptstyle\pm}\to t J_{j,l}^{\scriptscriptstyle\pm}$. One gets
\begin{align}\nonumber
&[\hat\rho_{a,b}^{\scriptscriptstyle+}(t,\vec x)+\hat\rho_{a+b,b}^{\scriptscriptstyle-}(t,\vec x)+\hat\rho_{a+b,a}^{\scriptscriptstyle-}(t,\vec x),\hat\varphi^{\dagger}(t,\vec y)]\\\nonumber&+[\hat\varphi(t,\vec x),\left(\hat\rho_{a,b}^{\scriptscriptstyle+}(t,\vec y)+\hat\rho_{a+b,b}^{\scriptscriptstyle-}(t,\vec y)+\hat\rho_{a+b,a}^{\scriptscriptstyle-}(t,\vec y)\right)^{\dagger}]\\\nonumber
&=\frac{\omega t}{L^{3}}\Biggl(\Bigl(e^{\frac{i}{\hbar}(\vec k_{a}+\vec k_{b})\vec x}e^{-\frac{i}{\hbar}\vec k_{b}\vec y}e^{-\frac{i}{\hbar}\gamma_{a}t}\hat a_{a}^{}\left(L_{a,b}^{\scriptscriptstyle+}c_{b}-L_{a+b,a}^{\scriptscriptstyle-}c_{a+b}\right)\\\nonumber&+
e^{-\frac{i}{\hbar}(\vec k_{a}+\vec k_{b})\vec x}e^{\frac{i}{\hbar}\vec k_{b}\vec y}e^{\frac{i}{\hbar}\gamma_{a}t}\hat a_{a}^{\dagger}\left(J_{a,b}^{\scriptscriptstyle+}d_{b}-J_{a+b,a}^{\scriptscriptstyle-}d_{a+b}\right)+\left[a\leftrightarrow b\right]\Bigr)\\\nonumber
&+e^{\frac{i}{\hbar}\vec k_{a}\vec x}e^{\frac{i}{\hbar}\vec k_{b}\vec y}e^{-\frac{i}{\hbar}(\gamma_{a}+\gamma_{b})t}\hat a_{a+b}^{}\left(L_{a+b,b}^{\scriptscriptstyle-}d_{b}-J_{a+b,a}^{\scriptscriptstyle-}c_{a}\right)\\\label{commTterms3}
&+e^{-\frac{i}{\hbar}\vec k_{a}\vec x}e^{-\frac{i}{\hbar}\vec k_{b}\vec y}e^{\frac{i}{\hbar}(\gamma_{a}+\gamma_{b})t}\hat a_{a+b}^{\dagger}\left(J_{a+b,b}^{\scriptscriptstyle-}c_{b}-L_{a+b,a}^{\scriptscriptstyle-}d_{a}\right)
+\left[\vec x\leftrightarrow\vec y\right]^{\dagger}\Biggr).
\end{align}
Analogously, using relations \eqref{JLtrs1} and \eqref{JLtrs2}, the sum of the commutators in \eqref{commTterms3} can be rewritten as
\begin{align}\nonumber
&[\hat\rho_{a,b}^{\scriptscriptstyle+}(t,\vec x)+\hat\rho_{a+b,b}^{\scriptscriptstyle-}(t,\vec x)+\hat\rho_{a+b,a}^{\scriptscriptstyle-}(t,\vec x),\hat\varphi^{\dagger}(t,\vec y)]\\\nonumber&+[\hat\varphi(t,\vec x),\left(\hat\rho_{a,b}^{\scriptscriptstyle+}(t,\vec y)+\hat\rho_{a+b,b}^{\scriptscriptstyle-}(t,\vec y)+\hat\rho_{a+b,a}^{\scriptscriptstyle-}(t,\vec y)\right)^{\dagger}]\\\nonumber
&=\frac{\omega t}{L^{3}}\Biggl(\Biggl(e^{\frac{i}{\hbar}(\vec k_{a}+\vec k_{b})\vec x}e^{-\frac{i}{\hbar}\vec k_{b}\vec y}e^{-\frac{i}{\hbar}\gamma_{a}t}\hat a_{a}^{}\left(L_{a,b}^{\scriptscriptstyle+}c_{b}-L_{a+b,a}^{\scriptscriptstyle-}c_{a+b}\right)\\\nonumber&+
e^{-\frac{i}{\hbar}(\vec k_{a}+\vec k_{b})\vec x}e^{\frac{i}{\hbar}\vec k_{b}\vec y}e^{\frac{i}{\hbar}\gamma_{a}t}\hat a_{a}^{\dagger}\frac{\sqrt{\left(\gamma_{a}+\gamma_{b}\right)^{2}+\omega^{2}}-(\gamma_{a}+\gamma_{b})}
{\sqrt{\gamma_{b}^{2}+\omega^{2}}+\gamma_{b}}\left(L_{a,b}^{\scriptscriptstyle+}c_{b}-L_{a+b,a}^{\scriptscriptstyle-}c_{a+b}\right)+\left[a\leftrightarrow b\right]\Biggr)\\\nonumber
&+e^{\frac{i}{\hbar}\vec k_{a}\vec x}e^{\frac{i}{\hbar}\vec k_{b}\vec y}e^{-\frac{i}{\hbar}(\gamma_{a}+\gamma_{b})t}\hat a_{a+b}^{}\frac{\sqrt{\gamma_{b}^{2}+\omega^{2}}-\gamma_{b}}{\omega}\left(L_{a+b,b}^{\scriptscriptstyle-}c_{b}
-L_{a+b,a}^{\scriptscriptstyle-}c_{a}\right)\\\label{commTterms4}
&+e^{-\frac{i}{\hbar}\vec k_{a}\vec x}e^{-\frac{i}{\hbar}\vec k_{b}\vec y}e^{\frac{i}{\hbar}(\gamma_{a}+\gamma_{b})t}\hat a_{a+b}^{\dagger}\frac{\sqrt{\gamma_{a}^{2}+\omega^{2}}-\gamma_{a}}{\omega}
\left(L_{a+b,b}^{\scriptscriptstyle-}c_{b}-L_{a+b,a}^{\scriptscriptstyle-}c_{a}\right)
+\left[\vec x\leftrightarrow\vec y\right]^{\dagger}\Biggr).
\end{align}
Here we also get coefficients \eqref{Lcoeff1} and \eqref{Lcoeff2}. The last step is to check whether these coefficients are equal to zero. Using \eqref{L+def}, \eqref{L-def} and definitions \eqref{cddef}, \eqref{qdef1}--\eqref{qdef4} together with
\begin{equation}
\frac{\vec k_{a}^{2}}{2m}=\sqrt{\gamma_{a}^{2}+\omega^{2}}-\omega,\qquad \frac{\vec k_{b}^{2}}{2m}=\sqrt{\gamma_{b}^{2}+\omega^{2}}-\omega,\qquad\frac{(\vec k_{a}+\vec k_{b})^{2}}{2m}=\sqrt{\left(\gamma_{a}+\gamma_{b}\right)^{2}+\omega^{2}}-\omega,
\end{equation}
one can check that
\begin{align}\label{lastbut1eqcoeff}
&L_{a,b}^{\scriptscriptstyle+}c_{b}-L_{a+b,a}^{\scriptscriptstyle-}c_{a+b}=0,\\\label{lasteqcoeff}
&L_{a+b,b}^{\scriptscriptstyle-}c_{b}-L_{a+b,a}^{\scriptscriptstyle-}c_{a}=0
\end{align}
for arbitrary nonzero values of $\gamma_{a}$ and $\gamma_{b}$. Evaluation of coefficients \eqref{Lcoeff1} and \eqref{Lcoeff2} is easier than of those in the previous cases, so the explicit analytical check of equalities \eqref{lastbut1eqcoeff} and \eqref{lasteqcoeff} is presented in Appendix~D. Thus, commutation relations \eqref{CCR1} and \eqref{CCR2} are satisfied for the terms with $\hat\rho_{j,l}^{\scriptscriptstyle\pm}(t,\vec x)$.

\subsection{Extra contribution}\label{ExtraCont}
Finally, let us consider the combination
\begin{equation}\label{phiEX}
\hat\phi_{\textrm{ex}}(t,\vec x)=\frac{1}{\sqrt{L^{3}}}\left(\sum\limits_{j\neq 0}\left(\frac{1}{2}-\frac{i\omega t}{\hbar}\right)\epsilon_{j}\hat a_{j}^{\dagger}\hat a_{j}^{}-\sum\limits_{j\neq 0}\frac{\epsilon_{j}}{2q}e^{\frac{i}{\hbar}\vec k_{j}\vec x}\left(e^{-\frac{i}{\hbar}\gamma_{j}t}c_{j}\hat P_{j}-e^{\frac{i}{\hbar}\gamma_{j}t}d_{j}\hat P_{-j}^{\dagger}\right)\right)
\end{equation}
with $\hat P_{j}=\left(\hat a_{0}-\hat a_{0}^{\dagger}\right)\hat a_{j}$ and with $\epsilon_{-j}=\epsilon_{j}$ being free dimensionless parameters. The operator $\hat\phi_{\textrm{ex}}(t,\vec x)$ is a solution of the homogeneous part of Eq.~\eqref{eqnlcquant2}, so one can add it to the solution $\hat\phi(t,\vec x)$ obtained above. It is not difficult to check that
\begin{align}
&[\hat\phi_{\textrm{ex}}(t,\vec x),\hat\varphi(t,\vec y)]+[\hat\varphi(t,\vec x),\hat\phi_{\textrm{ex}}(t,\vec y)]=0,\\
&[\hat\phi_{\textrm{ex}}^{}(t,\vec x),\hat\varphi^{\dagger}(t,\vec y)]+[\hat\varphi(t,\vec x),\hat\phi_{\textrm{ex}}^{\dagger}(t,\vec y)]=0,
\end{align}
so \eqref{phiEX} satisfies commutation relations \eqref{CCR1} and \eqref{CCR2}. The necessity for this extra contribution will become clear in the next section. Note that in order to have in our solution the terms $\sim\epsilon_{j}\hat a_{j}^{\dagger}\hat a_{j}^{}$, which contain only the operators related to the oscillation modes, again the terms containing the operators related to the nonoscillation modes (i.e., those with the operators $\hat P_{j}$) are necessary to fulfill the canonical commutation relations.

\section{Operators of the integrals of motion}
Let us write the whole solution for the operator $\hat\Psi(t,\vec x)$ using the results presented in the previous sections:
\begin{align}\nonumber
\hat\Psi(t,\vec x)&=e^{-\frac{i}{\hbar}\omega t}\left(\rule{0cm}{1cm}\right.\sqrt{\frac{\omega}{g}}+\hat\varphi(t,\vec x)+\beta\Biggl(\hat\phi_{\textrm{no}}(t)+\hat\phi_{\times}(t,\vec x)+
\hat\phi_{\textrm{t}}(t)\\\label{solfull}&+\frac{1}{2}\underset{\substack{j\neq 0, l\neq 0\\j\neq-l}}{\sum\sum}\hat\phi_{j,l}^{\scriptscriptstyle+}(t,\vec x)+\frac{1}{2}\underset{\substack{j\neq 0, l\neq 0\\j\neq l}}{\sum\sum}\hat\phi_{j,l}^{\scriptscriptstyle-}(t,\vec x)+\hat\phi_{\textrm{ex}}(t,\vec x)\Biggr)\left.\rule{0cm}{1cm}\right),
\end{align}
where $\hat\varphi(t,\vec x)$ is defined by \eqref{linsol2}; $\hat\phi_{\textrm{no}}(t)$ is defined by \eqref{phinosolfull}; $\hat\phi_{\times}(t,\vec x)$ is defined by \eqref{phiX2def} with \eqref{substX}, \eqref{Xcoeff1}--\eqref{Xcoeff4} and \eqref{deltaphiX}, \eqref{deltaphiXQ}; $\hat\phi_{\textrm{t}}(t)$ is defined by \eqref{phitosc}; $\hat\phi_{j,l}^{\scriptscriptstyle+}(t,\vec x)$ is defined by \eqref{phijldef+} with \eqref{coefftildeM+}, \eqref{coefftildeN+} or by \eqref{phijldef+1} with \eqref{L+def}--\eqref{Kab+}, \eqref{Qabzero} (depending on whether relation \eqref{transform1} holds or not); $\hat\phi_{j,l}^{\scriptscriptstyle-}(t,\vec x)$ is defined by \eqref{phijldef-} with \eqref{coefftildeM-}, \eqref{coefftildeN-} or by \eqref{phijldef-1} with \eqref{L-def}--\eqref{Kab-}, \eqref{Qab+def}, \eqref{Qab-def} (depending on whether relation \eqref{transform2} holds or not); and $\hat\phi_{\textrm{ex}}(t,\vec x)$ is defined by \eqref{phiEX}. As was demonstrated above, the canonical commutation relations \eqref{commrel1} and \eqref{commrel2} are satisfied for solution \eqref{solfull} in the approximation used above (i.e., up to and including the terms $\sim \beta$). Substituting \eqref{solfull} into \eqref{NpquadrQNL}, \eqref{EpquadrQNL} and into the momentum operator
\begin{equation}
\hat{\vec{P}}=-i\hbar\int d^{3}x\hat\Psi^{\dagger}(t,\vec x)\nabla\hat\Psi(t,\vec x),
\end{equation}
using the fact that
\begin{equation}\label{zerointegral}
\int d^{3}x\,e^{\pm\frac{i}{\hbar}\vec k_{j}\vec x}=0
\end{equation}
for any $\vec k_{j}\neq 0$ and using \eqref{NpNO2}, \eqref{EpNO2}, one can get
\begin{align}\label{Npfinal0}
\hat N_{p}&=\frac{1}{2}\left(1+\sum\limits_{j\neq 0}\left(1-\frac{{\vec k_{j}}^{2}}{2m\gamma_{j}}\right)\right)+\frac{dN_{0}}{d\omega}\,\tilde q\left(\hat a_{0}+\hat a_{0}^{\dagger}\right)+\sum\limits_{j\neq 0}\left(\epsilon_{j}-\frac{{\vec k_{j}}^{2}}{2m\gamma_{j}}\right)\hat a_{j}^{\dagger}\hat a_{j}^{},
\\\nonumber
\hat E_{p}&=\frac{1}{2}\sum\limits_{j\neq 0}\frac{{\vec k_{j}}^{2}}{2m\gamma_{j}}\left(\frac{{\vec k_{j}}^{2}}{2m}+\omega-\gamma_{j}\right)+
\frac{dE_{0}}{d\omega}\,\tilde q\left(\hat a_{0}+\hat a_{0}^{\dagger}\right)+\frac{1}{2}\frac{d^{2}E_{0}}{d\omega^{2}}\,{\tilde q}^{2}\left(\hat a_{0}+\hat a_{0}^{\dagger}\right)^{2}\\\label{Epfinal0}
&+\omega\sum\limits_{j\neq 0}\left(\epsilon_{j}-\frac{{\vec k_{j}}^{2}}{2m\gamma_{j}}\right)\hat a_{j}^{\dagger}\hat a_{j}^{}+\sum\limits_{j\neq 0}\gamma_{j}\hat a_{j}^{\dagger}\hat a_{j}^{},\\\label{Ppfinal0}
\hat{\vec{P_{p}}}&=\sum\limits_{j\neq 0}\vec k_{j}\hat a_{j}^{\dagger}\hat a_{j}^{}.
\end{align}
Here $\hat a_{j}^{\dagger}\hat a_{j}^{}$ stands for the number of quasi-particles with the momentum $\vec k_{j}$. As usual, the $c$-number terms (the first sums in \eqref{Npfinal0} and \eqref{Epfinal0}) arise when one passes from $\hat a_{j}^{}\hat a_{j}^{\dagger}$ to $\hat a_{j}^{\dagger}\hat a_{j}^{}$. We see that $\hat N_{p}$ does not depend on time and is diagonal. In particular, the time-dependent terms of \eqref{Npquadr2} (which are also nondiagonal) are compensated by the time-dependent terms coming from \eqref{phitosc}.

Note that the terms of \eqref{solfull} with $\hat \phi_{j,l}^{\scriptscriptstyle+}(t,\vec x)$ and $\hat\phi_{j,l}^{\scriptscriptstyle-}(t,\vec x)$ do not contribute to \eqref{Npfinal0}--\eqref{Ppfinal0} because of \eqref{zerointegral}. In this connection, the question arises: is it necessary to calculate these terms explicitly or one can skip this calculation? To answer this question, let us take $\hat\Psi'(t,\vec x)$ such that
\begin{equation}\label{psiprimedef}
\hat\Psi'(0,\vec x)=\sqrt{\frac{\omega}{g}}+\hat\varphi(0,\vec x)+\beta\Bigl(\hat\phi_{\textrm{no}}(0)+\hat\phi_{\times}(0,\vec x)+\hat\phi_{\textrm{t}}(0)+\hat\phi_{\textrm{ex}}(0,\vec x)\Bigr)
\end{equation}
instead of \eqref{solfull}. The canonical commutation relations for $\hat\Psi'(0,\vec x)$ look like
\begin{eqnarray}\label{commrelpsiprime1}
&&[\hat\Psi'(0,\vec x),{\hat\Psi}'^{\dagger}(0,\vec y)]=\delta^{(3)}(\vec x-\vec y)+O\left(\beta^{2}\frac{1}{L^{3}}\right),\\\label{commrelpsiprime2}
&&[\hat\Psi'(0,\vec x),\hat\Psi'(0,\vec y)]=O\left(\beta^{2}\frac{1}{L^{3}}\right),
\end{eqnarray}
and for the integrals of motion of the perturbation one gets \eqref{Npfinal0}--\eqref{Ppfinal0} at $t=0$. Suppose that for governing the evolution of $\hat\Psi'(t,\vec x)$ Eq.~\eqref{Heiseq} with $\hat H$ obtained by substituting \eqref{psiprimedef} into \eqref{initHamiltonian2} is used. In such a case, we can be sure that the terms $\sim\beta\frac{1}{L^{3}}$ do not appear in $[\hat\Psi'(t,\vec x),{\hat\Psi}'^{\dagger}(t,\vec y)]$ and $[\hat\Psi'(t,\vec x),\hat\Psi'(t,\vec y)]$ for $t\neq 0$. However, since the canonical commutation relations are not satisfied exactly and Eqs.~\eqref{Heiseq}, \eqref{eqmotgeneral} are not equivalent, we cannot be sure that the extra terms quadratic in the operators $\hat a_{j}$ and $\hat a_{j}^{\dagger}$ (including time-dependent terms) do not appear in $\hat N_{p}$ for $t\neq 0$ (see the discussions in Sections~\ref{section3}, \ref{section4}). If, on the contrary, Eq.~\eqref{eqmotgeneral} is used for governing the evolution of $\hat\Psi'(t,\vec x)$, then we can be sure that $\hat N_{p}$ has the form \eqref{Npfinal0} for $t\neq 0$, but we cannot be sure that the terms $\sim\beta\frac{1}{L^{3}}$ do not appear in $[\hat\Psi'(t,\vec x),{\hat\Psi}'^{\dagger}(t,\vec y)]$ and $[\hat\Psi'(t,\vec x),\hat\Psi'(t,\vec y)]$ for $t\neq 0$. Meanwhile, solution \eqref{solfull} guarantees that $\hat N_{p}$ defined by \eqref{Npfinal0} does not change in time {\em and} the canonical commutation relations do not contain the terms $\sim\beta\frac{1}{L^{3}}$ {\em for any} $t$ (recall that only the special choice \eqref{Qab+def}, \eqref{Qab-def} of the coefficients $Q_{j,l}^{\scriptscriptstyle-}$ ensures that commutation relations \eqref{CCR1}, \eqref{CCR2} are satisfied).

Now let us discuss formulas \eqref{Npfinal0}--\eqref{Ppfinal0}. The operator terms related to quasi-particles have the form
\begin{align}\label{NpQP}
&\hat N_{QP}=\sum\limits_{j\neq 0}\left(\epsilon_{j}-\frac{{\vec k_{j}}^{2}}{2m\gamma_{j}}\right)\hat a_{j}^{\dagger}\hat a_{j}^{},
\\\label{EpQP}
&\hat E_{QP}=\sum\limits_{j\neq 0}\left(\omega\left(\epsilon_{j}-\frac{{\vec k_{j}}^{2}}{2m\gamma_{j}}\right)+\gamma_{j}\right)\hat a_{j}^{\dagger}\hat a_{j}^{},\\\label{PpQP}
&\hat{{\vec P}}_{QP}=\sum\limits_{j\neq 0}\vec k_{j}\hat a_{j}^{\dagger}\hat a_{j}^{}.
\end{align}
One can see that the momentum of the $j$-th quasi-particle is $\vec k_{j}$, its particle number is $\epsilon_{j}-\frac{{\vec k_{j}}^{2}}{2m\gamma_{j}}$ and its energy is $\omega\left(\epsilon_{j}-\frac{{\vec k_{j}}^{2}}{2m\gamma_{j}}\right)+\gamma_{j}$. In the processes involving quasi-particles the momentum, the particle number and the energy must conserve, which is impossible for general $\epsilon_{j}$ --- the corresponding system of equations turns out to be overconstrained (see, for example, the case presented in Appendix~A, where the two conditions uniquely define the momentum $\vec p$ for the given momentum $\vec k$), the obvious exception is $\epsilon_{j}-\frac{{\vec k_{j}}^{2}}{2m\gamma_{j}}\sim\gamma_{j}$. However, $\epsilon_{j}$ are the free parameters of the theory, and one can choose their values according to physical demands. The most natural choice is just
\begin{equation}\label{epsilondef}
\epsilon_{j}=\frac{{\vec k_{j}}^{2}}{2m\gamma_{j}},
\end{equation}
leading to
\begin{equation}
\hat N_{QP}=0,\qquad\hat E_{QP}=\sum\limits_{j\neq 0}\gamma_{j}\hat a_{j}^{\dagger}\hat a_{j}^{},\qquad\hat{{\vec P}}_{QP}=\sum\limits_{j\neq 0}\vec k_{j}\hat a_{j}^{\dagger}\hat a_{j}^{}.
\end{equation}
Now the three conservation laws do not overconstrain the resulting system of equations, because for any quasi-particle its particle number is just zero.

Finally, skipping the irrelevant $c$-number terms, we get for \eqref{Npfinal0}--\eqref{Ppfinal0}
\begin{align}\label{Npfinal}
&\hat N_{p}=\frac{dN_{0}}{d\omega}\,\tilde q\left(\hat a_{0}+\hat a_{0}^{\dagger}\right),\\\label{Epfinal}
&\hat E_{p}=\frac{dE_{0}}{d\omega}\,\tilde q\left(\hat a_{0}+\hat a_{0}^{\dagger}\right)+\frac{1}{2}\frac{d^{2}E_{0}}{d\omega^{2}}\,{\tilde q}^{2}\left(\hat a_{0}+\hat a_{0}^{\dagger}\right)^{2}+\sum\limits_{j\neq 0}\gamma_{j}\hat a_{j}^{\dagger}\hat a_{j}^{},\\\label{Ppfinal}
&\hat{\vec{P_{p}}}=\sum\limits_{j\neq 0}\vec k_{j}\hat a_{j}^{\dagger}\hat a_{j}^{}.
\end{align}
These formulas imply that, in the approximation used in the present paper, at zero temperature all particles reside in the condensate. According to \eqref{Npfinal}, the particle number of the system can be changed only by adding particles to the condensate or removing them from the condensate, the latter can be done by generating the corresponding nonoscillation mode. As was mentioned above, in the classical theory this nonoscillation mode corresponds to a change of the frequency $\omega$ of the background solution \eqref{backgrsol0}. Thus, at least free quasi-particles do not change the particle number of the system, which is natural from the physical point of view.

An important point is that one gets a classical behavior of the particle number of the system, i.e., the number of particles in the condensate can take an arbitrary value and it is not quantized, see \eqref{backgrN} and \eqref{Npfinal}. Even though adding particles to the system or removing them from the system is performed by means of the nonoscillation quantum mode with $[\hat a_{0},\hat a_{0}^{\dagger}]=1$, formally we still can add, say, ``a half of the particle'' because the spectrum of the operator $\hat a_{0}+\hat a_{0}^{\dagger}$ is continuous (see Section~\ref{section4}). This happens because of our choice of the ``basis'' for examining physical consequences of the theory --- since from the very beginning the condensate is described by the ordinary (non-operator) function \eqref{backgrsol0}, providing the $c$-number value of $N_{0}$, even the quantum mode \eqref{quantmode} with \eqref{quantmode2} does not violate this behaviour of the system.

\section{Conclusion}
As was noted in Introduction, it is well known that the Bogolyubov approach results in nonconservation of the particle number in the effective theory describing quasi-particles. However, the origin of this nonconservation is not just the use of the approximation that replaces the creation and annihilation operators of condensed particles by $c$-numbers, but the fact that, in the case of a stationary time-dependent background solution, the use of only the linear approximation within the second quantization formalism is not sufficient for a correct description of even free quasi-particles. Indeed, it is possible to recover the canonical commutation relations in the linear approximation, but even with such a modification we still get \eqref{Npquadr3}. Meanwhile, the use of the first nonlinear correction, together with the use of the additional nonoscillation modes in the linear approximation, allows one to modify the Bogolyubov approach and to solve the problem of nonconserved particle number while keeping all the key steps and ideas proposed in \cite{Bogolyubov}. Note that the particle number is conserved automatically with the solution presented above and additional methods ensuring the particle number conservation turn out to be completely unnecessary. It is interesting that the nonoscillation modes which recover the canonical commutation relations at the linear order also provide the fulfillment of these relations at the next order (see Sections~\ref{sectCompensation} and \ref{ExtraCont}). Besides, due to these modes operator \eqref{Npfinal} provides classical (i.e., non-quantized) values for the number of particles. This behaviour is correct, it is just a consequence of the approximation that was chosen at the initial stage to describe the condensate.

Of course, solution \eqref{solfull} is not exact. Although nothing forbids the existence of solutions for all $\hat\phi_{n}(t,\vec x)$ in the infinite series \eqref{seriesPhiinf}, satisfying the canonical commutation relations \eqref{commrel1} and \eqref{commrel2} (in fact, this series would represent an {\em exact} solution of the Heisenberg equation \eqref{Heiseq}, although obtained using the methods of perturbation theory), I have no any rigorous proof that such an infinite series corresponding to a physically reasonable solution indeed exists. Moreover, it is not clear how to practically calculate even the second nonlinear correction --- the calculation of the first nonlinear correction demands bulky analytical calculations, so calculation of the second correction seems to be much more difficult.

However, even as an approximation the solution presented in this paper is more accurate with respect to the canonical commutation relations than those resulting in \eqref{CCRviolat} or \eqref{CCR1violat}, \eqref{CCR2violat}. Indeed, the contributions of the commutators discussed in this paper to the canonical commutation relations are of the order of $\beta\frac{1}{L^{3}}=\frac{1}{\sqrt{N_{0}}}\frac{1}{L^{3}}$ (compare with the estimates in Footnote~1). All these contributions compensate each other and the overall term $\sim\beta\frac{1}{L^{3}}$ vanishes. Thus, the canonical commutation relations are violated at least by the terms $\sim\beta^{2}\frac{1}{L^{3}}=\frac{1}{N_{0}}\frac{1}{L^{3}}$.

Note that since $\hat\Psi(0,\vec x)$ defined by \eqref{solfull} is such that
\begin{equation}
\hat\Psi(0,\vec x)\not\equiv\Psi_{0}(0,\vec x)+\hat\varphi_{\textrm{osc}}^{}(0,\vec x),
\end{equation}
where $\hat\varphi_{\textrm{osc}}^{}(0,\vec x)$ is defined by \eqref{linsol}, this $\hat\Psi(0,\vec x)$ leads to the interaction Hamiltonian that differs from the one following from the original approach based on the use of only the linear approximation.\footnote{Since for $\hat\Psi(0,\vec x)$ defined by \eqref{solfull} the canonical commutation relations are violated, even though by the terms $\sim\frac{1}{N_{0}}\frac{1}{L^{3}}$, one should keep in mind that the use of formula \eqref{Heisevol} with this $\hat\Psi(0,\vec x)$ may result in time-dependent terms in $\hat N$ (and probably in $\hat{\vec{P}}$) at the higher orders.} Moreover, as was noted in Section~\ref{section4}, there may arise the terms describing interaction of quasi-particles with the modes containing the operators $\hat a_{0}^{}$ and $\hat a_{0}^{\dagger}$ (i.e., with particles incoming to the system (namely, to the condensate) or outgoing from the system), which may lead to additional effects.\footnote{In some sense, these terms also describe interaction of quasi-particles with the condensate. However, this interaction differs from the interaction between non-condensed particles and condensed ones like the one considered, for example, in \cite{NZG,ZNG} within the hydrodynamic approach using the Hartree-Fock mean field approximation. Contrary to the case of \cite{NZG,ZNG}, here the interaction terms arise in a natural way.} Calculation of this interaction Hamiltonian and analysis of possible effects produced by the new terms in it, as well as generalization of the proposed method to an arbitrary interaction potential $V(|\vec x-\vec y|)$, call for further detailed investigation.

\section*{Acknowledgements}
The work was supported by the Grant 16-12-10494 of the Russian Science Foundation.

\section*{Appendix~A}
In the limit $L\to\infty$, let us take a mode with $\vec k_{a}=\frac{1}{2}\vec k+\vec p$, where $\vec k \bot \vec p$, a mode with $\vec k_{b}=\frac{1}{2}\vec k-\vec p$, and a mode with $\vec k_{c}=\vec k$. It is clear that
\begin{equation}
\vec k_{a}+\vec k_{b}=\vec k_{c}.
\end{equation}
It is easy to check that for
\begin{equation}
{\vec p}^{2}=\frac{\sqrt{\left(\frac{{\vec k}^{2}}{2}+4\omega m\right)^{2}+\frac{3}{4}\left({\vec k}^{2}\right)^{2}}-\left(\frac{{\vec k}^{2}}{2}+4\omega m\right)}{2}
\end{equation}
the relation $\gamma_{a}+\gamma_{b}=\gamma_{c}$ holds.

\section*{Appendix~B}
After straightforward calculations, for $[\hat{\tilde\phi}_{j,l}^{\scriptscriptstyle+}(t,\vec x),\hat\varphi(t,\vec y)]+[\hat\varphi(t,\vec x),\hat{\tilde\phi}_{j,l}^{\scriptscriptstyle+}(t,\vec y)]$ one can get
\begin{align}\nonumber
&[\hat{\tilde\phi}_{j,l}^{\scriptscriptstyle+}(t,\vec x),\hat\varphi(t,\vec y)]+[\hat\varphi(t,\vec x),\hat{\tilde\phi}_{j,l}^{\scriptscriptstyle+}(t,\vec y)]\\\nonumber
=\frac{\omega}{L^{3}}\Biggl(&\tilde M_{j,l}^{\scriptscriptstyle+}e^{\frac{i}{\hbar}(\vec k_{j}+\vec k_{l})\vec y}\left(\hat a_{j}^{}d_{l}e^{-\frac{i}{\hbar}\gamma_{j}t}e^{-\frac{i}{\hbar}\vec k_{l}\vec x}+\hat a_{l}^{}d_{j}e^{-\frac{i}{\hbar}\gamma_{l}t}e^{-\frac{i}{\hbar}\vec k_{j}\vec x}\right)\\\nonumber
-&\tilde M_{j,l}^{\scriptscriptstyle+}e^{\frac{i}{\hbar}(\vec k_{j}+\vec k_{l})\vec x}\left(\hat a_{j}^{}d_{l}e^{-\frac{i}{\hbar}\gamma_{j}t}e^{-\frac{i}{\hbar}\vec k_{l}\vec y}+\hat a_{l}^{}d_{j}e^{-\frac{i}{\hbar}\gamma_{l}t}e^{-\frac{i}{\hbar}\vec k_{j}\vec y}\right)\\\nonumber
+&\tilde N_{j,l}^{\scriptscriptstyle+}e^{-\frac{i}{\hbar}(\vec k_{j}+\vec k_{l})\vec y}\left(\hat a_{j}^{\dagger}c_{l}e^{\frac{i}{\hbar}\gamma_{j}t}e^{\frac{i}{\hbar}\vec k_{l}\vec x}+\hat a_{l}^{\dagger}c_{j}e^{\frac{i}{\hbar}\gamma_{l}t}e^{\frac{i}{\hbar}\vec k_{j}\vec x}\right)\\\label{commjl+}
-&\tilde N_{j,l}^{\scriptscriptstyle+}e^{-\frac{i}{\hbar}(\vec k_{j}+\vec k_{l})\vec x}\left(\hat a_{j}^{\dagger}c_{l}e^{\frac{i}{\hbar}\gamma_{j}t}e^{\frac{i}{\hbar}\vec k_{l}\vec y}+\hat a_{l}^{\dagger}c_{j}e^{\frac{i}{\hbar}\gamma_{l}t}e^{\frac{i}{\hbar}\vec k_{j}\vec y}\right)\Biggr).
\end{align}
Let us define (see \eqref{tildephidef})
\begin{equation}\label{tildephidefapp}
\hat{\tilde\phi}^{\scriptscriptstyle+}(t,\vec x)=\frac{1}{2}\underset{\substack{j\neq 0, l\neq 0\\j\neq-l}}{\sum\sum}\hat{\tilde\phi}_{j,l}^{\scriptscriptstyle+}(t,\vec x),\qquad\hat{\tilde\phi}^{\scriptscriptstyle-}(t,\vec x)=\frac{1}{2}\underset{\substack{j\neq 0, l\neq 0\\j\neq l}}{\sum\sum}\hat{\tilde\phi}_{j,l}^{\scriptscriptstyle-}(t,\vec x).
\end{equation}
For $\hat{\tilde\phi}^{\scriptscriptstyle+}(t,\vec x)$ defined by \eqref{tildephidefapp}, we get from \eqref{commjl+}
\begin{align}\nonumber
&[\hat{\tilde\phi}^{\scriptscriptstyle+}(t,\vec x),\hat\varphi(t,\vec y)]+[\hat\varphi(t,\vec x),\hat{\tilde\phi}^{\scriptscriptstyle+}(t,\vec y)]\\\nonumber
=\frac{\omega}{2L^{3}}\underset{\substack{j\neq 0, l\neq 0\\j\neq-l}}{\sum\sum}\Biggl(&\tilde M_{j,l}^{\scriptscriptstyle+}e^{\frac{i}{\hbar}(\vec k_{j}+\vec k_{l})\vec y}\left(\hat a_{j}^{}d_{l}e^{-\frac{i}{\hbar}\gamma_{j}t}e^{-\frac{i}{\hbar}\vec k_{l}\vec x}+\hat a_{l}^{}d_{j}e^{-\frac{i}{\hbar}\gamma_{l}t}e^{-\frac{i}{\hbar}\vec k_{j}\vec x}\right)\\\nonumber
-&\tilde M_{j,l}^{\scriptscriptstyle+}e^{\frac{i}{\hbar}(\vec k_{j}+\vec k_{l})\vec x}\left(\hat a_{j}^{}d_{l}e^{-\frac{i}{\hbar}\gamma_{j}t}e^{-\frac{i}{\hbar}\vec k_{l}\vec y}+\hat a_{l}^{}d_{j}e^{-\frac{i}{\hbar}\gamma_{l}t}e^{-\frac{i}{\hbar}\vec k_{j}\vec y}\right)\\\nonumber
+&\tilde N_{j,l}^{\scriptscriptstyle+}e^{-\frac{i}{\hbar}(\vec k_{j}+\vec k_{l})\vec y}\left(\hat a_{j}^{\dagger}c_{l}e^{\frac{i}{\hbar}\gamma_{j}t}e^{\frac{i}{\hbar}\vec k_{l}\vec x}+\hat a_{l}^{\dagger}c_{j}e^{\frac{i}{\hbar}\gamma_{l}t}e^{\frac{i}{\hbar}\vec k_{j}\vec x}\right)\\\label{AppB2}
-&\tilde N_{j,l}^{\scriptscriptstyle+}e^{-\frac{i}{\hbar}(\vec k_{j}+\vec k_{l})\vec x}\left(\hat a_{j}^{\dagger}c_{l}e^{\frac{i}{\hbar}\gamma_{j}t}e^{\frac{i}{\hbar}\vec k_{l}\vec y}+\hat a_{l}^{\dagger}c_{j}e^{\frac{i}{\hbar}\gamma_{l}t}e^{\frac{i}{\hbar}\vec k_{j}\vec y}\right)\Biggr).
\end{align}
Let us change $j\leftrightarrow l$ in the terms with $\hat a_{l}^{}$ and $\hat a_{l}^{\dagger}$ in \eqref{AppB2}. We get
\begin{align}\nonumber
&[\hat{\tilde\phi}^{\scriptscriptstyle+}(t,\vec x),\hat\varphi(t,\vec y)]+[\hat\varphi(t,\vec x),\hat{\tilde\phi}^{\scriptscriptstyle+}(t,\vec y)]\\\nonumber
=\frac{\omega}{L^{3}}\underset{\substack{j\neq 0, l\neq 0\\j\neq-l}}{\sum\sum}\Biggl(&\hat a_{j}^{}e^{-\frac{i}{\hbar}\gamma_{j}t}\left(
\frac{\tilde M_{j,l}^{\scriptscriptstyle+}+\tilde M_{l,j}^{\scriptscriptstyle+}}{2}d_{l}e^{\frac{i}{\hbar}(\vec k_{j}+\vec k_{l})\vec y}e^{-\frac{i}{\hbar}\vec k_{l}\vec x}-\frac{\tilde M_{j,l}^{\scriptscriptstyle+}+\tilde M_{l,j}^{\scriptscriptstyle+}}{2}d_{l}e^{\frac{i}{\hbar}(\vec k_{j}+\vec k_{l})\vec x}e^{-\frac{i}{\hbar}\vec k_{l}\vec y}\right)\\\label{AppB3}
+&\hat a_{j}^{\dagger}e^{\frac{i}{\hbar}\gamma_{j}t}\left(\frac{\tilde N_{j,l}^{\scriptscriptstyle+}+\tilde N_{l,j}^{\scriptscriptstyle+}}{2}c_{l}e^{-\frac{i}{\hbar}(\vec k_{j}+\vec k_{l})\vec y}e^{\frac{i}{\hbar}\vec k_{l}\vec x}-\frac{\tilde N_{j,l}^{\scriptscriptstyle+}+\tilde N_{l,j}^{\scriptscriptstyle+}}{2}c_{l}e^{-\frac{i}{\hbar}(\vec k_{j}+\vec k_{l})\vec x}e^{\frac{i}{\hbar}\vec k_{l}\vec y}\right)\Biggr).
\end{align}
Finally, let us change $l\to -j-l$ in the terms $\sim e^{\frac{i}{\hbar}(\vec k_{j}+\vec k_{l})\vec x}e^{-\frac{i}{\hbar}\vec k_{l}\vec y}$ and $\sim e^{-\frac{i}{\hbar}(\vec k_{j}+\vec k_{l})\vec x}e^{\frac{i}{\hbar}\vec k_{l}\vec y}$ (according to definition \eqref{kdef}, $\vec k_{-j-l}=-\vec k_{-j}-\vec k_{-l}$) in \eqref{AppB3}, resulting in
\begin{align}\nonumber
&[\hat{\tilde\phi}^{\scriptscriptstyle+}(t,\vec x),\hat\varphi(t,\vec y)]+[\hat\varphi(t,\vec x),\hat{\tilde\phi}^{\scriptscriptstyle+}(t,\vec y)]\\\nonumber
=\frac{\omega}{L^{3}}\underset{\substack{j\neq 0, l\neq 0\\j\neq-l}}{\sum\sum}\Biggl(&\hat a_{j}^{}e^{-\frac{i}{\hbar}\gamma_{j}t}e^{\frac{i}{\hbar}(\vec k_{j}+\vec k_{l})\vec y}e^{-\frac{i}{\hbar}\vec k_{l}\vec x}\left(\frac{\tilde M_{j,l}^{\scriptscriptstyle+}+\tilde M_{l,j}^{\scriptscriptstyle+}}{2}d_{l}-
\frac{\tilde M_{j,-j-l}^{\scriptscriptstyle+}+\tilde M_{-j-l,j}^{\scriptscriptstyle+}}{2}d_{j+l}\right)\\\label{commtildefinal+}
+&\hat a_{j}^{\dagger}e^{\frac{i}{\hbar}\gamma_{j}t}e^{-\frac{i}{\hbar}(\vec k_{j}+\vec k_{l})\vec y}e^{\frac{i}{\hbar}\vec k_{l}\vec x}\left(\frac{\tilde N_{j,l}^{\scriptscriptstyle+}+\tilde N_{l,j}^{\scriptscriptstyle+}}{2}c_{l}-
\frac{\tilde N_{j,-j-l}^{\scriptscriptstyle+}+\tilde N_{-j-l,j}^{\scriptscriptstyle+}}{2}c_{j+l}\right)\Biggr).
\end{align}

Analogously, for $[\hat{\tilde\phi}_{j,l}^{\scriptscriptstyle-}(t,\vec x),\hat\varphi(t,\vec y)]+[\hat\varphi(t,\vec x),\hat{\tilde\phi}_{j,l}^{\scriptscriptstyle-}(t,\vec y)]$ one can get
\begin{align}\nonumber
&[\hat{\tilde\phi}_{j,l}^{\scriptscriptstyle-}(t,\vec x),\hat\varphi(t,\vec y)]+[\hat\varphi(t,\vec x),\hat{\tilde\phi}_{j,l}^{\scriptscriptstyle-}(t,\vec y)]\\\nonumber
=\frac{\omega}{L^{3}}\Biggl(&\tilde M_{j,l}^{\scriptscriptstyle-}e^{\frac{i}{\hbar}(\vec k_{j}-\vec k_{l})\vec y}\left(\hat a_{j}^{}c_{l}e^{-\frac{i}{\hbar}\gamma_{j}t}e^{\frac{i}{\hbar}\vec k_{l}\vec x}+\hat a_{l}^{\dagger}d_{j}e^{\frac{i}{\hbar}\gamma_{l}t}e^{-\frac{i}{\hbar}\vec k_{j}\vec x}\right)\\\nonumber
-&\tilde M_{j,l}^{\scriptscriptstyle-}e^{\frac{i}{\hbar}(\vec k_{j}-\vec k_{l})\vec x}\left(\hat a_{j}^{}c_{l}e^{-\frac{i}{\hbar}\gamma_{j}t}e^{\frac{i}{\hbar}\vec k_{l}\vec y}+\hat a_{l}^{\dagger}d_{j}e^{\frac{i}{\hbar}\gamma_{l}t}e^{-\frac{i}{\hbar}\vec k_{j}\vec y}\right)\\\nonumber
+&\tilde N_{j,l}^{\scriptscriptstyle-}e^{-\frac{i}{\hbar}(\vec k_{j}-\vec k_{l})\vec y}\left(\hat a_{l}^{}c_{j}e^{-\frac{i}{\hbar}\gamma_{l}t}e^{\frac{i}{\hbar}\vec k_{j}\vec x}+\hat a_{j}^{\dagger}d_{l}e^{\frac{i}{\hbar}\gamma_{j}t}e^{-\frac{i}{\hbar}\vec k_{l}\vec x}\right)\\\label{commjl-}
-&\tilde N_{j,l}^{\scriptscriptstyle-}e^{-\frac{i}{\hbar}(\vec k_{j}-\vec k_{l})\vec x}\left(\hat a_{l}^{}c_{j}e^{-\frac{i}{\hbar}\gamma_{l}t}e^{\frac{i}{\hbar}\vec k_{j}\vec y}+\hat a_{j}^{\dagger}d_{l}e^{\frac{i}{\hbar}\gamma_{j}t}e^{-\frac{i}{\hbar}\vec k_{l}\vec y}\right)\Biggr).
\end{align}
For $\hat{\tilde\phi}^{\scriptscriptstyle-}(t,\vec x)$ defined by \eqref{tildephidefapp}, we get from \eqref{commjl-}
\begin{align}\nonumber
&[\hat{\tilde\phi}^{\scriptscriptstyle-}(t,\vec x),\hat\varphi(t,\vec y)]+[\hat\varphi(t,\vec x),\hat{\tilde\phi}^{\scriptscriptstyle-}(t,\vec y)]\\\nonumber
=\frac{\omega}{2L^{3}}\underset{\substack{j\neq 0, l\neq 0\\j\neq l}}{\sum\sum}\Biggl(&\tilde M_{j,l}^{\scriptscriptstyle-}e^{\frac{i}{\hbar}(\vec k_{j}-\vec k_{l})\vec y}\left(\hat a_{j}^{}c_{l}e^{-\frac{i}{\hbar}\gamma_{j}t}e^{\frac{i}{\hbar}\vec k_{l}\vec x}+\hat a_{l}^{\dagger}d_{j}e^{\frac{i}{\hbar}\gamma_{l}t}e^{-\frac{i}{\hbar}\vec k_{j}\vec x}\right)\\\nonumber
-&\tilde M_{j,l}^{\scriptscriptstyle-}e^{\frac{i}{\hbar}(\vec k_{j}-\vec k_{l})\vec x}\left(\hat a_{j}^{}c_{l}e^{-\frac{i}{\hbar}\gamma_{j}t}e^{\frac{i}{\hbar}\vec k_{l}\vec y}+\hat a_{l}^{\dagger}d_{j}e^{\frac{i}{\hbar}\gamma_{l}t}e^{-\frac{i}{\hbar}\vec k_{j}\vec y}\right)\\\nonumber
+&\tilde N_{j,l}^{\scriptscriptstyle-}e^{-\frac{i}{\hbar}(\vec k_{j}-\vec k_{l})\vec y}\left(\hat a_{l}^{}c_{j}e^{-\frac{i}{\hbar}\gamma_{l}t}e^{\frac{i}{\hbar}\vec k_{j}\vec x}+\hat a_{j}^{\dagger}d_{l}e^{\frac{i}{\hbar}\gamma_{j}t}e^{-\frac{i}{\hbar}\vec k_{l}\vec x}\right)\\\label{AppB4}
-&\tilde N_{j,l}^{\scriptscriptstyle-}e^{-\frac{i}{\hbar}(\vec k_{j}-\vec k_{l})\vec x}\left(\hat a_{l}^{}c_{j}e^{-\frac{i}{\hbar}\gamma_{l}t}e^{\frac{i}{\hbar}\vec k_{j}\vec y}+\hat a_{j}^{\dagger}d_{l}e^{\frac{i}{\hbar}\gamma_{j}t}e^{-\frac{i}{\hbar}\vec k_{l}\vec y}\right)\Biggr).
\end{align}
Let us change $j\leftrightarrow l$ in the terms with $\hat a_{l}^{}$ and $\hat a_{l}^{\dagger}$ in \eqref{AppB4}. We get
\begin{align}\nonumber
&[\hat{\tilde\phi}^{\scriptscriptstyle-}(t,\vec x),\hat\varphi(t,\vec y)]+[\hat\varphi(t,\vec x),\hat{\tilde\phi}^{\scriptscriptstyle-}(t,\vec y)]\\\nonumber
=\frac{\omega}{L^{3}}\underset{\substack{j\neq 0, l\neq 0\\j\neq l}}{\sum\sum}\Biggl(&\hat a_{j}^{}e^{-\frac{i}{\hbar}\gamma_{j}t}\left(\frac{\tilde M_{j,l}^{\scriptscriptstyle-}+\tilde N_{l,j}^{\scriptscriptstyle-}}{2}c_{l}e^{\frac{i}{\hbar}(\vec k_{j}-\vec k_{l})\vec y}e^{\frac{i}{\hbar}\vec k_{l}\vec x}-\frac{\tilde M_{j,l}^{\scriptscriptstyle-}+\tilde N_{l,j}^{\scriptscriptstyle-}}{2}c_{l}e^{\frac{i}{\hbar}(\vec k_{j}-\vec k_{l})\vec x}e^{\frac{i}{\hbar}\vec k_{l}\vec y}\right)\\\label{AppB5}
+&\hat a_{j}^{\dagger}e^{\frac{i}{\hbar}\gamma_{j}t}\left(\frac{\tilde N_{j,l}^{\scriptscriptstyle-}+\tilde M_{l,j}^{\scriptscriptstyle-}}{2}d_{l}e^{-\frac{i}{\hbar}(\vec k_{j}-\vec k_{l})\vec y}e^{-\frac{i}{\hbar}\vec k_{l}\vec x}-\frac{\tilde N_{j,l}^{\scriptscriptstyle-}+\tilde M_{l,j}^{\scriptscriptstyle-}}{2}d_{l}e^{-\frac{i}{\hbar}(\vec k_{j}-\vec k_{l})\vec x}e^{-\frac{i}{\hbar}\vec k_{l}\vec y}\right)\Biggr).
\end{align}
Let us change $l\to j-l$ in the terms $\sim e^{\frac{i}{\hbar}(\vec k_{j}-\vec k_{l})\vec x}e^{\frac{i}{\hbar}\vec k_{l}\vec y}$ and $\sim e^{-\frac{i}{\hbar}(\vec k_{j}-\vec k_{l})\vec x}e^{-\frac{i}{\hbar}\vec k_{l}\vec y}$ in \eqref{AppB5}, resulting in
\begin{align}\nonumber
&[\hat{\tilde\phi}^{\scriptscriptstyle-}(t,\vec x),\hat\varphi(t,\vec y)]+[\hat\varphi(t,\vec x),\hat{\tilde\phi}^{\scriptscriptstyle-}(t,\vec y)]\\\nonumber
=\frac{\omega}{L^{3}}\underset{\substack{j\neq 0, l\neq 0\\j\neq l}}{\sum\sum}\Biggl(&\hat a_{j}^{}e^{-\frac{i}{\hbar}\gamma_{j}t}e^{\frac{i}{\hbar}(\vec k_{j}-\vec k_{l})\vec y}e^{\frac{i}{\hbar}\vec k_{l}\vec x}\left(\frac{\tilde M_{j,l}^{\scriptscriptstyle-}+\tilde N_{l,j}^{\scriptscriptstyle-}}{2}c_{l}-\frac{\tilde M_{j,j-l}^{\scriptscriptstyle-}+\tilde N_{j-l,j}^{\scriptscriptstyle-}}{2}c_{j-l}\right)\\\label{commtilde-}
+&\hat a_{j}^{\dagger}e^{\frac{i}{\hbar}\gamma_{j}t}e^{-\frac{i}{\hbar}(\vec k_{j}-\vec k_{l})\vec y}e^{-\frac{i}{\hbar}\vec k_{l}\vec x}\left(\frac{\tilde N_{j,l}^{\scriptscriptstyle-}+\tilde M_{l,j}^{\scriptscriptstyle-}}{2}d_{l}-\frac{\tilde N_{j,j-l}^{\scriptscriptstyle-}+\tilde M_{j-l,j}^{\scriptscriptstyle-}}{2}d_{j-l}\right)\Biggr).
\end{align}
And finally, let us change $l\to-l$ in \eqref{commtilde-}, resulting in
\begin{align}\nonumber
&[\hat{\tilde\phi}^{\scriptscriptstyle-}(t,\vec x),\hat\varphi(t,\vec y)]+[\hat\varphi(t,\vec x),\hat{\tilde\phi}^{\scriptscriptstyle-}(t,\vec y)]\\\nonumber
=\frac{\omega}{L^{3}}\underset{\substack{j\neq 0, l\neq 0\\j\neq -l}}{\sum\sum}\Biggl(&\hat a_{j}^{}e^{-\frac{i}{\hbar}\gamma_{j}t}e^{\frac{i}{\hbar}(\vec k_{j}+\vec k_{l})\vec y}e^{-\frac{i}{\hbar}\vec k_{l}\vec x}\left(\frac{\tilde M_{j,-l}^{\scriptscriptstyle-}+\tilde N_{-l,j}^{\scriptscriptstyle-}}{2}c_{l}-\frac{\tilde M_{j,j+l}^{\scriptscriptstyle-}+\tilde N_{j+l,j}^{\scriptscriptstyle-}}{2}c_{j+l}\right)\\\label{commtildefinal-}
+&\hat a_{j}^{\dagger}e^{\frac{i}{\hbar}\gamma_{j}t}e^{-\frac{i}{\hbar}(\vec k_{j}+\vec k_{l})\vec y}e^{\frac{i}{\hbar}\vec k_{l}\vec x}\left(\frac{\tilde N_{j,-l}^{\scriptscriptstyle-}+\tilde M_{-l,j}^{\scriptscriptstyle-}}{2}d_{l}-\frac{\tilde N_{j,j+l}^{\scriptscriptstyle-}+\tilde M_{j+l,j}^{\scriptscriptstyle-}}{2}d_{j+l}\right)\Biggr).
\end{align}
Combining \eqref{commtildefinal+} and \eqref{commtildefinal-}, for the operator $\hat{\tilde\phi}(t,\vec x)=\hat{\tilde\phi}^{\scriptscriptstyle+}(t,\vec x)+\hat{\tilde\phi}^{\scriptscriptstyle-}(t,\vec x)$ we get \eqref{commtildefinal}.

\section*{Appendix~C}
After straightforward calculations, for $[\hat{\tilde\phi}_{j,l}^{\scriptscriptstyle+}(t,\vec x),\hat\varphi^{\dagger}(t,\vec y)]+[\hat\varphi(t,\vec x),\hat{\tilde\phi}_{j,l}^{\scriptscriptstyle+{\dagger}}(t,\vec y)]$ one can get
\begin{align}\nonumber
&[\hat{\tilde\phi}_{j,l}^{\scriptscriptstyle+}(t,\vec x),\hat\varphi^{\dagger}(t,\vec y)]+[\hat\varphi(t,\vec x),\hat{\tilde\phi}_{j,l}^{\scriptscriptstyle+{\dagger}}(t,\vec y)]\\\nonumber
=\frac{\omega}{L^{3}}\Biggl(&\tilde M_{j,l}^{\scriptscriptstyle+}e^{-\frac{i}{\hbar}(\vec k_{j}+\vec k_{l})\vec y}\left(\hat a_{j}^{\dagger}c_{l}e^{\frac{i}{\hbar}\gamma_{j}t}e^{\frac{i}{\hbar}\vec k_{l}\vec x}+\hat a_{l}^{\dagger}c_{j}e^{\frac{i}{\hbar}\gamma_{l}t}e^{\frac{i}{\hbar}\vec k_{j}\vec x}\right)\\\nonumber
+&\tilde M_{j,l}^{\scriptscriptstyle+}e^{\frac{i}{\hbar}(\vec k_{j}+\vec k_{l})\vec x}\left(\hat a_{j}^{}c_{l}e^{-\frac{i}{\hbar}\gamma_{j}t}e^{-\frac{i}{\hbar}\vec k_{l}\vec y}+\hat a_{l}^{}c_{j}e^{-\frac{i}{\hbar}\gamma_{l}t}e^{-\frac{i}{\hbar}\vec k_{j}\vec y}\right)\\\nonumber
+&\tilde N_{j,l}^{\scriptscriptstyle+}e^{\frac{i}{\hbar}(\vec k_{j}+\vec k_{l})\vec y}\left(\hat a_{j}^{}d_{l}e^{-\frac{i}{\hbar}\gamma_{j}t}e^{-\frac{i}{\hbar}\vec k_{l}\vec x}+\hat a_{l}^{}d_{j}e^{-\frac{i}{\hbar}\gamma_{l}t}e^{-\frac{i}{\hbar}\vec k_{j}\vec x}\right)\\\label{commjldagg+}
+&\tilde N_{j,l}^{\scriptscriptstyle+}e^{-\frac{i}{\hbar}(\vec k_{j}+\vec k_{l})\vec x}\left(\hat a_{j}^{\dagger}d_{l}e^{\frac{i}{\hbar}\gamma_{j}t}e^{\frac{i}{\hbar}\vec k_{l}\vec y}+\hat a_{l}^{\dagger}d_{j}e^{\frac{i}{\hbar}\gamma_{l}t}e^{\frac{i}{\hbar}\vec k_{j}\vec y}\right)\Biggr).
\end{align}
For $\hat{\tilde\phi}^{\scriptscriptstyle+}(t,\vec x)$ defined by \eqref{tildephidefapp}, we get from \eqref{commjldagg+}
\begin{align}\nonumber
&[\hat{\tilde\phi}^{\scriptscriptstyle+}(t,\vec x),\hat\varphi^{\dagger}(t,\vec y)]+[\hat\varphi(t,\vec x),\hat{\tilde\phi}^{\scriptscriptstyle+{\dagger}}(t,\vec y)]\\\nonumber
=\frac{\omega}{2L^{3}}\underset{\substack{j\neq 0, l\neq 0\\j\neq-l}}{\sum\sum}\Biggl(&\tilde M_{j,l}^{\scriptscriptstyle+}e^{-\frac{i}{\hbar}(\vec k_{j}+\vec k_{l})\vec y}\left(\hat a_{j}^{\dagger}c_{l}e^{\frac{i}{\hbar}\gamma_{j}t}e^{\frac{i}{\hbar}\vec k_{l}\vec x}+\hat a_{l}^{\dagger}c_{j}e^{\frac{i}{\hbar}\gamma_{l}t}e^{\frac{i}{\hbar}\vec k_{j}\vec x}\right)\\\nonumber
+&\tilde M_{j,l}^{\scriptscriptstyle+}e^{\frac{i}{\hbar}(\vec k_{j}+\vec k_{l})\vec x}\left(\hat a_{j}^{}c_{l}e^{-\frac{i}{\hbar}\gamma_{j}t}e^{-\frac{i}{\hbar}\vec k_{l}\vec y}+\hat a_{l}^{}c_{j}e^{-\frac{i}{\hbar}\gamma_{l}t}e^{-\frac{i}{\hbar}\vec k_{j}\vec y}\right)\\\nonumber
+&\tilde N_{j,l}^{\scriptscriptstyle+}e^{\frac{i}{\hbar}(\vec k_{j}+\vec k_{l})\vec y}\left(\hat a_{j}^{}d_{l}e^{-\frac{i}{\hbar}\gamma_{j}t}e^{-\frac{i}{\hbar}\vec k_{l}\vec x}+\hat a_{l}^{}d_{j}e^{-\frac{i}{\hbar}\gamma_{l}t}e^{-\frac{i}{\hbar}\vec k_{j}\vec x}\right)\\\label{AppC2}
+&\tilde N_{j,l}^{\scriptscriptstyle+}e^{-\frac{i}{\hbar}(\vec k_{j}+\vec k_{l})\vec x}\left(\hat a_{j}^{\dagger}d_{l}e^{\frac{i}{\hbar}\gamma_{j}t}e^{\frac{i}{\hbar}\vec k_{l}\vec y}+\hat a_{l}^{\dagger}d_{j}e^{\frac{i}{\hbar}\gamma_{l}t}e^{\frac{i}{\hbar}\vec k_{j}\vec y}\right)\Biggr).
\end{align}
In full analogy with the calculation of the first commutation relation, let us change $j\leftrightarrow l$ in the terms with $\hat a_{l}^{}$ and $\hat a_{l}^{\dagger}$ in \eqref{AppC2}. We get
\begin{align}\nonumber
&[\hat{\tilde\phi}^{\scriptscriptstyle+}(t,\vec x),\hat\varphi^{\dagger}(t,\vec y)]+[\hat\varphi(t,\vec x),\hat{\tilde\phi}^{\scriptscriptstyle+{\dagger}}(t,\vec y)]\\\nonumber
=\frac{\omega}{L^{3}}\underset{\substack{j\neq 0, l\neq 0\\j\neq-l}}{\sum\sum}\Biggl(&\hat a_{j}^{\dagger}e^{\frac{i}{\hbar}\gamma_{j}t}\left(
\frac{\tilde M_{j,l}^{\scriptscriptstyle+}+\tilde M_{l,j}^{\scriptscriptstyle+}}{2}c_{l}e^{-\frac{i}{\hbar}(\vec k_{j}+\vec k_{l})\vec y}e^{\frac{i}{\hbar}\vec k_{l}\vec x}+
\frac{\tilde N_{j,l}^{\scriptscriptstyle+}+\tilde N_{l,j}^{\scriptscriptstyle+}}{2}d_{l}e^{-\frac{i}{\hbar}(\vec k_{j}+\vec k_{l})\vec x}e^{\frac{i}{\hbar}\vec k_{l}\vec y}\right)\\\label{AppC3}
+&\hat a_{j}^{}e^{-\frac{i}{\hbar}\gamma_{j}t}\left(\frac{\tilde M_{j,l}^{\scriptscriptstyle+}+\tilde M_{l,j}^{\scriptscriptstyle+}}{2}c_{l}e^{\frac{i}{\hbar}(\vec k_{j}+\vec k_{l})\vec x}e^{-\frac{i}{\hbar}\vec k_{l}\vec y}+\frac{\tilde N_{j,l}^{\scriptscriptstyle+}+\tilde N_{l,j}^{\scriptscriptstyle+}}{2}d_{l}e^{\frac{i}{\hbar}(\vec k_{j}+\vec k_{l})\vec y}e^{-\frac{i}{\hbar}\vec k_{l}\vec x}\right)\Biggr).
\end{align}
Finally, let us change $l\to -j-l$ in the terms $\sim e^{-\frac{i}{\hbar}(\vec k_{j}+\vec k_{l})\vec x}e^{\frac{i}{\hbar}\vec k_{l}\vec y}$ and $\sim e^{\frac{i}{\hbar}(\vec k_{j}+\vec k_{l})\vec y}e^{-\frac{i}{\hbar}\vec k_{l}\vec x}$ in \eqref{AppC3}, resulting in
\begin{align}\nonumber
&[\hat{\tilde\phi}^{\scriptscriptstyle+}(t,\vec x),\hat\varphi^{\dagger}(t,\vec y)]+[\hat\varphi(t,\vec x),\hat{\tilde\phi}^{\scriptscriptstyle+{\dagger}}(t,\vec y)]\\\nonumber
=\frac{\omega}{L^{3}}\underset{\substack{j\neq 0, l\neq 0\\j\neq-l}}{\sum\sum}\Biggl(&\hat a_{j}^{\dagger}e^{\frac{i}{\hbar}\gamma_{j}t}e^{-\frac{i}{\hbar}(\vec k_{j}+\vec k_{l})\vec y}e^{\frac{i}{\hbar}\vec k_{l}\vec x}\left(\frac{\tilde M_{j,l}^{\scriptscriptstyle+}+\tilde M_{l,j}^{\scriptscriptstyle+}}{2}c_{l}+
\frac{\tilde N_{j,-j-l}^{\scriptscriptstyle+}+\tilde N_{-j-l,j}^{\scriptscriptstyle+}}{2}d_{j+l}\right)
\\\label{commdagger+}
+&\hat a_{j}^{}e^{-\frac{i}{\hbar}\gamma_{j}t}e^{\frac{i}{\hbar}(\vec k_{j}+\vec k_{l})\vec x}e^{-\frac{i}{\hbar}\vec k_{l}\vec y}\left(\frac{\tilde M_{j,l}^{\scriptscriptstyle+}+\tilde M_{l,j}^{\scriptscriptstyle+}}{2}c_{l}+
\frac{\tilde N_{j,-j-l}^{\scriptscriptstyle+}+\tilde N_{-j-l,j}^{\scriptscriptstyle+}}{2}d_{j+l}\right)\Biggr).
\end{align}

Analogously, for $[\hat{\tilde\phi}_{j,l}^{\scriptscriptstyle-}(t,\vec x),\hat\varphi^{\dagger}(t,\vec y)]+[\hat\varphi(t,\vec x),\hat{\tilde\phi}_{j,l}^{\scriptscriptstyle-{\dagger}}(t,\vec y)]$ one can get
\begin{align}\nonumber
&[\hat{\tilde\phi}_{j,l}^{\scriptscriptstyle-}(t,\vec x),\hat\varphi^{\dagger}(t,\vec y)]+[\hat\varphi(t,\vec x),\hat{\tilde\phi}_{j,l}^{\scriptscriptstyle-{\dagger}}(t,\vec y)]\\\nonumber
=\frac{\omega}{L^{3}}\Biggl(&\tilde M_{j,l}^{\scriptscriptstyle-}e^{-\frac{i}{\hbar}(\vec k_{j}-\vec k_{l})\vec y}\left(\hat a_{j}^{\dagger}d_{l}e^{\frac{i}{\hbar}\gamma_{j}t}e^{-\frac{i}{\hbar}\vec k_{l}\vec x}+\hat a_{l}^{}c_{j}e^{-\frac{i}{\hbar}\gamma_{l}t}e^{\frac{i}{\hbar}\vec k_{j}\vec x}\right)\\\nonumber
+&\tilde M_{j,l}^{\scriptscriptstyle-}e^{\frac{i}{\hbar}(\vec k_{j}-\vec k_{l})\vec x}\left(\hat a_{j}^{}d_{l}e^{-\frac{i}{\hbar}\gamma_{j}t}e^{\frac{i}{\hbar}\vec k_{l}\vec y}+\hat a_{l}^{\dagger}c_{j}e^{\frac{i}{\hbar}\gamma_{l}t}e^{-\frac{i}{\hbar}\vec k_{j}\vec y}\right)\\\nonumber
+&\tilde N_{j,l}^{\scriptscriptstyle-}e^{\frac{i}{\hbar}(\vec k_{j}-\vec k_{l})\vec y}\left(\hat a_{j}^{}c_{l}e^{-\frac{i}{\hbar}\gamma_{j}t}e^{\frac{i}{\hbar}\vec k_{l}\vec x}+\hat a_{l}^{\dagger}d_{j}e^{\frac{i}{\hbar}\gamma_{l}t}e^{-\frac{i}{\hbar}\vec k_{j}\vec x}\right)\\\label{commjldagg-}
+&\tilde N_{j,l}^{\scriptscriptstyle-}e^{-\frac{i}{\hbar}(\vec k_{j}-\vec k_{l})\vec x}\left(\hat a_{j}^{\dagger}c_{l}e^{\frac{i}{\hbar}\gamma_{j}t}e^{-\frac{i}{\hbar}\vec k_{l}\vec y}+\hat a_{l}^{}d_{j}e^{-\frac{i}{\hbar}\gamma_{l}t}e^{\frac{i}{\hbar}\vec k_{j}\vec y}\right)\Biggr).
\end{align}
For $\hat{\tilde\phi}^{\scriptscriptstyle-}(t,\vec x)$ defined by \eqref{tildephidefapp}, we get from \eqref{commjldagg-}
\begin{align}\nonumber
&[\hat{\tilde\phi}^{\scriptscriptstyle-}(t,\vec x),\hat\varphi^{\dagger}(t,\vec y)]+[\hat\varphi(t,\vec x),\hat{\tilde\phi}^{\scriptscriptstyle-{\dagger}}(t,\vec y)]\\\nonumber
=\frac{\omega}{2L^{3}}\underset{\substack{j\neq 0, l\neq 0\\j\neq l}}{\sum\sum}\Biggl(&\tilde M_{j,l}^{\scriptscriptstyle-}e^{-\frac{i}{\hbar}(\vec k_{j}-\vec k_{l})\vec y}\left(\hat a_{j}^{\dagger}d_{l}e^{\frac{i}{\hbar}\gamma_{j}t}e^{-\frac{i}{\hbar}\vec k_{l}\vec x}+\hat a_{l}^{}c_{j}e^{-\frac{i}{\hbar}\gamma_{l}t}e^{\frac{i}{\hbar}\vec k_{j}\vec x}\right)\\\nonumber
+&\tilde M_{j,l}^{\scriptscriptstyle-}e^{\frac{i}{\hbar}(\vec k_{j}-\vec k_{l})\vec x}\left(\hat a_{j}^{}d_{l}e^{-\frac{i}{\hbar}\gamma_{j}t}e^{\frac{i}{\hbar}\vec k_{l}\vec y}+\hat a_{l}^{\dagger}c_{j}e^{\frac{i}{\hbar}\gamma_{l}t}e^{-\frac{i}{\hbar}\vec k_{j}\vec y}\right)\\\nonumber
+&\tilde N_{j,l}^{\scriptscriptstyle-}e^{\frac{i}{\hbar}(\vec k_{j}-\vec k_{l})\vec y}\left(\hat a_{j}^{}c_{l}e^{-\frac{i}{\hbar}\gamma_{j}t}e^{\frac{i}{\hbar}\vec k_{l}\vec x}+\hat a_{l}^{\dagger}d_{j}e^{\frac{i}{\hbar}\gamma_{l}t}e^{-\frac{i}{\hbar}\vec k_{j}\vec x}\right)\\\label{AppC4}
+&\tilde N_{j,l}^{\scriptscriptstyle-}e^{-\frac{i}{\hbar}(\vec k_{j}-\vec k_{l})\vec x}\left(\hat a_{j}^{\dagger}c_{l}e^{\frac{i}{\hbar}\gamma_{j}t}e^{-\frac{i}{\hbar}\vec k_{l}\vec y}+\hat a_{l}^{}d_{j}e^{-\frac{i}{\hbar}\gamma_{l}t}e^{\frac{i}{\hbar}\vec k_{j}\vec y}\right)\Biggr).
\end{align}
Let us change $j\leftrightarrow l$ in the terms with $\hat a_{l}^{}$ and $\hat a_{l}^{\dagger}$ in \eqref{AppC4}. We get
\begin{align}\nonumber
&[\hat{\tilde\phi}^{\scriptscriptstyle-}(t,\vec x),\hat\varphi^{\dagger}(t,\vec y)]+[\hat\varphi(t,\vec x),\hat{\tilde\phi}^{\scriptscriptstyle-{\dagger}}(t,\vec y)]\\\nonumber
=\frac{\omega}{L^{3}}\underset{\substack{j\neq 0, l\neq 0\\j\neq l}}{\sum\sum}\Biggl(&\hat a_{j}^{\dagger}e^{\frac{i}{\hbar}\gamma_{j}t}\left(\frac{\tilde M_{j,l}^{\scriptscriptstyle-}+\tilde N_{l,j}^{\scriptscriptstyle-}}{2}d_{l}e^{-\frac{i}{\hbar}(\vec k_{j}-\vec k_{l})\vec y}e^{-\frac{i}{\hbar}\vec k_{l}\vec x}+\frac{\tilde N_{j,l}^{\scriptscriptstyle-}+\tilde M_{l,j}^{\scriptscriptstyle-}}{2}c_{l}e^{-\frac{i}{\hbar}(\vec k_{j}-\vec k_{l})\vec x}e^{-\frac{i}{\hbar}\vec k_{l}\vec y}\right)\\\label{AppC5}
+&\hat a_{j}^{}e^{-\frac{i}{\hbar}\gamma_{j}t}\left(\frac{\tilde M_{j,l}^{\scriptscriptstyle-}+\tilde N_{l,j}^{\scriptscriptstyle-}}{2}d_{l}e^{\frac{i}{\hbar}(\vec k_{j}-\vec k_{l})\vec x}e^{\frac{i}{\hbar}\vec k_{l}\vec y}+\frac{\tilde N_{j,l}^{\scriptscriptstyle-}+\tilde M_{l,j}^{\scriptscriptstyle-}}{2}c_{l}e^{\frac{i}{\hbar}(\vec k_{j}-\vec k_{l})\vec y}e^{\frac{i}{\hbar}\vec k_{l}\vec x}\right)\Biggr).
\end{align}
Let us change $l\to j-l$ in the terms $\sim e^{-\frac{i}{\hbar}(\vec k_{j}-\vec k_{l})\vec x}e^{-\frac{i}{\hbar}\vec k_{l}\vec y}$ and $\sim e^{\frac{i}{\hbar}(\vec k_{j}-\vec k_{l})\vec y}e^{\frac{i}{\hbar}\vec k_{l}\vec x}$ in \eqref{AppC5}, resulting in
\begin{align}\nonumber
&[\hat{\tilde\phi}^{\scriptscriptstyle-}(t,\vec x),\hat\varphi^{\dagger}(t,\vec y)]+[\hat\varphi(t,\vec x),\hat{\tilde\phi}^{\scriptscriptstyle-{\dagger}}(t,\vec y)]\\\nonumber
=\frac{\omega}{L^{3}}\underset{\substack{j\neq 0, l\neq 0\\j\neq l}}{\sum\sum}\Biggl(&\hat a_{j}^{\dagger}e^{\frac{i}{\hbar}\gamma_{j}t}e^{-\frac{i}{\hbar}(\vec k_{j}-\vec k_{l})\vec y}e^{-\frac{i}{\hbar}\vec k_{l}\vec x}\left(\frac{\tilde M_{j,l}^{\scriptscriptstyle-}+\tilde N_{l,j}^{\scriptscriptstyle-}}{2}d_{l}+
\frac{\tilde N_{j,j-l}^{\scriptscriptstyle-}+\tilde M_{j-l,j}^{\scriptscriptstyle-}}{2}c_{j-l}\right)\\\label{AppC6}
+&\hat a_{j}^{}e^{-\frac{i}{\hbar}\gamma_{j}t}e^{\frac{i}{\hbar}(\vec k_{j}-\vec k_{l})\vec x}e^{\frac{i}{\hbar}\vec k_{l}\vec y}\left(\frac{\tilde M_{j,l}^{\scriptscriptstyle-}+\tilde N_{l,j}^{\scriptscriptstyle-}}{2}d_{l}+
\frac{\tilde N_{j,j-l}^{\scriptscriptstyle-}+\tilde M_{j-l,j}^{\scriptscriptstyle-}}{2}c_{j-l}\right)\Biggr).
\end{align}
And finally, let us change $l\to-l$ in \eqref{AppC6}, resulting in
\begin{align}\nonumber
&[\hat{\tilde\phi}^{\scriptscriptstyle-}(t,\vec x),\hat\varphi^{\dagger}(t,\vec y)]+[\hat\varphi(t,\vec x),\hat{\tilde\phi}^{\scriptscriptstyle-{\dagger}}(t,\vec y)]\\\nonumber
=\frac{\omega}{L^{3}}\underset{\substack{j\neq 0, l\neq 0\\j\neq-l}}{\sum\sum}\Biggl(&\hat a_{j}^{\dagger}e^{\frac{i}{\hbar}\gamma_{j}t}e^{-\frac{i}{\hbar}(\vec k_{j}+\vec k_{l})\vec y}e^{\frac{i}{\hbar}\vec k_{l}\vec x}\left(\frac{\tilde M_{j,-l}^{\scriptscriptstyle-}+\tilde N_{-l,j}^{\scriptscriptstyle-}}{2}d_{l}+
\frac{\tilde N_{j,j+l}^{\scriptscriptstyle-}+\tilde M_{j+l,j}^{\scriptscriptstyle-}}{2}c_{j+l}\right)\\\label{commdagger-}
+&\hat a_{j}^{}e^{-\frac{i}{\hbar}\gamma_{j}t}e^{\frac{i}{\hbar}(\vec k_{j}+\vec k_{l})\vec x}e^{-\frac{i}{\hbar}\vec k_{l}\vec y}\left(\frac{\tilde M_{j,-l}^{\scriptscriptstyle-}+\tilde N_{-l,j}^{\scriptscriptstyle-}}{2}d_{l}+
\frac{\tilde N_{j,j+l}^{\scriptscriptstyle-}+\tilde M_{j+l,j}^{\scriptscriptstyle-}}{2}c_{j+l}\right)\Biggr).
\end{align}
Combining \eqref{commdagger+} and \eqref{commdagger-}, for the operator $\hat{\tilde\phi}(t,\vec x)=\hat{\tilde\phi}^{\scriptscriptstyle+}(t,\vec x)+\hat{\tilde\phi}^{\scriptscriptstyle-}(t,\vec x)$ we get \eqref{commdagger}.

\section*{Appendix~D}
Let us start with equality \eqref{lasteqcoeff}. In explicit form (with \eqref{L-def} and definitions \eqref{qdef3}, \eqref{qdef4}), the expression in the l.h.s. of Eq.~\eqref{lasteqcoeff} looks like
\begin{align}\nonumber
&-\frac{i}{\hbar\gamma_{a}}\left(\left(\sqrt{\gamma_{a}^{2}+\omega^{2}}+\gamma_{a}\right)\left(d_{a+b}d_{b}+c_{a+b}c_{b}-c_{a+b}d_{b}\right)
-\omega\left(d_{a+b}d_{b}+c_{a+b}c_{b}-d_{a+b}c_{b}\right)\right)c_{b}\\\label{AppDexpr}
&+\frac{i}{\hbar\gamma_{b}}\left(\left(\sqrt{\gamma_{b}^{2}+\omega^{2}}+\gamma_{b}\right)\left(d_{a+b}d_{a}+c_{a+b}c_{a}-c_{a+b}d_{a}\right)
-\omega\left(d_{a+b}d_{a}+c_{a+b}c_{a}-d_{a+b}c_{a}\right)\right)c_{a}.
\end{align}
Using the relations
\begin{equation}\label{AppDrels}
c_{j}d_{j}=\frac{\omega}{2\gamma_{j}},\qquad c_{j}^{2}=\frac{\omega^{2}}{2\gamma_{j}\left(\sqrt{\gamma_{j}^{2}+\omega^{2}}-\gamma_{j}\right)}
=\frac{\sqrt{\gamma_{j}^{2}+\omega^{2}}+\gamma_{j}}{2\gamma_{j}}
\end{equation}
following from \eqref{cddef}, one can get for \eqref{AppDexpr}
\begin{align}\nonumber
-\frac{ic_{a+b}}{2\hbar\gamma_{a}\gamma_{b}}\Biggl[&\left(\sqrt{\gamma_{a}^{2}+\omega^{2}}+\gamma_{a}\right)\left(\omega\left(\frac{d_{a+b}}{c_{a+b}}-1\right)+\sqrt{\gamma_{b}^{2}
+\omega^{2}}+\gamma_{b}\right)\\\nonumber&-\omega\left(\left(\sqrt{\gamma_{b}^{2}+\omega^{2}}
+\gamma_{b}\right)\left(1-\frac{d_{a+b}}{c_{a+b}}\right)+\omega\frac{d_{a+b}}{c_{a+b}}\right)\\\nonumber
&-\left(\sqrt{\gamma_{b}^{2}+\omega^{2}}+\gamma_{b}\right)\left(\omega\left(\frac{d_{a+b}}{c_{a+b}}-1\right)+\sqrt{\gamma_{a}^{2}
+\omega^{2}}+\gamma_{a}\right)\\&+\omega\left(
\left(\sqrt{\gamma_{a}^{2}+\omega^{2}}+\gamma_{a}\right)\left(1-\frac{d_{a+b}}{c_{a+b}}\right)+\omega\frac{d_{a+b}}{c_{a+b}}\right)
\Biggr].
\end{align}
Combining the similar terms, one can easily check that the latter expression is equal to zero identically. Thus, Eq.~\eqref{lasteqcoeff} indeed holds.

Now we turn to equality \eqref{lastbut1eqcoeff}. In explicit form (with \eqref{L+def}, \eqref{L-def} and definitions \eqref{qdef1}--\eqref{qdef4}), the expression in the l.h.s. of Eq.~\eqref{lastbut1eqcoeff} looks like
\begin{align}\nonumber
&-\frac{i}{\hbar(\gamma_{a}+\gamma_{b})}\left(\left(\sqrt{\left(\gamma_{a}+\gamma_{b}\right)^{2}+\omega^{2}}+\gamma_{a}+\gamma_{b}\right)\left(c_{a}c_{b}-c_{a}d_{b}-d_{a}c_{b}\right)
-\omega\left(d_{a}d_{b}-c_{a}d_{b}-d_{a}c_{b}\right)\right)c_{b}\\\label{AppDexpr2}
&+\frac{i}{\hbar\gamma_{b}}\left(\left(\sqrt{\gamma_{b}^{2}+\omega^{2}}+\gamma_{b}\right)\left(d_{a+b}d_{a}+c_{a+b}c_{a}-c_{a+b}d_{a}\right)
-\omega\left(d_{a+b}d_{a}+c_{a+b}c_{a}-d_{a+b}c_{a}\right)\right)c_{a+b}.
\end{align}
Using relations \eqref{AppDrels}, one can get for \eqref{AppDexpr2}
\begin{align}\nonumber
-\frac{ic_{a}}{2\hbar(\gamma_{a}+\gamma_{b})\gamma_{b}}\Biggl[&\left(\sqrt{\left(\gamma_{a}+\gamma_{b}\right)^{2}+\omega^{2}}+\gamma_{a}+\gamma_{b}\right)
\left(\left(\sqrt{\gamma_{b}^{2}+\omega^{2}}+\gamma_{b}\right)\left(1-\frac{d_{a}}{c_{a}}\right)-\omega\right)\\\nonumber
&-\omega\left(-\left(\sqrt{\gamma_{b}^{2}+\omega^{2}}
+\gamma_{b}\right)\frac{d_{a}}{c_{a}}+\omega\left(\frac{d_{a}}{c_{a}}-1\right)\right)\\\nonumber
&-\left(\sqrt{\gamma_{b}^{2}+\omega^{2}}+\gamma_{b}\right)\left(\left(\sqrt{\left(\gamma_{a}+\gamma_{b}\right)^{2}+\omega^{2}}+\gamma_{a}+\gamma_{b}\right)
\left(1-\frac{d_{a}}{c_{a}}\right)+\omega\frac{d_{a}}{c_{a}}\right)\\&+\omega\left(\sqrt{\left(\gamma_{a}+\gamma_{b}\right)^{2}+\omega^{2}}
+\gamma_{a}+\gamma_{b}+\omega\left(\frac{d_{a}}{c_{a}}-1\right)\right)
\Biggr].
\end{align}
Combining the similar terms, one can easily check that the latter expression is equal to zero identically. Thus, Eq.~\eqref{lastbut1eqcoeff} also holds.

\section*{Supplementary material}
Supplementary material contains the .xml file for the computer algebra system Maxima with the wxMaxima interface (https://maxima.sourceforge.io), which was used to compute some coefficients in the commutators presented in this paper; and its .pdf version allowing one to look at the result of computation without installing Maxima.

\end{document}